\documentclass{ws-ijmpa}

\usepackage{amsthm}
\usepackage{amsmath}
\usepackage{amsfonts}
\usepackage{xspace}

\usepackage{graphicx}
\usepackage{multirow}
\usepackage{lineno}
\usepackage{hyperref}
\usepackage[capitalise]{cleveref}
\usepackage[normalem]{ulem}
\hypersetup{
	colorlinks=true,
	linkcolor=blue,
	citecolor=blue,
	urlcolor=blue
}
\newcommand{\customUL}{\bgroup\markoverwith{\textcolor{blue}{\rule[-0.3ex]{2pt}{0.6pt}}}\ULon}

\usepackage{booktabs}

\usepackage{hyperref}
\newcommand{\orcid}[1]{\href{https://orcid.org/#1}{ORCID: #1}}

\usepackage{subcaption}

\usepackage{placeins}

\usepackage{lmodern}

\usepackage[super,compress]{cite}

\def\nol {$\ensuremath{0\ell$\xspace}}
\def\l {$\ensuremath{1\ell$\xspace}}
\def\ll {$\ensuremath{2\ell$\xspace}}
\def\llSC {$\ensuremath{2\ell^\text{ss}$\xspace}}
\def\lll {$\ensuremath{3\ell}$\xspace}
\def\llll {$\ensuremath{4\ell}$\xspace}

\def \sbsbModel{$\ensuremath{\tilde{b}_{1} \rightarrow t W \tilde{\chi}_{1}^{0}$\xspace}}

\newcommand{\pt}{\ensuremath{p_{T}}\xspace}

\newcommand{\met}{\ensuremath{E_\mathrm{T}^\mathrm{miss}}\xspace}

\newcommand{\meff}{\ensuremath{m_\mathrm{eff}}\xspace}

\def\sbottom{\ensuremath{\tilde{b}}}

\def\ggino{\ensuremath{\mathchoice%
      {\displaystyle\raise.4ex\hbox{$\displaystyle\tilde\chi$}}%
         {\textstyle\raise.4ex\hbox{$\textstyle\tilde\chi$}}%
       {\scriptstyle\raise.3ex\hbox{$\scriptstyle\tilde\chi$}}%
 {\scriptscriptstyle\raise.3ex\hbox{$\scriptscriptstyle\tilde\chi$}}}}

\def\chinop{\ensuremath{\mathchoice%
      {\displaystyle\raise.4ex\hbox{$\displaystyle\tilde\chi^+$}}%
         {\textstyle\raise.4ex\hbox{$\textstyle\tilde\chi^+$}}%
       {\scriptstyle\raise.3ex\hbox{$\scriptstyle\tilde\chi^+$}}%
 {\scriptscriptstyle\raise.3ex\hbox{$\scriptscriptstyle\tilde\chi^+$}}}}
\def\chinom{\ensuremath{\mathchoice%
      {\displaystyle\raise.4ex\hbox{$\displaystyle\tilde\chi^-$}}%
         {\textstyle\raise.4ex\hbox{$\textstyle\tilde\chi^-$}}%
       {\scriptstyle\raise.3ex\hbox{$\scriptstyle\tilde\chi^-$}}%
 {\scriptscriptstyle\raise.3ex\hbox{$\scriptscriptstyle\tilde\chi^-$}}}}
\def\chinopm{\ensuremath{\mathchoice%
      {\displaystyle\raise.4ex\hbox{$\displaystyle\tilde\chi^\pm$}}%
         {\textstyle\raise.4ex\hbox{$\textstyle\tilde\chi^\pm$}}%
       {\scriptstyle\raise.3ex\hbox{$\scriptstyle\tilde\chi^\pm$}}%
 {\scriptscriptstyle\raise.3ex\hbox{$\scriptscriptstyle\tilde\chi^\pm$}}}}
\def\chinomp{\ensuremath{\mathchoice%
      {\displaystyle\raise.4ex\hbox{$\displaystyle\tilde\chi^\mp$}}%
         {\textstyle\raise.4ex\hbox{$\textstyle\tilde\chi^\mp$}}%
       {\scriptstyle\raise.3ex\hbox{$\scriptstyle\tilde\chi^\mp$}}%
 {\scriptscriptstyle\raise.3ex\hbox{$\scriptscriptstyle\tilde\chi^\mp$}}}}

\def\chinoonep{\ensuremath{\mathchoice%
      {\displaystyle\raise.4ex\hbox{$\displaystyle\tilde\chi^+_1$}}%
         {\textstyle\raise.4ex\hbox{$\textstyle\tilde\chi^+_1$}}%
       {\scriptstyle\raise.3ex\hbox{$\scriptstyle\tilde\chi^+_1$}}%
 {\scriptscriptstyle\raise.3ex\hbox{$\scriptscriptstyle\tilde\chi^+_1$}}}}
\def\chinoonem{\ensuremath{\mathchoice%
      {\displaystyle\raise.4ex\hbox{$\displaystyle\tilde\chi^-_1$}}%
         {\textstyle\raise.4ex\hbox{$\textstyle\tilde\chi^-_1$}}%
       {\scriptstyle\raise.3ex\hbox{$\scriptstyle\tilde\chi^-_1$}}%
 {\scriptscriptstyle\raise.3ex\hbox{$\scriptscriptstyle\tilde\chi^-_1$}}}}
\def\chinoonepm{\ensuremath{\mathchoice%
      {\displaystyle\raise.4ex\hbox{$\displaystyle\tilde\chi^\pm_1$}}%
         {\textstyle\raise.4ex\hbox{$\textstyle\tilde\chi^\pm_1$}}%
       {\scriptstyle\raise.3ex\hbox{$\scriptstyle\tilde\chi^\pm_1$}}%
 {\scriptscriptstyle\raise.3ex\hbox{$\scriptscriptstyle\tilde\chi^\pm_1$}}}}

\def\chinotwop{\ensuremath{\mathchoice%
      {\displaystyle\raise.4ex\hbox{$\displaystyle\tilde\chi^+_2$}}%
         {\textstyle\raise.4ex\hbox{$\textstyle\tilde\chi^+_2$}}%
       {\scriptstyle\raise.3ex\hbox{$\scriptstyle\tilde\chi^+_2$}}%
 {\scriptscriptstyle\raise.3ex\hbox{$\scriptscriptstyle\tilde\chi^+_2$}}}}
\def\chinotwom{\ensuremath{\mathchoice%
      {\displaystyle\raise.4ex\hbox{$\displaystyle\tilde\chi^-_2$}}%
         {\textstyle\raise.4ex\hbox{$\textstyle\tilde\chi^-_2$}}%
       {\scriptstyle\raise.3ex\hbox{$\scriptstyle\tilde\chi^-_2$}}%
 {\scriptscriptstyle\raise.3ex\hbox{$\scriptscriptstyle\tilde\chi^-_2$}}}}
\def\chinotwopm{\ensuremath{\mathchoice%
      {\displaystyle\raise.4ex\hbox{$\displaystyle\tilde\chi^\pm_2$}}%
         {\textstyle\raise.4ex\hbox{$\textstyle\tilde\chi^\pm_2$}}%
       {\scriptstyle\raise.3ex\hbox{$\scriptstyle\tilde\chi^\pm_2$}}%
 {\scriptscriptstyle\raise.3ex\hbox{$\scriptscriptstyle\tilde\chi^\pm_2$}}}}

\def\nino{\ensuremath{\mathchoice%
      {\displaystyle\raise.4ex\hbox{$\displaystyle\tilde\chi^0$}}%
         {\textstyle\raise.4ex\hbox{$\textstyle\tilde\chi^0$}}%
       {\scriptstyle\raise.3ex\hbox{$\scriptstyle\tilde\chi^0$}}%
 {\scriptscriptstyle\raise.3ex\hbox{$\scriptscriptstyle\tilde\chi^0$}}}}

\def\ninoone{\ensuremath{\mathchoice%
      {\displaystyle\raise.4ex\hbox{$\displaystyle\tilde\chi^0_1$}}%
         {\textstyle\raise.4ex\hbox{$\textstyle\tilde\chi^0_1$}}%
       {\scriptstyle\raise.3ex\hbox{$\scriptstyle\tilde\chi^0_1$}}%
 {\scriptscriptstyle\raise.3ex\hbox{$\scriptscriptstyle\tilde\chi^0_1$}}}}
\def\ninotwo{\ensuremath{\mathchoice%
      {\displaystyle\raise.4ex\hbox{$\displaystyle\tilde\chi^0_2$}}%
         {\textstyle\raise.4ex\hbox{$\textstyle\tilde\chi^0_2$}}%
       {\scriptstyle\raise.3ex\hbox{$\scriptstyle\tilde\chi^0_2$}}%
 {\scriptscriptstyle\raise.3ex\hbox{$\scriptscriptstyle\tilde\chi^0_2$}}}}
\def\ninothree{\ensuremath{\mathchoice%
      {\displaystyle\raise.4ex\hbox{$\displaystyle\tilde\chi^0_3$}}%
         {\textstyle\raise.4ex\hbox{$\textstyle\tilde\chi^0_3$}}%
       {\scriptstyle\raise.3ex\hbox{$\scriptstyle\tilde\chi^0_3$}}%
 {\scriptscriptstyle\raise.3ex\hbox{$\scriptscriptstyle\tilde\chi^0_3$}}}}
\def\ninofour{\ensuremath{\mathchoice%
      {\displaystyle\raise.4ex\hbox{$\displaystyle\tilde\chi^0_4$}}%
         {\textstyle\raise.4ex\hbox{$\textstyle\tilde\chi^0_4$}}%
       {\scriptstyle\raise.3ex\hbox{$\scriptstyle\tilde\chi^0_4$}}%
 {\scriptscriptstyle\raise.3ex\hbox{$\scriptscriptstyle\tilde\chi^0_4$}}}}

\def\sbottom{\ensuremath{\tilde{b}}}

\graphicspath{{Figures/}}

\begin{document}
\markboth{A. Tudorache, O. Ducu}{Search potential for $\tilde{b}_{1} \rightarrow t W \tilde{\chi}_{1}^{0}$ via $\tilde{\chi}^{\pm}_{1}$ at the LHC and HL-LHC, with multi-lepton signatures
}

%
%

\title{Search potential for $\tilde{b}_{1} \rightarrow tW \tilde{\chi}_{1}^{0}$ via $\tilde{\chi}^{\pm}_{1}$ at the LHC and HL-LHC, with multi-lepton signatures
}

\author{Alexandra Tudorache~\orcid{0000-0001-6307-1437}, Otilia Ducu~\orcid{0000-0001-5914-0524}$^{*}$
}

\address{
	\vspace{0.2cm}
	Horia Hulubei National Institute of Physics and Nuclear Engineering (IFIN-HH)\\
	Department of Particle Physics\\
	Magurele, Ilfov, Romania (077125)\\
	\vspace{0.2cm}
	otilia.ducu@gmail.com ($^{*}$corresponding author)\\
	alexandra.tudorache@cern.ch
}

\maketitle


\begin{abstract}

The search potential for sbottom pair production in an R-parity conserving scenario is explored in multi-lepton final states at LHC Run-3 and the HL-LHC. In this model, the sbottom decays via a chargino, $\tilde{b}_1 \to t \tilde{\chi}_1^\pm$, with a branching ratio (BR) of 100\%. The chargino subsequently decays into a $W$ boson and the lightest neutralino, $\tilde{\chi}_1^\pm \to W \tilde{\chi}_1^0$, also with a BR of 100\%. Two mass configurations are considered for $\tilde{\chi}_1^0$: a fixed value of 50~GeV, and a scenario where $\tilde{\chi}_1^\pm$ is 100~GeV heavier than $\tilde{\chi}_1^0$. The study follows the ATLAS object definitions and event selection criteria as detailed in Refs.~\citen{ATLAS:2021yyr, Ducu:2024arr, ATLAS:2019fag}, extending the analysis presented in Ref.~\citen{Ducu:2024arr}. Results are presented as projected exclusion limits in the $\tilde{b}_1$--$\tilde{\chi}_1^0$ and $\tilde{b}_1$--$\tilde{\chi}_1^\pm$ mass planes for three center-of-mass energies (13~TeV, 13.6~TeV, and 14~TeV) and three integrated luminosity scenarios (139~fb$^{-1}$, 300~fb$^{-1}$, and 3000~fb$^{-1}$). Depending on the $\tilde{\chi}_1^0$ mass assumption, sbottom masses up to 1.28~TeV can be excluded at the HL-LHC with 3000~fb$^{-1}$. These projections highlight the impact of alternative mass configurations on the sensitivity of multi-lepton \sbsbModel searches and provide guidance for future strategies targeting challenging SUSY scenarios at the LHC.

\end{abstract}

\ccode{PACS numbers:11.30.Pb,12.60.Jv}


%
%

\section{Introduction}	
\label{sec:intro}

\begin{figure}[t!]
\centering
    \begin{subfigure}[t]{0.66\columnwidth}
        \centering
        \includegraphics[width=\linewidth]{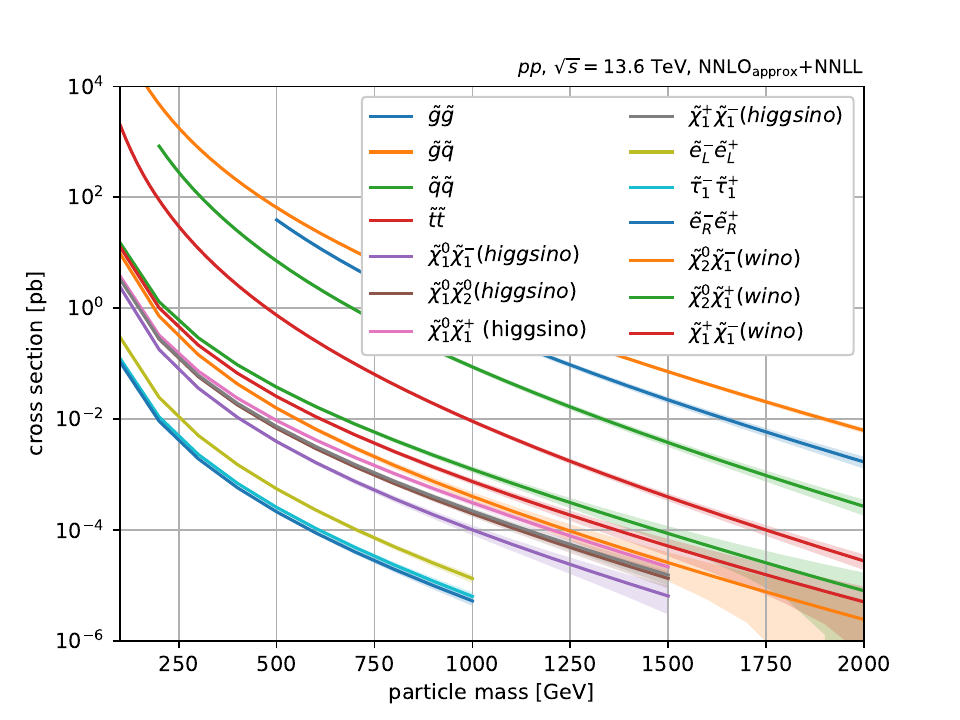}
        \caption{}
        \label{fig:SUSY_xsec}
    \end{subfigure}%
    \hfill
    \begin{subfigure}[t]{0.3\columnwidth}
        \centering
        \includegraphics[width=\linewidth]{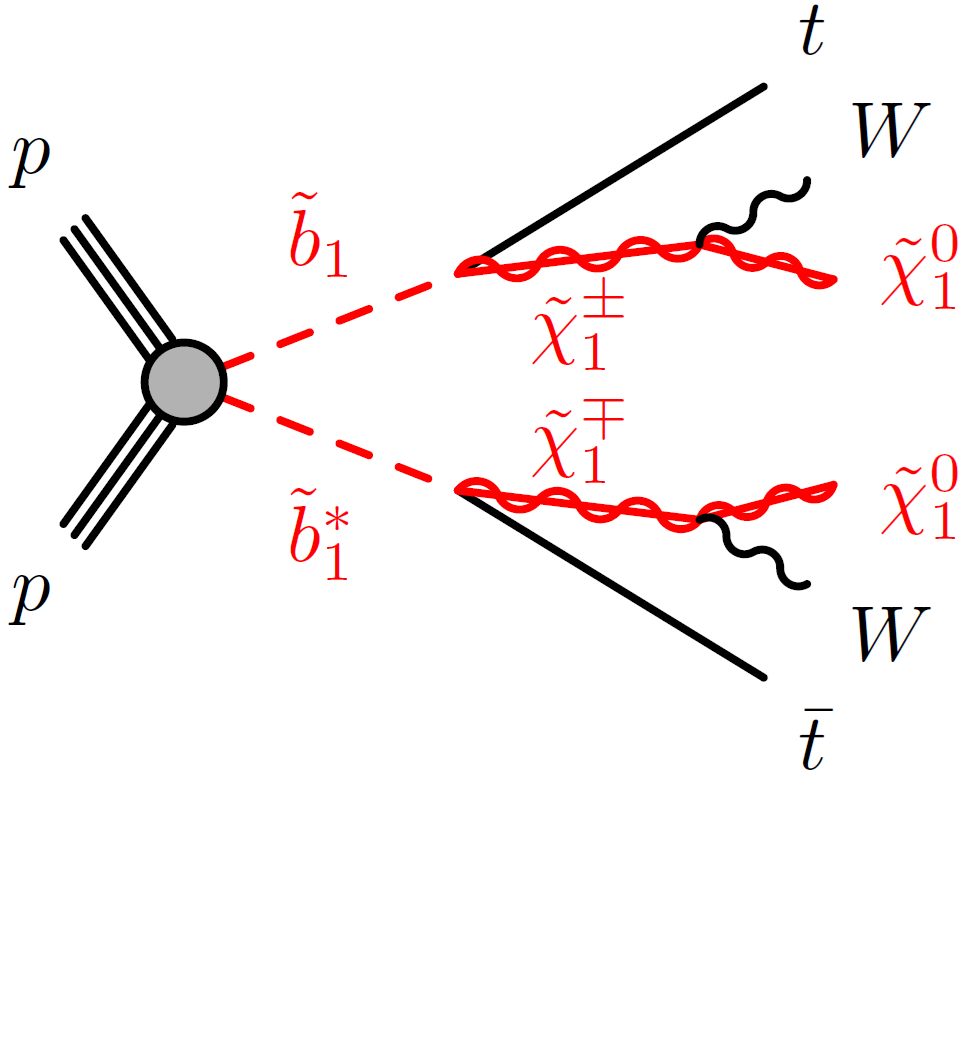}
        \caption{}
        \label{fig:sbottom_Diag}
    \end{subfigure}
\vspace{0.cm}
\caption{
    (a) Plot showing the cross-section values at $\sqrt s = 13.6$~TeV\cite{ATLAS:XSec,Borschensky:2014cia}. 
    (b) Feynman diagram for the sbottom pair production model\cite{ATLAS:2019fag}.
}
\label{fig:combined_figure}
\end{figure}

In high-energy physics, one of the most intriguing theoretical extensions of the Standard Model (SM) is Supersymmetry (SUSY)~\cite{Martin:1997ns}. By introducing a spectrum of superpartners, SUSY offers potential solutions to several major open problems in particle physics, including the SM gauge hierarchy problem and the nature of dark matter. Beyond high-energy physics, supersymmetry has found applications in fields ranging from quantum mechanics to cosmology. The Large Hadron Collider (LHC) program\cite{Evans:2008zzb} is actively searching for these predicted particles. In particular, the two general-purpose experiments, ATLAS\cite{ATLAS:2008xda} and CMS\cite{CMS:2008xjf}, have conducted extensive dedicated analyses targeting gluinos, squarks, and electroweakinos. Their typical production cross-sections at a center-of-mass energy of $\sqrt{s} = 13.6~\mathrm{TeV}$ are shown in~\cref{fig:SUSY_xsec}.

This paper focuses on sbottom pair production, illustrated in~\cref{fig:sbottom_Diag}. The R-parity~\cite{FARRAR1978575} is assumed to be conserved, and the sbottom has a one-step decay via a chargino, $\sbottom_1 \to t \chinoonepm$ (100\% BR), followed by $\chinoonepm \to W \ninoone$ (100\% BR). \ninoone\ is considered to be the lightest supersymmetric particle (LSP). To better understand the experimental search potential for sbottom quarks at LHC Run 3 and the HL-LHC, two sets of signal samples are considered: 
\begin{itemize}
    \item \textcolor{blue}{\bfseries\textit{Scenario~1}}: The chargino mass is set to be 100~GeV above the LSP mass, ensuring that the $W$ boson from its decay is always on-shell, as done in Ref.~\citen{Ducu:2024arr} and ATLAS Ref.~\citen{ATLAS:2019fag}. Only the two-body $\sbottom_1$ decays are considered in this case.
    \item \textcolor{blue}{\bfseries\textit{Scenario~2}}: The chargino mass is varied independently of the LSP mass, which is fixed at 50~GeV, as done in CMS Ref.~\citen{CMS:2020cpy}. In this scenario, both two-body and three-body $\sbottom_1$ decays are considered.
\end{itemize}

For the two-body decay $\sbottom_1 \to t \chinoonepm$ to occur, the sbottom mass must satisfy:
\begin{equation*}
m_{\sbottom_1} > m_t + m_{\chinoonepm} = 172.76~\text{GeV} + m_{\chinoonepm}.
\end{equation*}

In addition, the three-body decay $\sbottom_1 \to W b \chinoonepm$ is also kinematically allowed if:
\begin{equation*}
m_{\sbottom_1} > m_W + m_b + m_{\chinoonepm} = 80.38~\text{GeV} + 4.18~\text{GeV} + m_{\chinoonepm}.
\end{equation*}

Following the work in Ref.~\citen{Ducu:2024arr}, Monte Carlo (MC) samples for sbottom pair production are simulated using the \texttt{MadGraph}\cite{Alwall:2014hca} generator, version \verb|MG5_aMC_v3.5.5|~--~a framework that calculates particle interactions at the most fundamental level. The output from \texttt{MadGraph} is passed to \texttt{Pythia8}~\cite{Bierlich:2022pfr,Sjostrand:2014zea}, which handles the supersymmetric particle decays, parton showering, and hadronization. The $\sbottom_1$ mass is varied from 600 to 1700~GeV, and depending on the scenario considered, either the LSP or $\chinoonepm$ mass ranges from 50 to 1425~GeV~\cite{Ducu:2024arr,ATLAS:2019fag,CMS:2020cpy}. The masses of the first- and second-generation squarks, as well as the gluino, are decoupled and set to 4.5~TeV. Two sets of MC event samples are produced at three different proton-proton collision energies: $\sqrt{s} = 13$~TeV (LHC Run-2), 13.6~TeV (LHC Run-3), and 14~TeV (anticipated for the HL-LHC).

As in Ref.~\citen{Ducu:2024arr}, the signal event samples are processed via \texttt{DELPHES}~\cite{deFavereau:2013fsa}, a fast simulation framework for the ATLAS detector. The ATLAS parameter card from \texttt{DELPHES} is utilized with specific modifications to align with the following selection criteria: jets are reconstructed using the anti-$k_\mathrm{T}$ algorithm~\cite{Cacciari:2008gp} with $R = 0.4$, and $b$-tagged jets are selected with a 70\% efficiency. The selection efficiencies for electrons and muons are updated in line with the results published in ATLAS Refs.~\citen{ATLAS:2023dxj,ATLAS:2020auj}.

The \texttt{SimpleAnalysis}~\cite{ATLAS:2022yru} framework is employed to analyze the \texttt{DELPHES} root files. Two analyses are considered: first, the ATLAS Ref.~\citen{ATLAS:2021yyr} study, which targets signal regions defined by events with at least four leptons; and second, the analysis from ATLAS Ref.~\citen{ATLAS:2019fag}, implemented using the same framework as in Ref.~\citen{Ducu:2024arr}, which focuses on signal regions with same-sign leptons and three leptons. In both cases, leptons correspond to electrons and muons. A sequence of selection criteria is applied to extract event yields at various stages, enabling the calculation of the region acceptance \(A\)\footnote{The acceptance \(A\) is defined as the fraction of generated events that pass the region selection.} and the signal significance \(Z\). Following recommended procedures, each event is assigned a weight incorporating the \texttt{MadGraph} generator weight, the production cross-section, and the integrated luminosity from ATLAS. Three integrated luminosity scenarios are considered: 139~fb\(^{-1}\) (end of LHC Run-2), 300~fb\(^{-1}\) (projected for the end of LHC Run-3), and 3000~fb\(^{-1}\) (anticipated for the HL-LHC). The production cross-section values are taken from Ref.~\citen{ATLAS:XSec}.

Following Refs.~\citen{Cowan:2010js,Ducu:2024arr}, $Z$ is evaluated using the formula in the following ~\cref{eqn:Zn_function}:
\begin{equation}
	Z = \pm\sqrt{2} \times \sqrt{ n \mathrm{ln} \frac{n(b+\sigma^2)}{b^2+n\sigma^2} - \frac{b^2}{\sigma^2}\mathrm{ln}\frac{b^2+n\sigma^2}{b(b+\sigma^2)}},
	\label{eqn:Zn_function}
\end{equation}
where $n$ is the total number of observed events, and $b$ is the expected number of background events, with $\sigma$ representing its associated uncertainty. The $Z$ value is a key indicator in particle physics, used to quantify the likelihood of rejecting a background-only hypothesis. A value of 1.64 is assumed to be sufficient to exclude a signal hypothesis, while a value of 5 to discover it\cite{Cowan:2010js}.

\section{ATLAS search with four leptons final states} 
\label{sec:ATLAS_4Lep_Sbottom}

ATLAS Ref.~\citen{ATLAS:2021yyr} presents a search for SUSY with at least four leptons ($4\ell$). The considered models include electroweakino pair prodution, slepton pair production and gluino pair production~--~no sbottom pair production model is considered. Unlike in ATLAS Ref.~\citen{ATLAS:2019fag}, the signal leptons are defined using loose isolation working points and soft $\pt$ requirements: $>7$~GeV for electrons and $>5$~GeV for muons. For the results shown in this paper, the selections from ATLAS Ref.~\citen{ATLAS:2021yyr}  are implemented and further used with the \texttt{SimpleAnalysis} framework.

\begin{table}[t!]
\caption{
    Event counting and acceptance values after the lepton selection discussed in the text, for the four representative signal mass points. The statistical uncertainty is also shown, and the events are normalized to an integrated luminosity of 139~fb$^{-1}$ at $\sqrt{s} = 13.6$ TeV. The \textit{Scenario~2} \sbsbModel model, which assumes an LSP mass of 50~GeV, is considered.
}
\label{tab:Sbottom13TeVLepSel_SS3Lep4Lep_CMSGrid}
 
\begin{subtable}[t]{\linewidth}
    \caption{Results obtained using the signal lepton definitions from ATLAS Ref.~\citen{ATLAS:2021yyr} (4$\ell$) analysis.
    }
    \label{subtab:Sbottom13TeVLepSel_4Lep}
\centering
{\tiny
\def\arraystretch{1.5}
\setlength{\tabcolsep}{0.0pc}
\begin{tabular*}{\textwidth}{@{\extracolsep{\fill}}llcccc}
\toprule
& & (750, 625, 50) & (750, 550, 50) & (750, 400, 50) & (750, 50, 50)  \\
& Selection & N events ($A$) & N events ($A$) & N events ($A$) & N events ($A$) \\
\midrule
 & All & 8869.2 $\pm$ 47.5 (100.0 \%) & 8868.9 $\pm$ 47.4 (100.0 \%) & 8866.5 $\pm$ 47.6 (100.0 \%) &  8869.8 $\pm$ 47.6 (100.0 \%) \\
\midrule\midrule
\multirow[c]{6}{*}[0in]{\rotatebox{90}{$\ell$ selection}}
 & $=0\ell$               & 2792.2 $\pm$ 26.7 (31.5 \%)   &    2548.4 $\pm$ 25.4 (28.7 \%)    & 2648.9 $\pm$ 26.0 (29.9 \%)  &   4449.4 $\pm$ 33.7 (50.2 \%) \\ 
 & $=1\ell$               & 3565.8 $\pm$ 30.2 (40.2 \%)   &    3551.2 $\pm$ 30.0 (40.0 \%)    & 3569.6 $\pm$ 30.2 (40.3 \%)  &   3418.8 $\pm$ 29.5 (38.5 \%) \\ 
 & $=2\ell$               & 1815.3 $\pm$ 21.5 (20.5 \%)   &    1944.9 $\pm$ 22.2 (21.9 \%)    & 1875.9 $\pm$ 21.9 (21.2 \%)  &   841.8 $\pm$ 14.7 (9.5 \%)   \\ 
 & $=2\ell^{\mathrm{SS}}$ & 687.4 $\pm$ 13.2 (7.7 \%)     &    765.1 $\pm$ 13.9 (8.6 \%)      & 746.2 $\pm$ 13.8 (8.4 \%)    &   237.8 $\pm$ 7.8 (2.7 \%)    \\ 
 & $=3\ell$               & 446.4 $\pm$ 10.7 (5.0 \%)     &    554.1 $\pm$ 11.8 (6.2 \%)      & 533.3 $\pm$ 11.7 (6.0 \%)    &   83.2 $\pm$ 4.6 (0.9 \%)     \\ 
 & $\geq4\ell$            & 49.2 $\pm$ 3.5 (0.6 \%)       &    90.2 $\pm$ 4.8 (1.0 \%)        & 80.2 $\pm$ 4.5 (0.9 \%)      &   1.3 $\pm$ 0.6 (0.0 \%)      \\ 
\bottomrule
\end{tabular*}
}
\end{subtable}

\vspace{1em}

\begin{subtable}[t]{0.5\linewidth}
    \caption{Results obtained using the signal lepton definitions, including the requirement that the two leading leptons have transverse momentum of $\pt > 20$~GeV, as defined in ATLAS Ref.~\citen{ATLAS:2019fag} (\llSC/\lll).}
    \label{subtab:Sbottom13TeVLepSel_SS3Lep}
\centering
{\scriptsize
\def\arraystretch{1.5}
\setlength{\tabcolsep}{0.0pc}
\begin{tabular*}{\textwidth}{@{\extracolsep{\fill}}llc}
\toprule
& &  (750, 550, 50)  \\
& Selection & N events ($A$) \\
\midrule
 & All &8868.9 $\pm$ 47.4 (100.0 \%) \\
\midrule\midrule
\multirow[c]{6}{*}[0in]{\rotatebox{90}{$\ell$ selection}}
 & $=0\ell$ & 3010.0 $\pm$ 27.6 (33.9 \%)    \\
 & $=1\ell$ & 2945.7 $\pm$ 27.3 (33.2 \%)   \\
 & $=2\ell$ & 1124.3 $\pm$ 16.9 (12.7 \%)     \\
 & $=2\ell^{\mathrm{SS}}$ & 407.6 $\pm$ 10.2 (4.6 \%)  \\
 & $=3\ell$     & 399.5 $\pm$ 10.1 (4.5 \%) \\
 & $\geq 4\ell$ & 65.6 $\pm$ 4.1 (0.7 \%) \\
\bottomrule
\end{tabular*}
}
\end{subtable}
\end{table}

Various results are shown in the following using four benchmark signal mass points, ($\sbottom_1$, $\chinoonepm$, $\ninoone$): (750, 625, 50)~GeV, (750, 550, 50)~GeV, (750, 400, 50)~GeV and (750, 50, 50)~GeV. The \textit{Scenario~2} \sbsbModel model, which assumes an LSP mass of 50~GeV, is considered. \cref{tab:Sbottom13TeVLepSel_SS3Lep4Lep_CMSGrid} shows event counts at various selection stages: zero leptons (\nol), one lepton (\l), two leptons (\ll), two same-sign leptons (\llSC), three leptons (\lll), and at least four leptons (\llll). Leptons are required to satisfy the signal lepton criteria from ATLAS Ref.~\citen{ATLAS:2021yyr}. To illustrate the impact of relaxing these criteria, results for one benchmark point using the tighter lepton definitions from ATLAS Ref.~\citen{ATLAS:2019fag} are also provided. For completeness, the total number of events in the signal MC samples is listed under the ``\textit{All}” entry.

As expected, a notable increase in the event yields after the \llll\ requirement is observed when using the looser lepton selections from ATLAS Ref.~\citen{ATLAS:2021yyr} compared to the tighter criteria in ATLAS Ref.~\citen{ATLAS:2019fag}. These relaxed selections also lead to increased yields in the other categories; however, in those regions, the detector background~--~particularly from fake/non-prompt leptons and electron charge misidentification~--~are expected to be too large, making good signal-to-background separation difficult to achieve. In future analyses, this could potentially be improved using machine learning techniques.

\begin{figure}[t!]
\begin{center}
    \includegraphics[width=0.32\columnwidth]{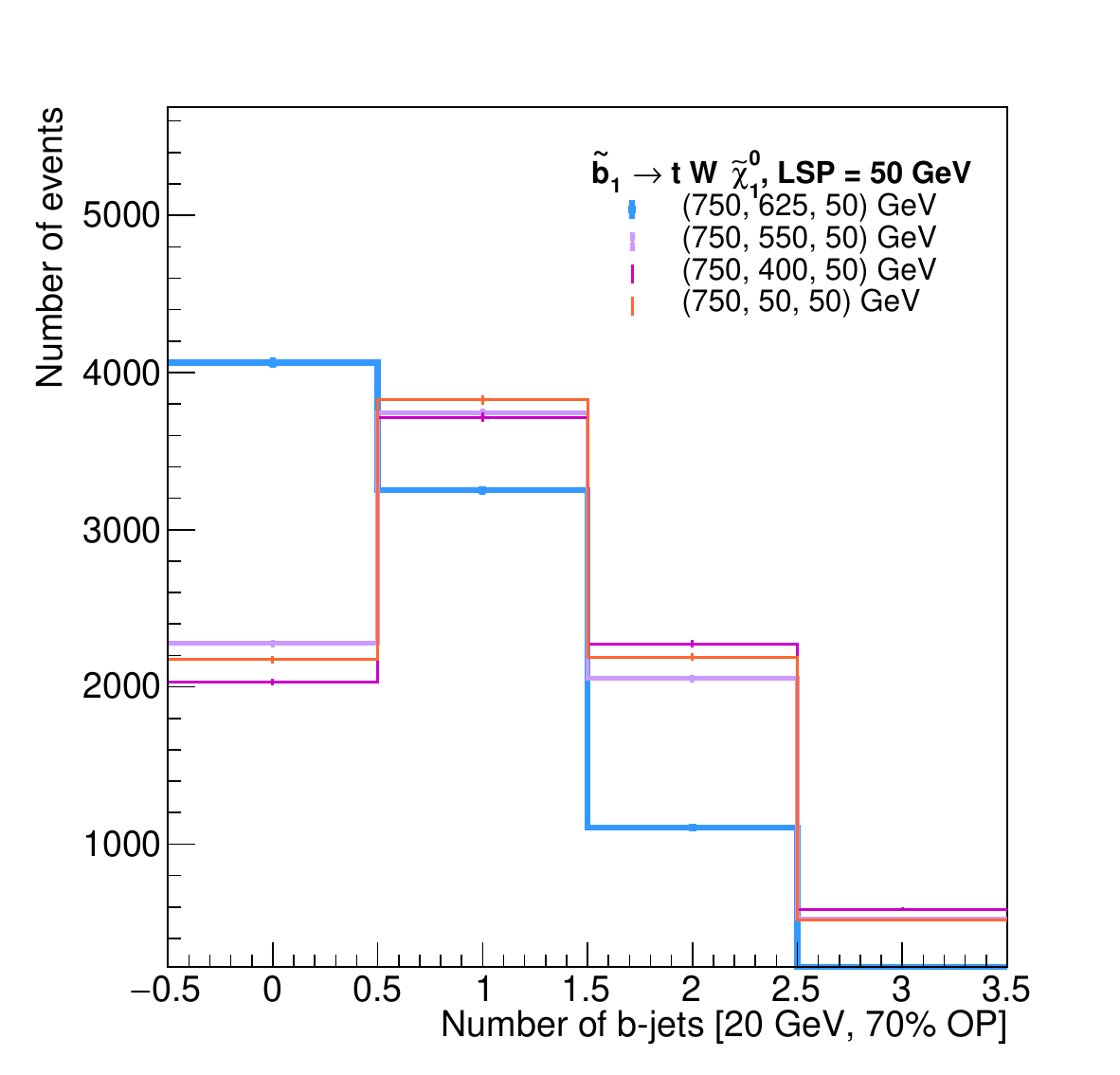}
    \includegraphics[width=0.32\columnwidth]{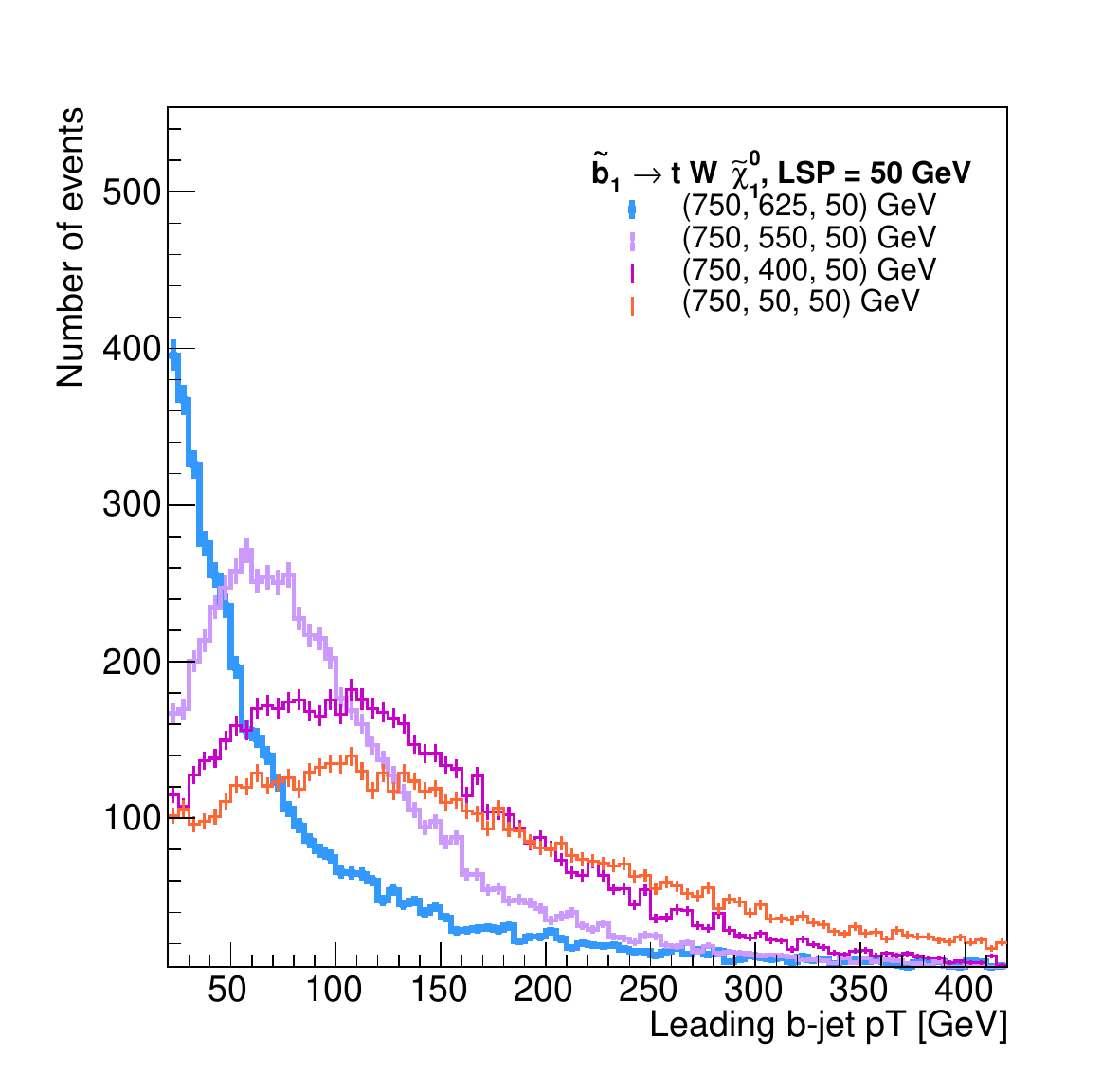}
    \includegraphics[width=0.32\columnwidth]{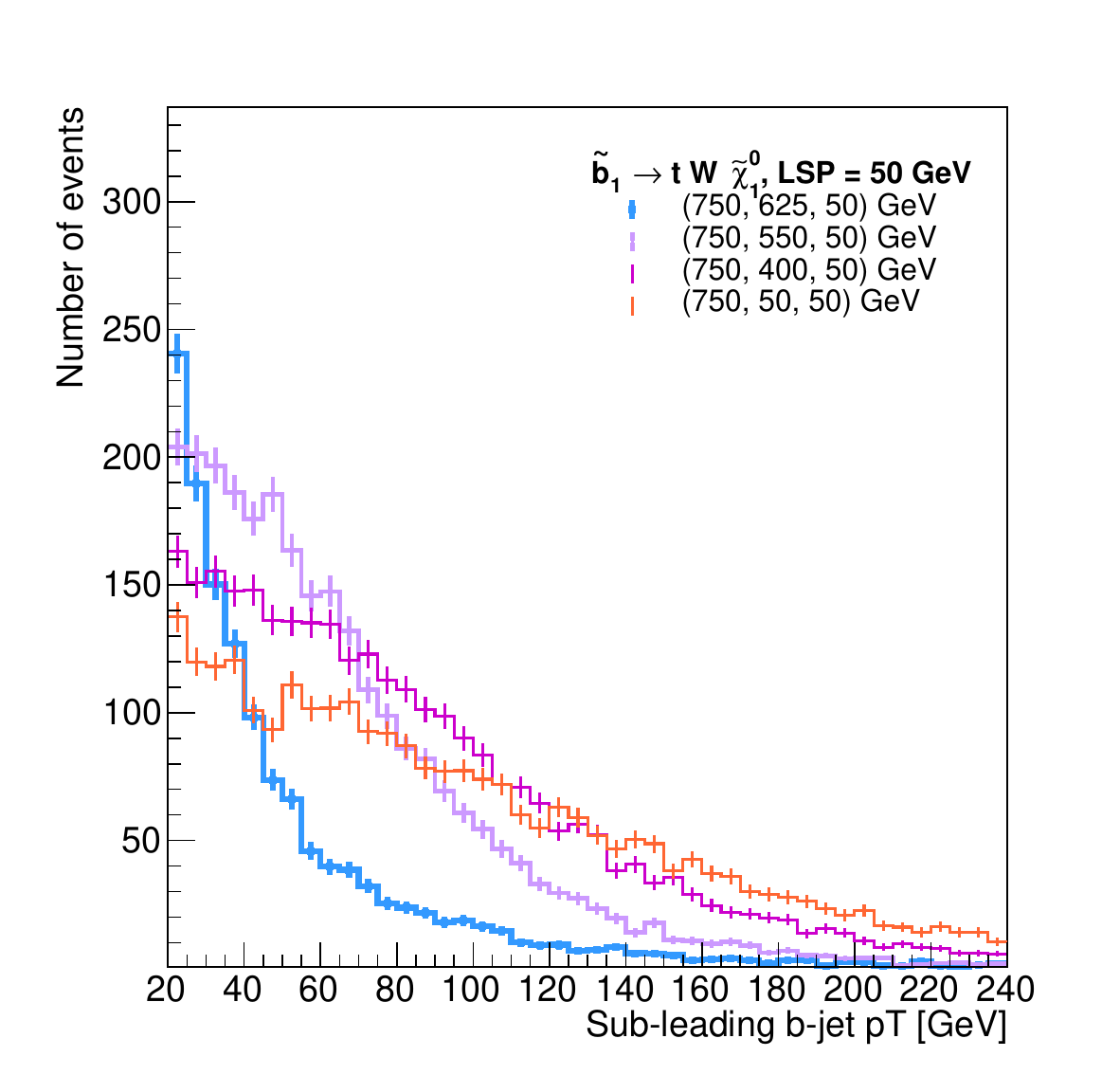}
    \includegraphics[width=0.32\columnwidth]{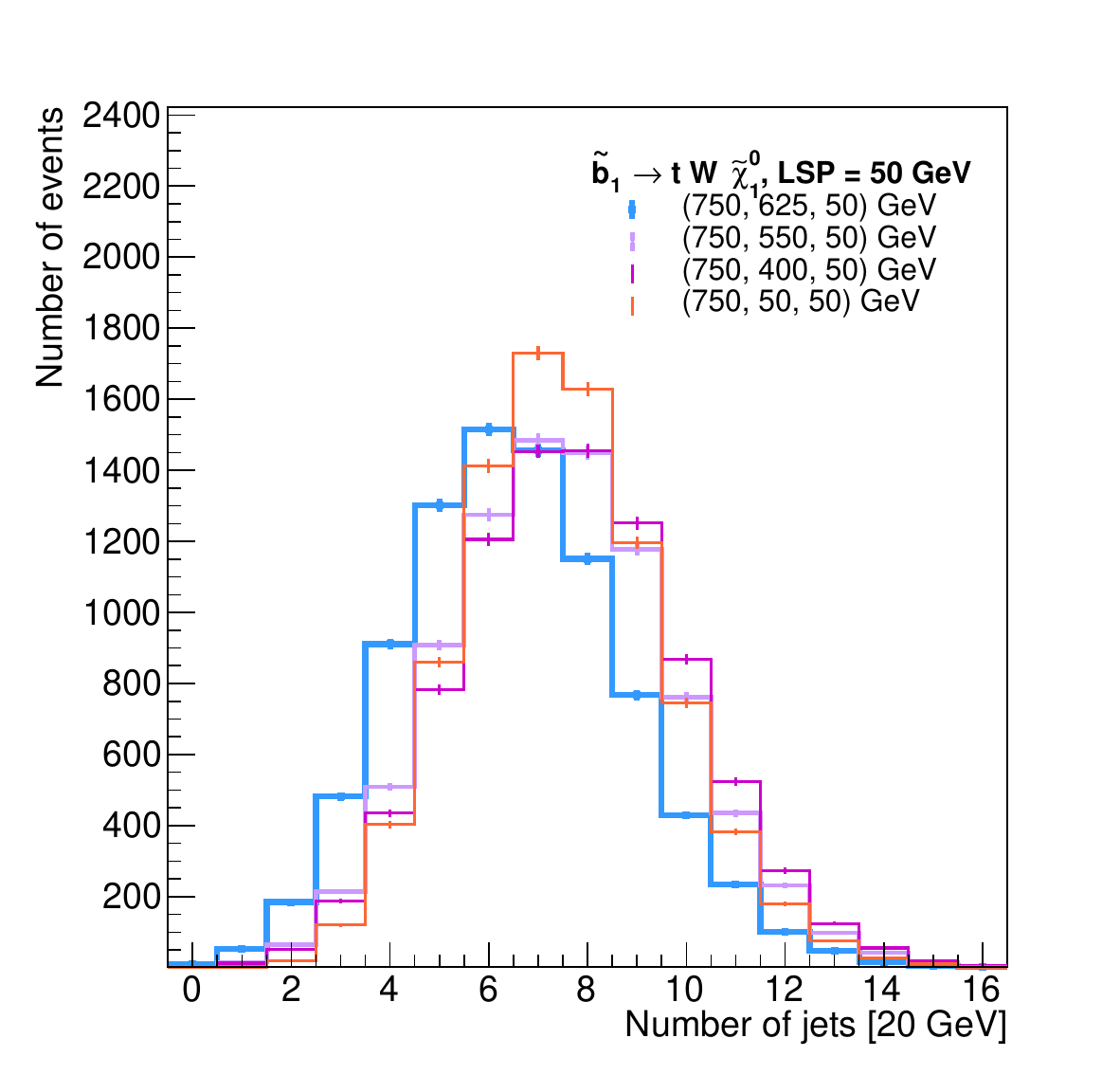}
    \includegraphics[width=0.32\columnwidth]{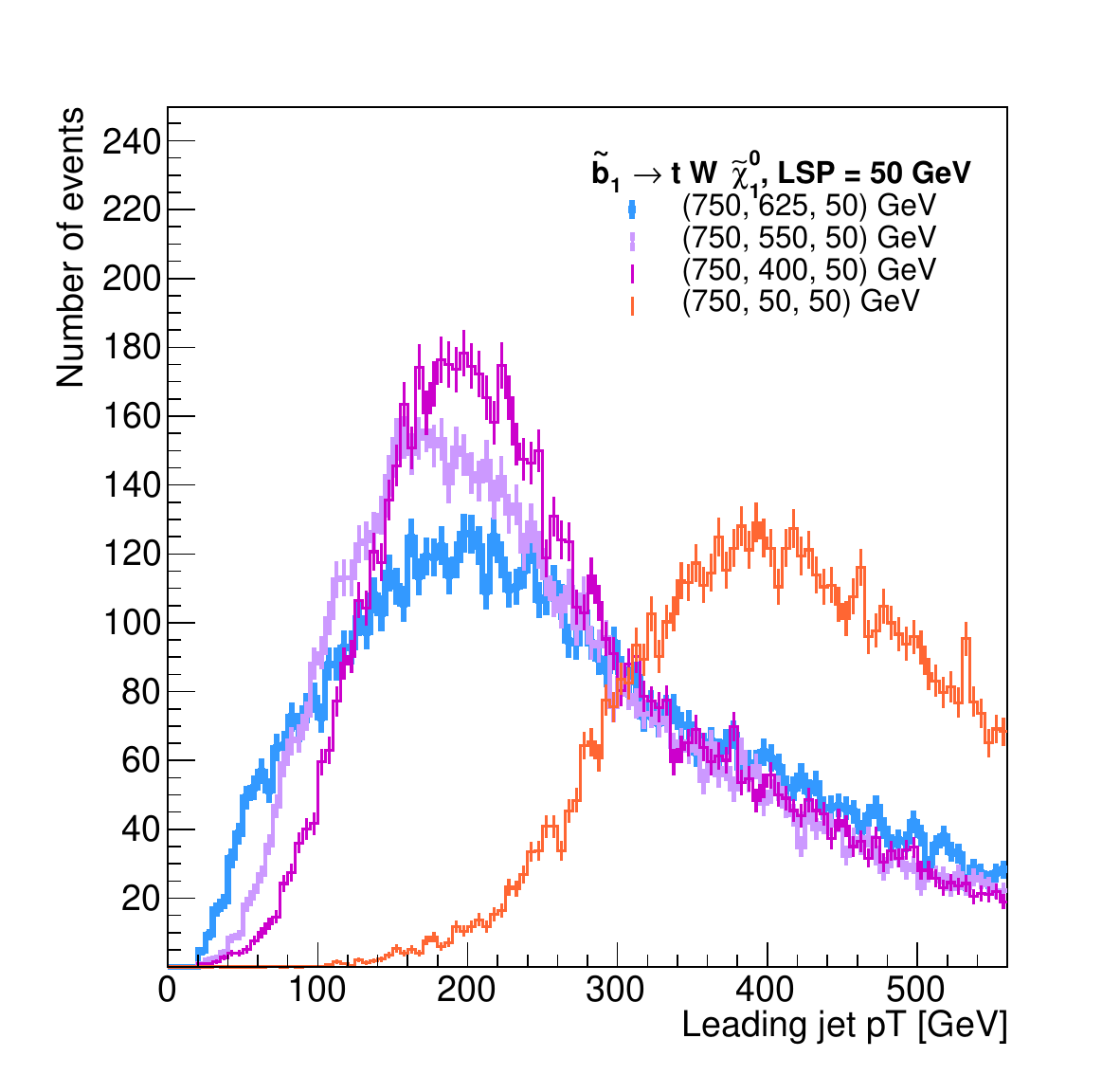}
    \includegraphics[width=0.32\columnwidth]{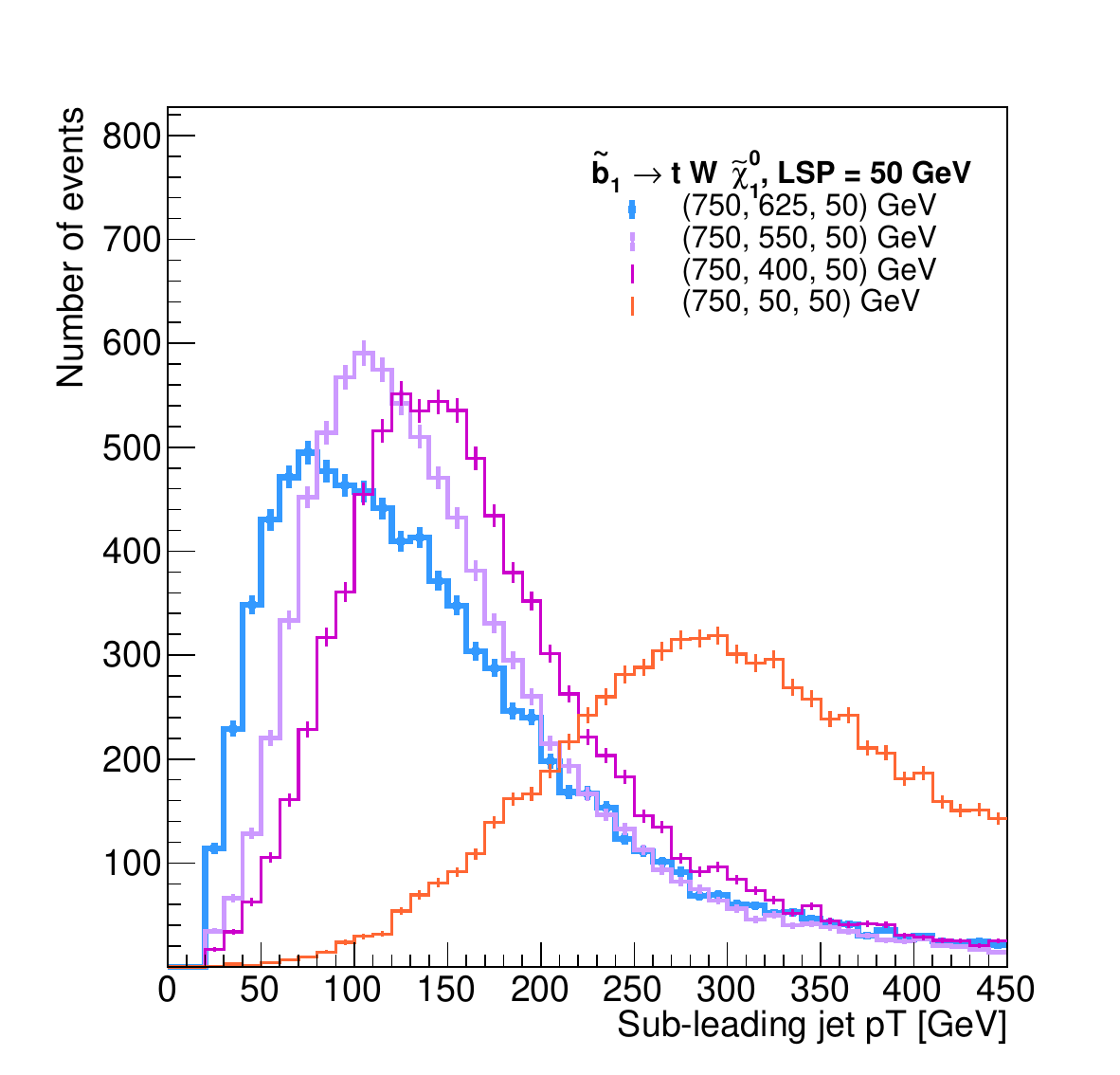}
    \includegraphics[width=0.32\columnwidth]{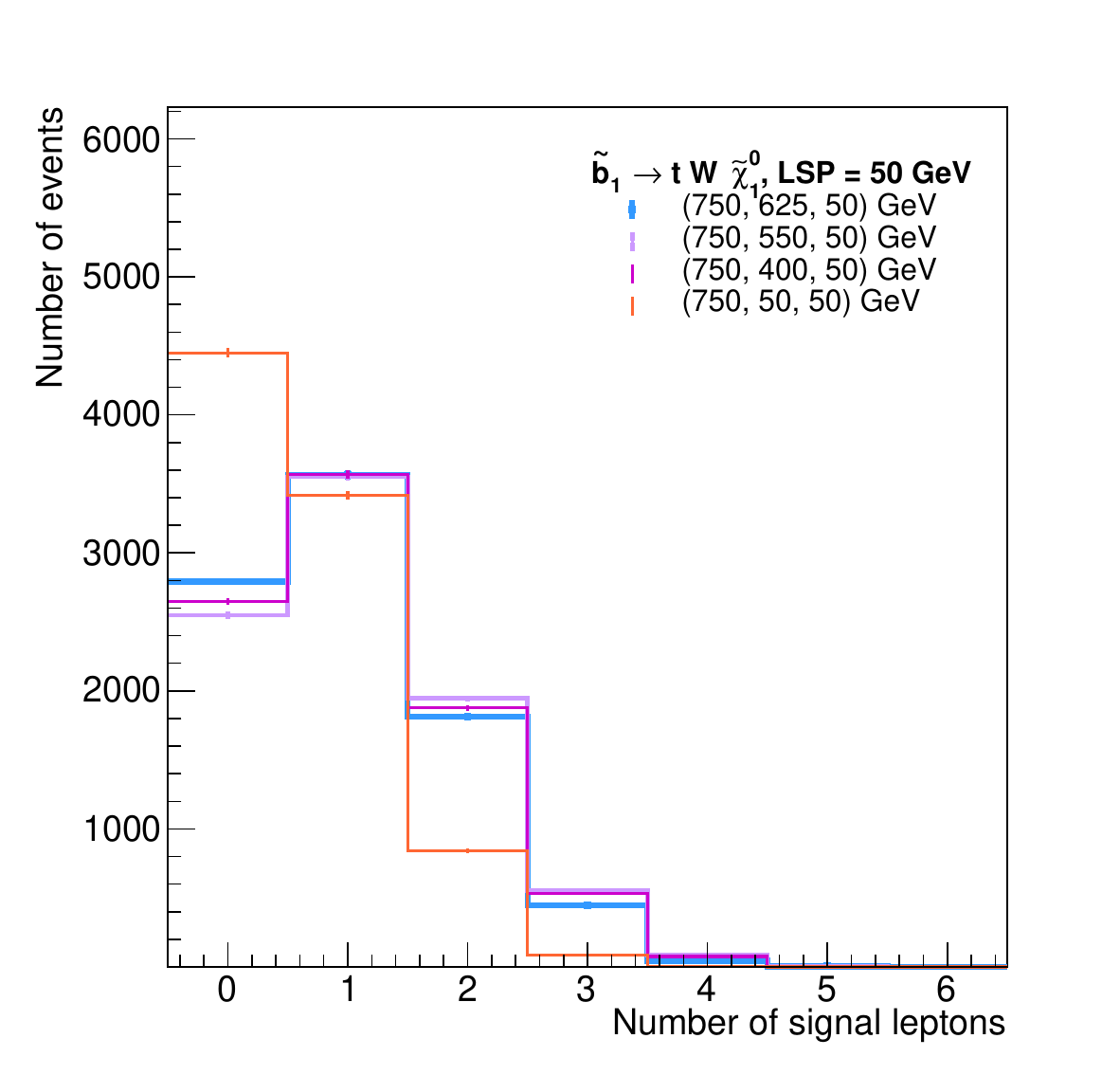}
    \includegraphics[width=0.32\columnwidth]{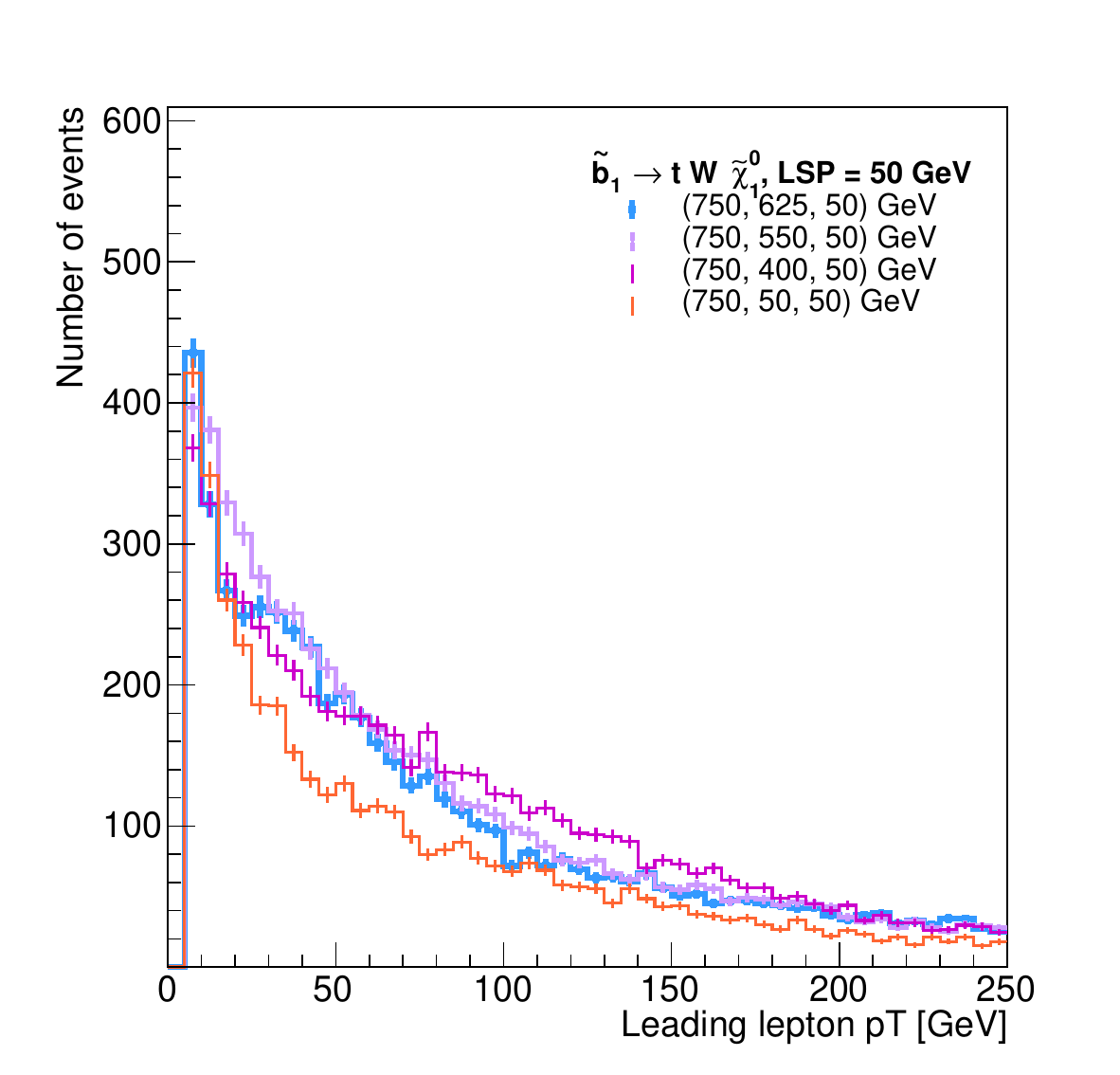}
    \includegraphics[width=0.32\columnwidth]{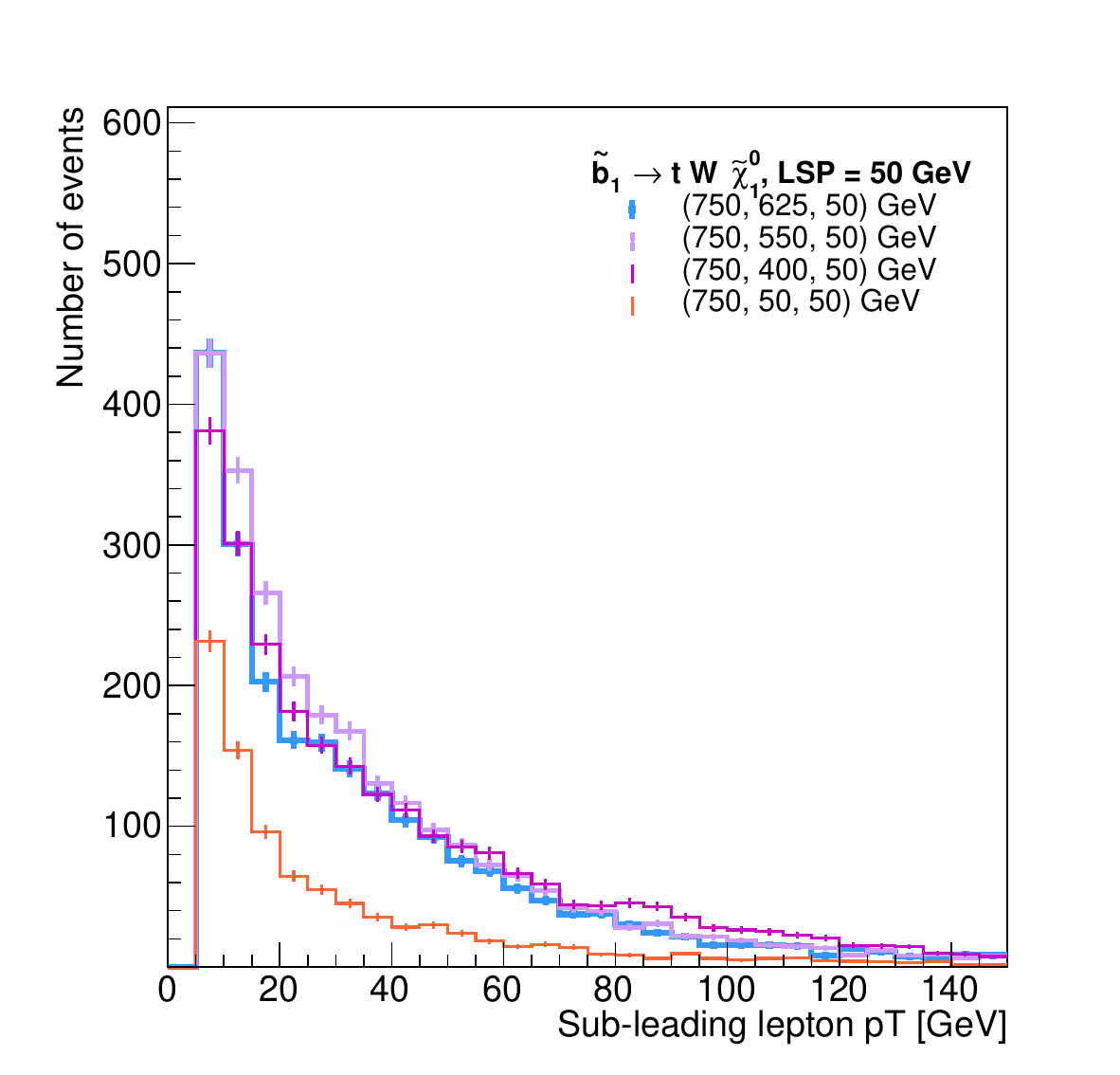}
    \includegraphics[width=0.32\columnwidth]{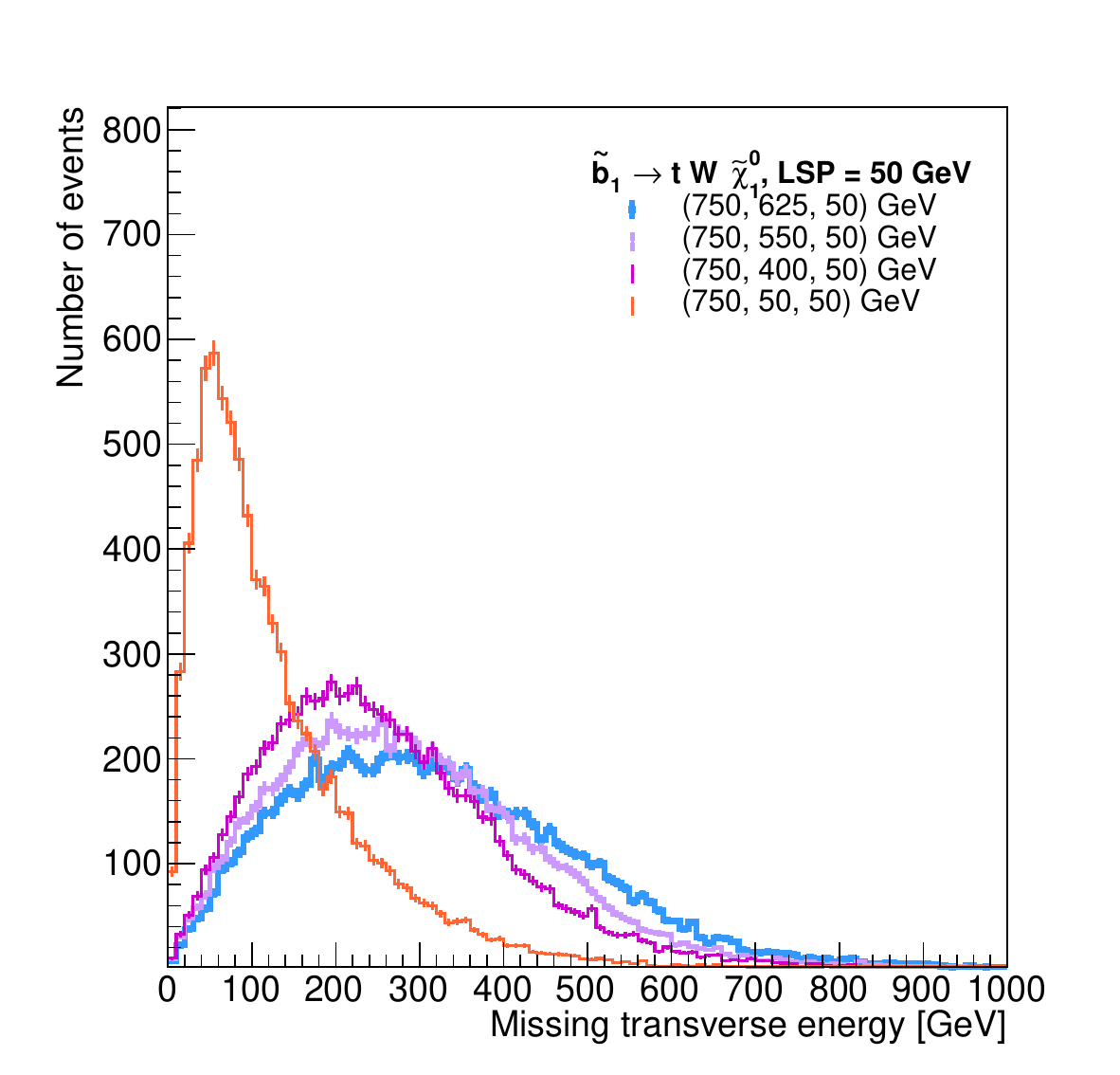}
    \includegraphics[width=0.32\columnwidth]{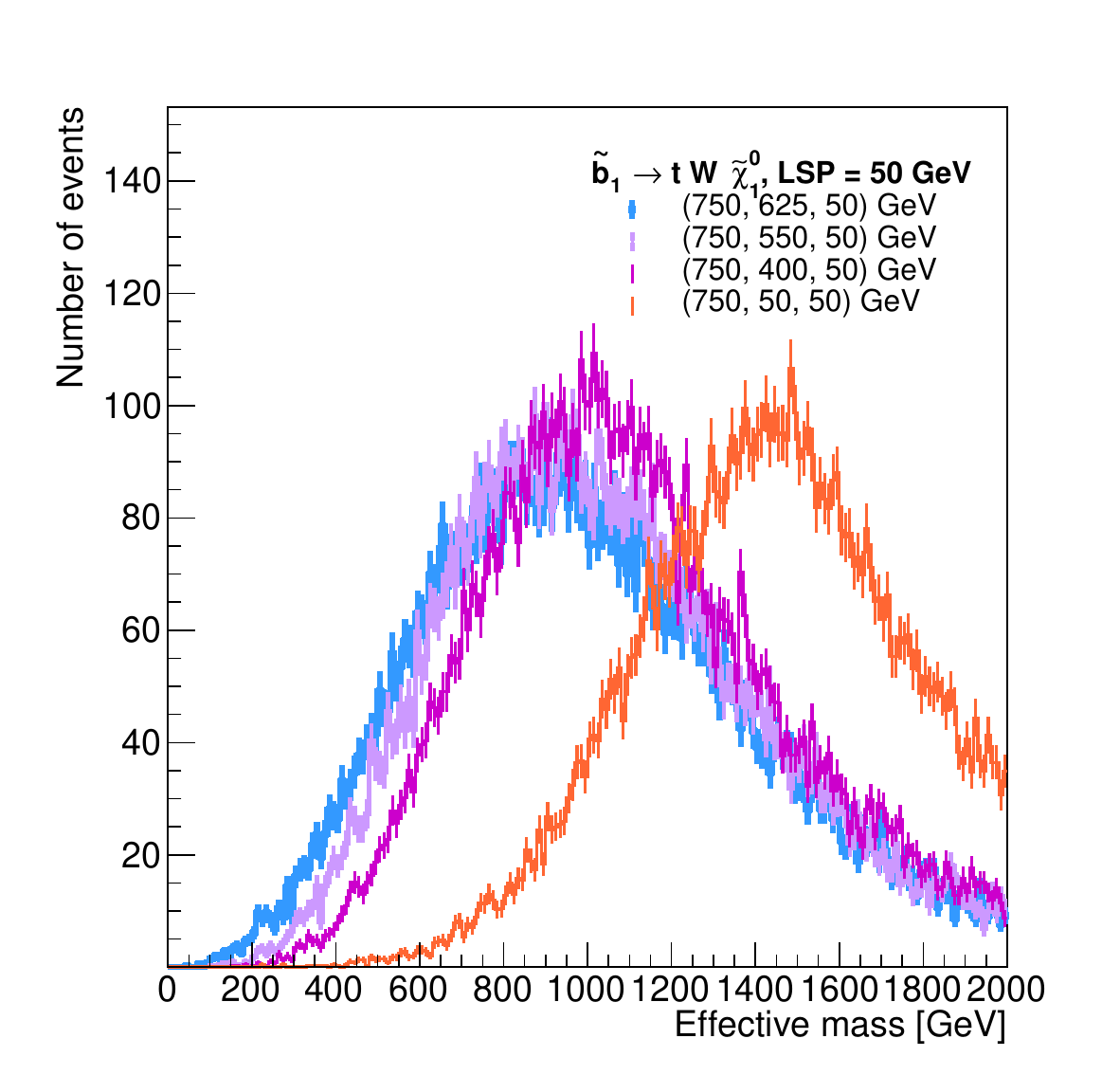}
    \includegraphics[width=0.32\columnwidth]{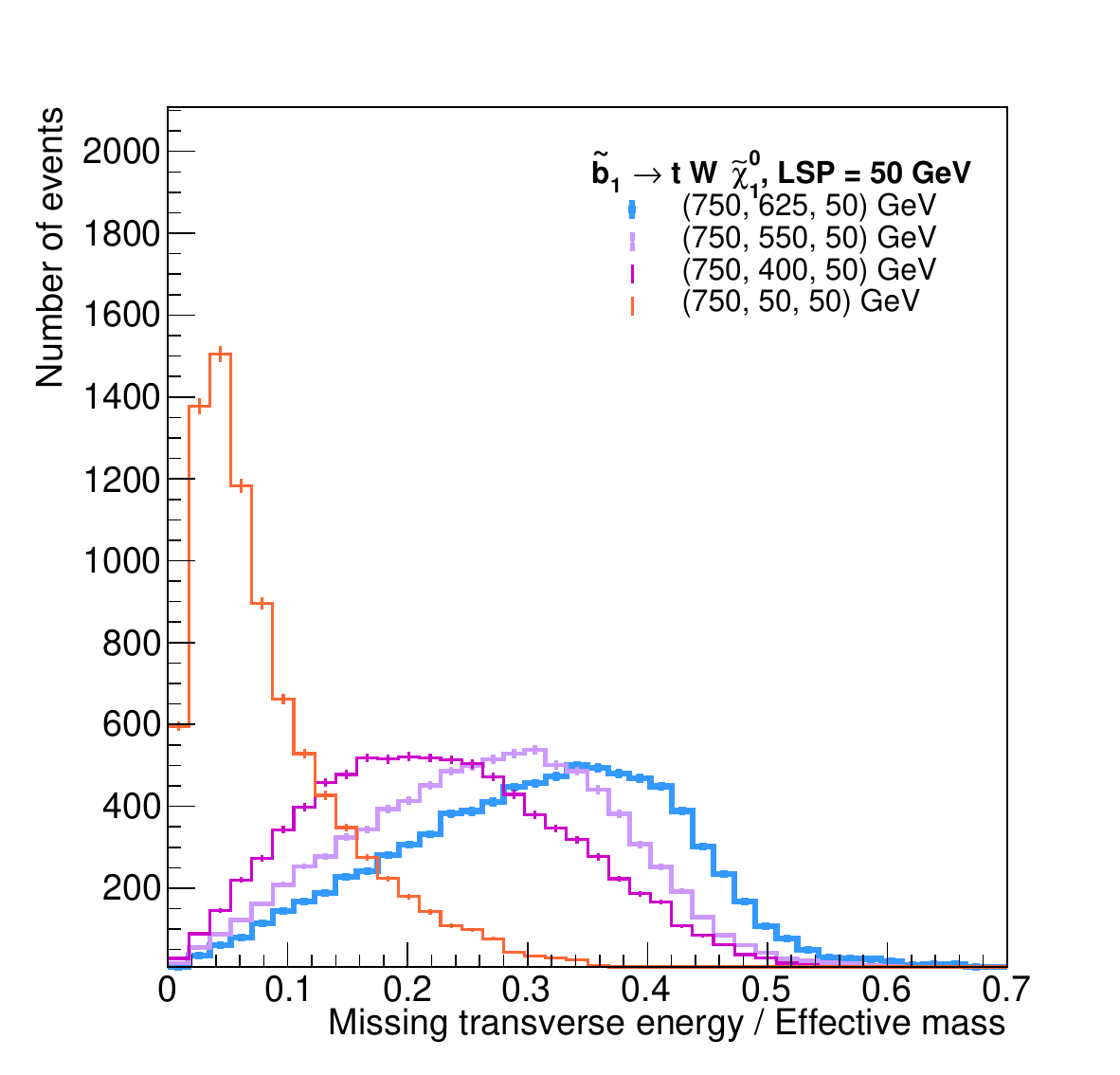}
    \vspace{-0.3cm}	
\end{center}
\caption{ 
    From left-to-right: distributions showing the number of $b$-tagged jets in the event, the leading $b$-tagged jet $\pt$, \met, number of jets in the event, the leading jet $\pt$ and the \met to \meff ratio in selected MC signal simulations. The statistical uncertainty is shown with a vertical line, and the distributions are normalized to 139~fb$^{-1}$ ($\sqrt s = 13.6$~TeV). \textit{Scenario~2} \sbsbModel model, which assumes an LSP mass of 50~GeV, is considered.
}
\label{fig:Distrib}
\end{figure}

\cref{fig:Distrib} shows selected distributions of the number of ($b$-tagged) jets in the event, the leading ($b$-tagged) jet \pt, the missing transverse energy \met, and the \met to \meff\footnote{\meff\ stands for the sum of the \pt\ of the signal leptons and jets, plus \met.} ratio for the four selected benchmark signal points. As expected, in the compressed mass spectrum~--~specifically at the (750, 625, 50)~GeV mass point~--~the leptons and $b$-tagged jets are very soft. It is interesting to note that for this point, many events appear to have zero $b$-tagged jets. This is very likely because the $b$-tagged jets present in these events are very soft, with $\pt < 20$~GeV~--~the \pt\ threshold used in ATLAS Ref.~\citen{ATLAS:2021yyr}. In contrast, in the very boosted region~--~represented by the (750, 50, 50)~GeV mass point~--~the jets are very energetic ($\pt > 300$–$400$~GeV for the leading two); here, the \met is generally low ($< 150$~GeV), as is the lepton \pt.

\begin{table}[t!]
\centering
\caption{
    Event counts and acceptance after the SR0$^{\mathrm{loose}}_{\mathrm{bveto}}$, SR0$^{\mathrm{tight}}_{\mathrm{bveto}}$, and SR0$_{\mathrm{breq}}$ signal region (pre-)selections discussed in the text, for the four representative signal mass points. Statistical uncertainties are also shown. Event yields are normalized to an integrated luminosity of 139~fb$^{-1}$ at $\sqrt{s} = 13.6$~TeV. The \textit{Scenario~2} \sbsbModel model, which assumes an LSP mass of 50~GeV, is considered. 
}
\label{tab:Sbottom13TeV_4LSRSel_CMSGrid}
{\scriptsize
\def\arraystretch{1.5}
\setlength{\tabcolsep}{0.0pc}
    \begin{tabular*}{\textwidth}{@{\extracolsep{\fill}}llcccc}
        \noalign{\smallskip}\hline\noalign{\smallskip}
        &   &  (750, 625, 50) & (750, 550, 50) & (750, 400, 50) & (750, 50, 50) \\
        & Selection & N events ($A$) & N events ($A$) & N events ($A$) & N events ($A$) \\
        \noalign{\smallskip}\hline\noalign{\smallskip}
\multirow[c]{3}{*}[0in]{\rotatebox{90}{SR0$^{\mathrm{loose/tight}}_{\mathrm{bveto}}$}}
 & Pre-sel 1 & 39.8 $\pm$ 3.2 (0.4 \%) & 71.4 $\pm$ 4.3 (0.8 \%)  & 65.4 $\pm$ 4.1 (0.7 \%) & 1.3 $\pm$ 0.6 (0.0 \%) \\
 & Pre-sel 2 & 29.1 $\pm$ 2.7 (0.3 \%) & 36.0 $\pm$ 3.0 (0.4 \%)  & 30.4 $\pm$ 2.8 (0.3 \%) & 0.8 $\pm$ 0.4 (0.0 \%) \\ 
 & {\color{blue} \textbf{SR0$^{\mathrm{loose}}_{\mathrm{bveto}}$}} & 21.2 $\pm$ 2.3 (0.2 \%) & 24.6 $\pm$ 2.5 (0.3 \%) & 23.0 $\pm$ 2.4 (0.3 \%) & 0.8 $\pm$ 0.4 (0.0 \%) \\ 
 & {\color{blue} \textbf{SR0$^{\mathrm{tight}}_{\mathrm{bveto}}$}} & 3.1 $\pm$ 0.9 (0.0 \%) & 4.8 $\pm$ 1.1 (0.1 \%) & 3.8 $\pm$ 1.0 (0.0 \%) & 0.5 $\pm$ 0.4 (0.0 \%) \\ 
         \noalign{\smallskip}\hline\noalign{\smallskip}
\multirow[c]{3}{*}[0in]{\rotatebox{90}{SR0$_{\mathrm{breq}}$}}
 & Pre-sel 1 &   39.8 $\pm$ 3.2 (0.4 \%)         & 71.4 $\pm$ 4.3 (0.8 \%) & 65.4 $\pm$ 4.1 (0.7 \%)   & 1.3 $\pm$ 0.6 (0.0 \%)  \\
 & Pre-sel 2 &   10.7 $\pm$ 1.7 (0.1 \%)         & 35.5 $\pm$ 3.0 (0.4 \%) & 35.0 $\pm$ 3.0 (0.4 \%)   & 0.5 $\pm$ 0.4 (0.0 \%)  \\
 & {\color{blue} \bf SR0$_{\mathrm{breq}}$} &  3.8 $\pm$ 1.0 (0.0 \%)          & 6.1 $\pm$ 1.2 (0.1 \%) & 7.4 $\pm$ 1.4 (0.1 \%)  & 0.3 $\pm$ 0.3 (0.0 \%) \\
        \noalign{\smallskip}\hline\hline\noalign{\smallskip}
\end{tabular*}
}	
\end{table}

From all the signal regions defined in ATLAS Ref.~\citen{ATLAS:2021yyr}, those relevant to the study presented in our paper are SR0$^{\mathrm{loose}}_{\mathrm{bveto}}$, SR0$^{\mathrm{tight}}_{\mathrm{bveto}}$, and SR0$_{\mathrm{breq}}$. Their definitions, along with some pre-selection steps, consist in:
\[
\left\{
\begin{array}{l}
\text{Pre-sel 1: } \geq \llll, \; Z \text{ boson mass veto}, \\
\text{Pre-sel 2: Pre-sel 1, Nr. } b\text{-tagged jets } = 0, \\
{\color{blue} \mathrm{SR0}^{\mathrm{loose}}_{\mathrm{bveto}}: \text{ Pre-sel 2, } \meff > 600 \text{~GeV.}} \\
{\color{blue} \mathrm{SR0}^{\mathrm{tight}}_{\mathrm{bveto}}: \text{ Pre-sel 2, } \meff > 1250 \text{~GeV.}}
\end{array}
\right.
\]\
\[
\left\{
\begin{array}{l}
\text{Pre-sel 1: } \geq \llll, \; Z \text{ boson mass veto}, \\
\text{Pre-sel 2: Pre-sel 1, Nr. } b\text{-tagged jets } \geq 1, \\
{\color{blue} \mathrm{SR0}_{\mathrm{breq}}: \text{ Pre-sel 2, } \meff > 1300 \text{~GeV.}}
\end{array}
\right.
\]
Due to the requirement on the number of $b$-tagged jets, the SR0$^{\mathrm{loose}}_{\mathrm{bveto}}$ and SR0$^{\mathrm{tight}}_{\mathrm{bveto}}$ regions are orthogonal to SR0$_{\mathrm{breq}}$, and can therefore be combined. The event counts and acceptance values corresponding to the various pre-selection and signal region requirements are shown in \cref{tab:Sbottom13TeV_4LSRSel_CMSGrid} for the selected benchmark signal mass points. The considered integrated luminosity is 139~fb$^{-1}$ at a center-of-mass energy of $\sqrt{s} = 13.6$~TeV. The signal region acceptance is quite small~--~reaching a maximum of 0.8\% at the \textit{Pre-sel 1} step and decreasing to below 0.1\% in the final signal regions. This behavior is expected, as stringent requirements are necessary to suppress the large Standard Model and detector-related backgrounds, as detailed in ATLAS Ref.~\citen{ATLAS:2021yyr}.

\begin{figure}[t!]
    \centering
    \includegraphics[width=0.45\linewidth]{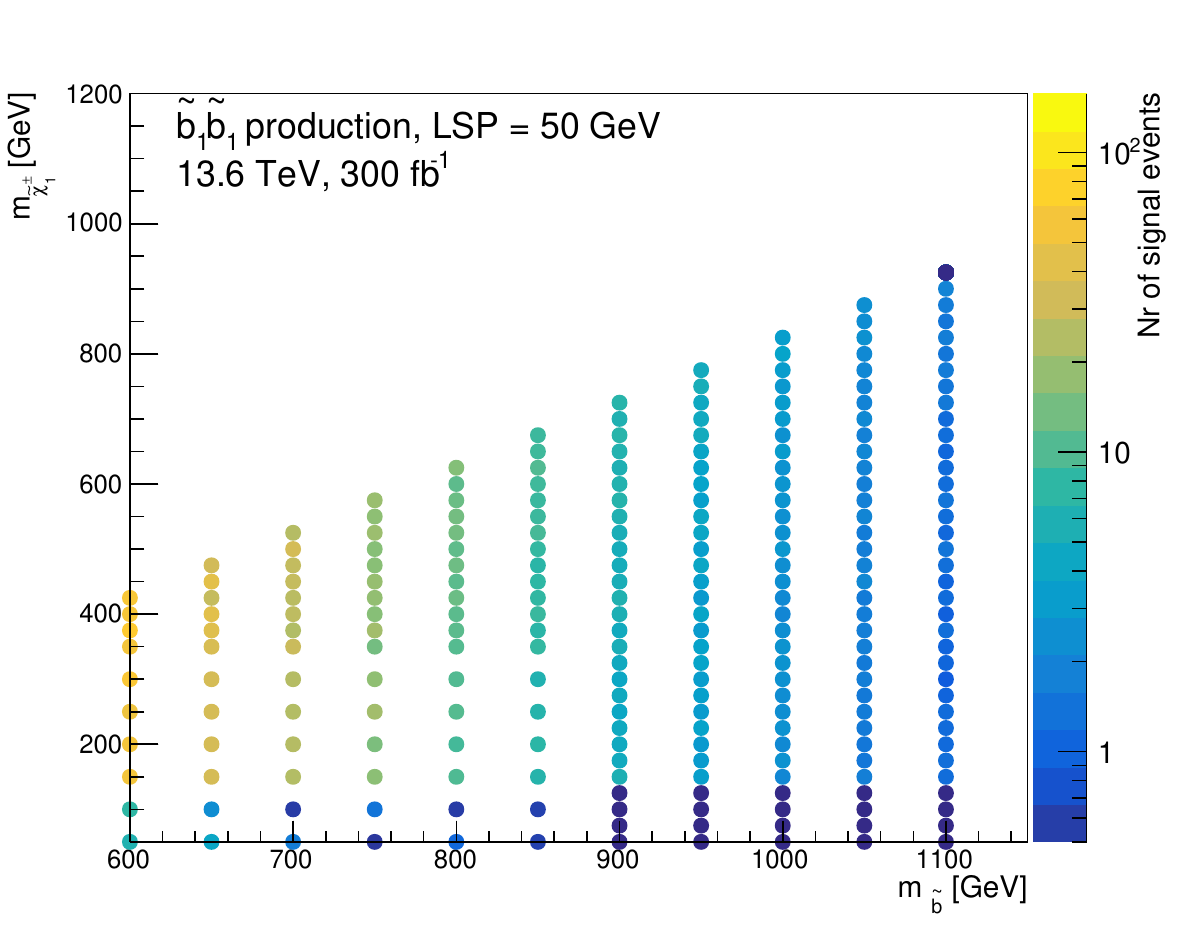}
    \includegraphics[width=0.45\linewidth]{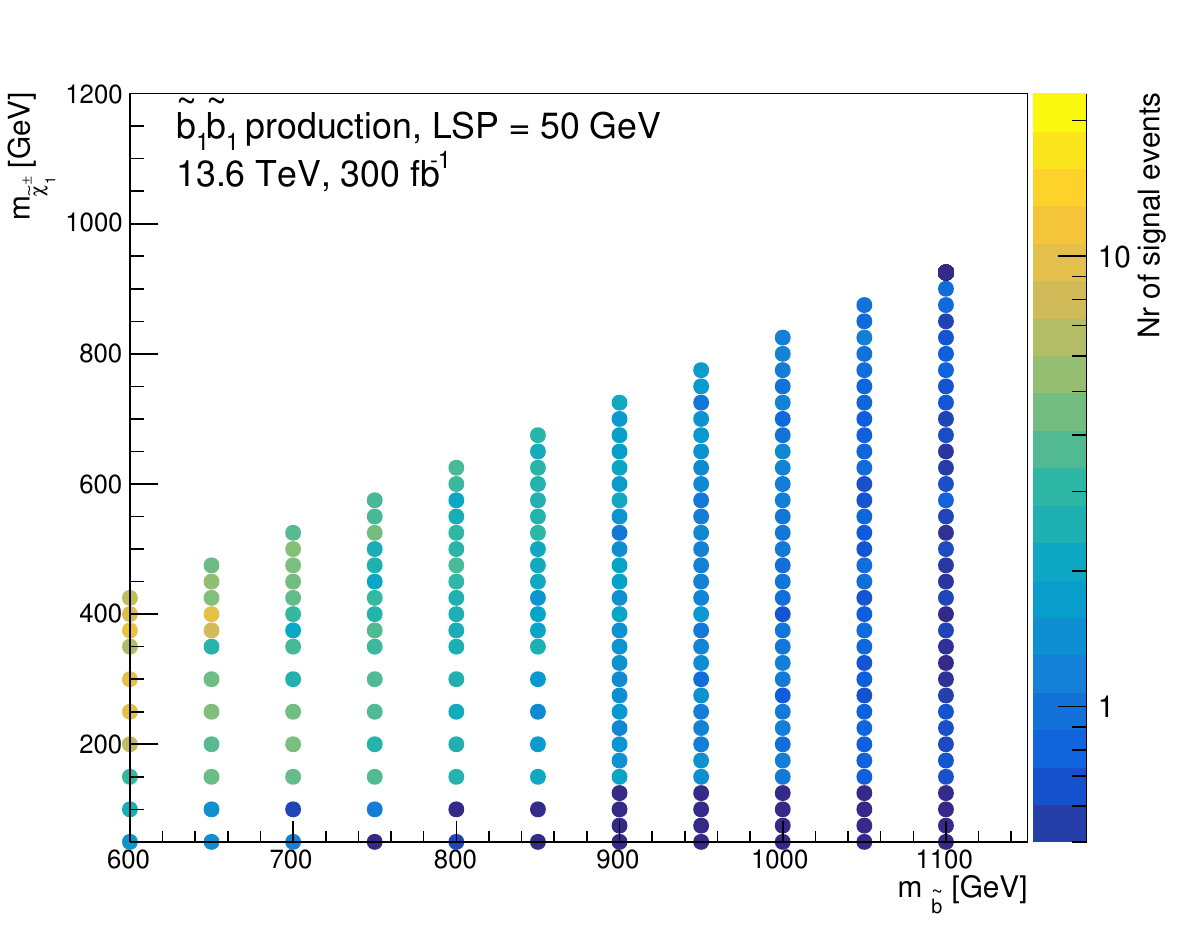}
    \includegraphics[width=0.45\linewidth]{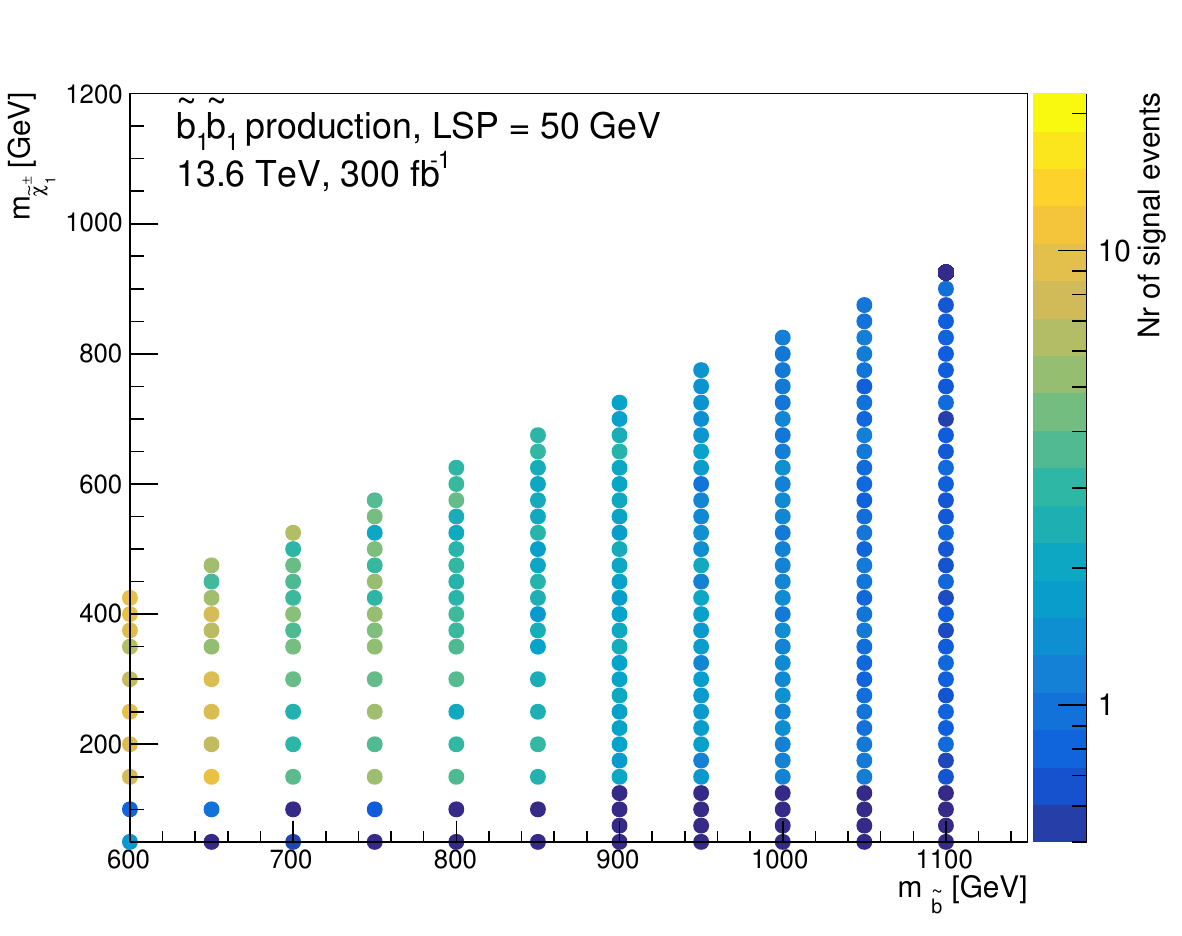}
    \includegraphics[width=0.45\linewidth]{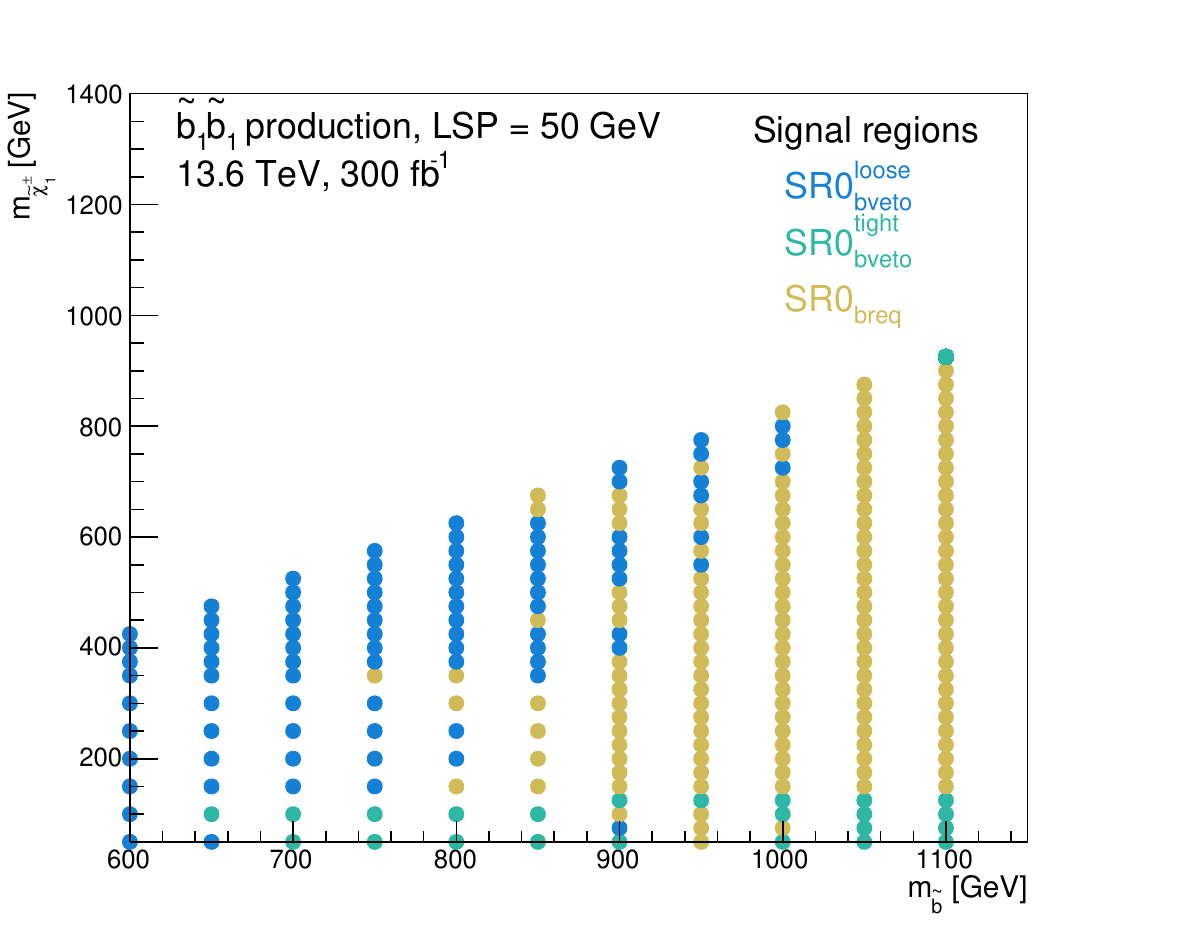}
    \caption{
        From left-to-right: number of events in the SR0$^{\mathrm{loose}}_{\mathrm{bveto}}$, SR0$^{\mathrm{tight}}_{\mathrm{bveto}}$, and SR0$_{\mathrm{breq}}$ signal regions, obtained using the \textit{Scenario~2} \sbsbModel signal samples (see $z$ axis), normalized to 300~fb$^{-1}$ at $\sqrt{s} = 13.6$~TeV. The bottom-right plot shows the signal region that yields the best signal significance $Z$ for each mass point.
    }
    \label{fig:SignalEvnt_13TeV_4L_ATLAS_Scenario2M}
\end{figure}

\cref{fig:SignalEvnt_13TeV_4L_ATLAS_Scenario2M} shows the number of events in the three signal regions for the \textit{Scenario~2} \sbsbModel generated signal samples, along with the signal region that yields the best signal significance $Z$ for each mass point. As expected, SR0$^{\mathrm{loose}}_{\mathrm{bveto}}$ consistently exhibits the highest acceptance, while SR0$^{\mathrm{tight}}_{\mathrm{bveto}}$ and SR0$_{\mathrm{breq}}$ have a comparable acceptance.

\section{ATLAS search with same-sign or three leptons final states} 
\label{sec:ATLAS_SS3L_Sbottom}

The search for the \sbsbModel with experimental signatures involving same-sign or three leptons (\llSC/\lll) is performed in ATLAS Ref.~\citen{ATLAS:2019fag}, and forms the basis for the exclusion limit results presented in Ref.~\citen{Ducu:2024arr}, which were derived under the \textit{Scenario~1} model, assuming $\chinoonepm = \text{LSP} + 100$~GeV. In the present study, the same ATLAS analysis is employed to reinterpret the data under the \textit{Scenario~2} model, which assumes an LSP mass of 50~GeV, and to compare the resulting sbottom exclusion range with that from \textit{Scenario~1}.

Building upon previous results, this study offers fresh motivation for sbottom searches in upcoming LHC runs. As the LHC continues to increase its integrated luminosity, sensitivity to scenarios with compressed mass spectra~--~such as those involving small mass splittings between the LSP and sbottom, or heavier electroweakinos~--~becomes increasingly important. In addition, exploring alternative mass configurations allows for probing more realistic or experimentally challenging SUSY models that remain unconstrained by current searches. The aim is therefore to improve understanding of how search performance depends on specific mass hypotheses, and to guide the design of future sbottom search strategies targeting multi-lepton and multi-jet final states during both LHC Run 3 and the HL-LHC.

As discussed also in Ref.~\citen{Ducu:2024arr}, 
the lepton isolation and identification working points used in \texttt{SimpleAnalysis} framework are taken from ATLAS Ref.~\citen{ATLAS:2019fag}. The two leading signal leptons are required to have $p_T > 20$~GeV, while the remaining leptons must satisfy the following criteria: electrons with $p_T > 10$~GeV and $|\eta| < 2.47$, and muons with $p_T > 10$~GeV and $|\eta| < 2.5$. Two signal regions are of interest: Rpc2L1b and Rpc2L2b. Their definition, along with some pre-selection steps are: 
\[
\left\{
\begin{array}{l}
\text{Pre-sel 1: } \geq \llSC, \text{ Nr. } b\text{-tagged jets } \geq 1, \\
\text{Pre-sel 2: Pre-sel 1, Nr. jets} \geq 6 \text{ (}p_T > 40~\text{GeV}), \\
{\color{blue} \mathrm{Rpc2L1b}: \text{Pre-sel 2, } \met / \meff > 0.25.}
\end{array}
\right.
\]
\[
\left\{
\begin{array}{l}
\text{Pre-sel 1: } \geq \llSC, \text{ Nr. } b\text{-tagged jets } \geq 2, \\
\text{Pre-sel 2: Pre-sel 1, Nr. jets}  \geq 6  \text{ (}p_T > 25~\text{GeV}), \\
\text{Pre-sel 3: Pre-sel 2, } \met > 300~\text{GeV}, \\
\text{Pre-sel 4: Pre-sel 3, } \meff > 1.4~\text{TeV}, \\
{\color{blue} \mathrm{Rpc2L2b}: \text{Pre-sel 4, } \met / \meff > 0.14. }
\end{array}
\right.
\]

\begin{table}[t!]
\centering
\caption{
      Event counts and acceptance after the Rpc2L1b and Rpc2L2b signal region (pre-)selections discussed in the text, for the four representative signal mass points. The statistical uncertainty is also shown, and the events are normalized to an integrated luminosity of 139~fb$^{-1}$ at $\sqrt{s} = 13.6$ TeV. The \textit{Scenario~2} \sbsbModel model, which assumes an LSP mass of 50~GeV, is considered.
}
\label{tab:Sbottom13TeV_SS3LSRSel_CMSGrid}
{\scriptsize
\def\arraystretch{1.5}
\setlength{\tabcolsep}{0.0pc}
    \begin{tabular*}{\textwidth}{@{\extracolsep{\fill}}llcccc}
        \noalign{\smallskip}\hline\noalign{\smallskip}
        &   &  (750, 625, 50) & (750, 550, 50) & (750, 400, 50) & (750, 50, 50) \\
        & Selection & N events ($A$) & N events ($A$) & N events ($A$) & N events ($A$) \\
        \noalign{\smallskip}\hline\noalign{\smallskip}
\multirow[c]{3}{*}[0in]{\rotatebox{90}{Rpc2L1b}}
 & Pre-sel 1 &           302.1 $\pm$ 8.8 (3.4 \%)  &   562.9 $\pm$ 11.9 (6.3 \%)  &   577.0 $\pm$ 12.1 (6.5 \%)  &    55.1 $\pm$ 3.7 (0.6 \%)      \\
 & Pre-sel 2 &           23.2 $\pm$ 2.4 (0.3 \%)   &   96.0 $\pm$ 4.9 (1.1 \%)    &   128.3 $\pm$ 5.7 (1.4 \%)   &    15.8 $\pm$ 2.0 (0.2 \%)      \\
 & {\color{blue} \bf Rpc2L1b} &       9.2 $\pm$ 1.5 (0.1 \%)    &   33.7 $\pm$ 2.9 (0.4 \%)    &   30.7 $\pm$ 2.8 (0.3 \%)    &    0.0 $\pm$ 0.0 (0.0 \%)       \\
  \noalign{\smallskip}\hline\noalign{\smallskip}
\multirow[c]{5}{*}[0in]{\rotatebox{90}{Rpc2L2b}}
 & Pre-sel 1 &           60.2 $\pm$ 3.9 (0.7 \%)   &   180.6 $\pm$ 6.8 (2.0 \%)   &   187.3 $\pm$ 6.9 (2.1 \%)   &    9.2 $\pm$ 1.5 (0.1 \%)       \\
 & Pre-sel 2 &           24.5 $\pm$ 2.5 (0.3 \%)   &   86.6 $\pm$ 4.7 (1.0 \%)    &   98.9 $\pm$ 5.0 (1.1 \%)    &    5.9 $\pm$ 1.2 (0.1 \%)       \\
 & Pre-sel 3 &           13.8 $\pm$ 1.9 (0.2 \%)   &   39.0 $\pm$ 3.1 (0.4 \%)    &   32.2 $\pm$ 2.9 (0.4 \%)    &    0.5 $\pm$ 0.4 (0.0 \%)       \\
 & Pre-sel 4 &           7.4 $\pm$ 1.4 (0.1 \%)    &   23.1 $\pm$ 2.4 (0.3 \%)    &   17.6 $\pm$ 2.1 (0.2 \%)    &    0.5 $\pm$ 0.4 (0.0 \%)       \\
 & {\color{blue} \bf Rpc2L2b} &       7.4 $\pm$ 1.4 (0.1 \%)    &   23.1 $\pm$ 2.4 (0.3 \%)    &   17.4 $\pm$ 2.1 (0.2 \%)    &    0.5 $\pm$ 0.4 (0.0 \%)       \\
\noalign{\smallskip}\hline\hline\noalign{\smallskip}
\end{tabular*}
}	
\end{table}

\begin{figure}[t!]
\centering
    \begin{subfigure}[t]{0.45\columnwidth}
        \centering
        \includegraphics[width=\linewidth]{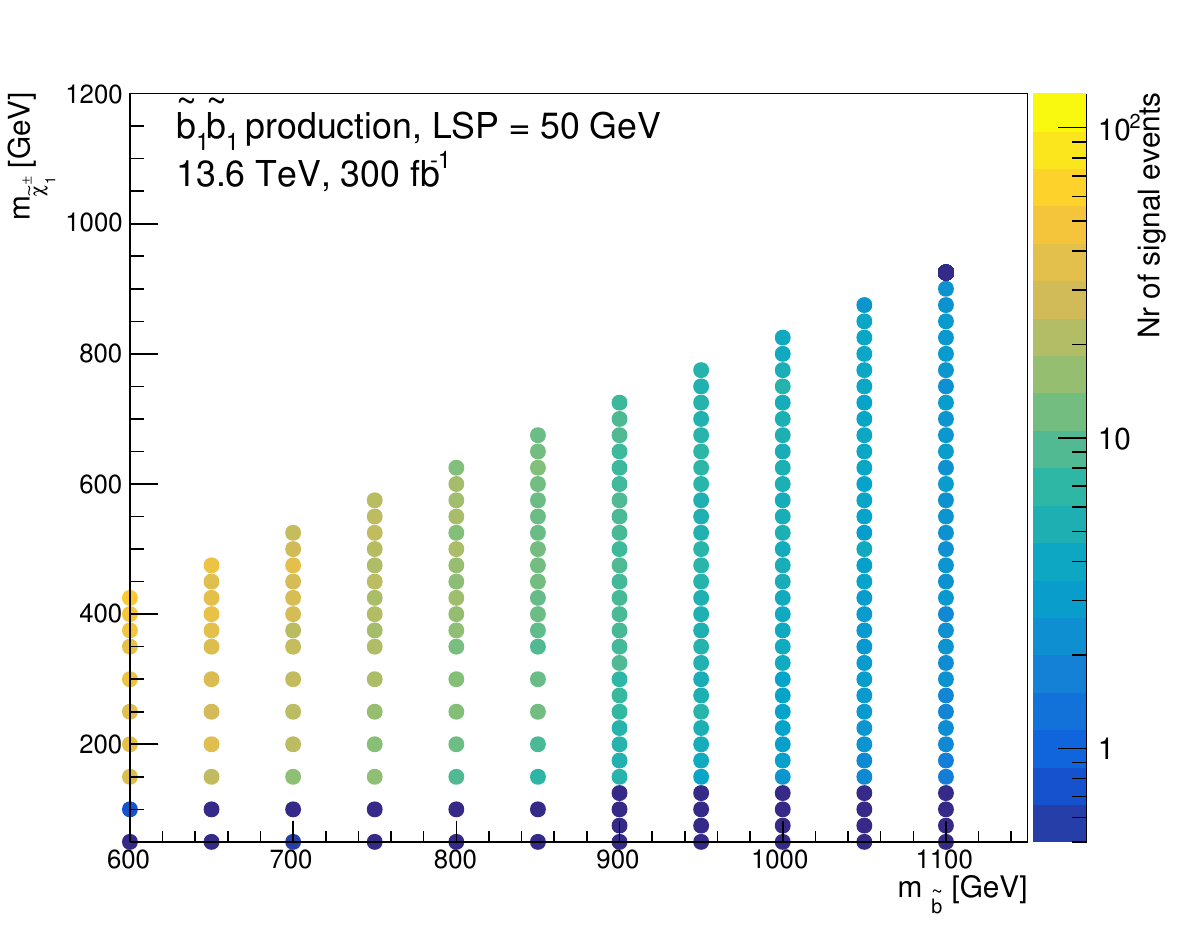}
        \caption{}
    \end{subfigure}%
    \hfill
    \begin{subfigure}[t]{0.45\columnwidth}
        \centering
        \includegraphics[width=\linewidth]{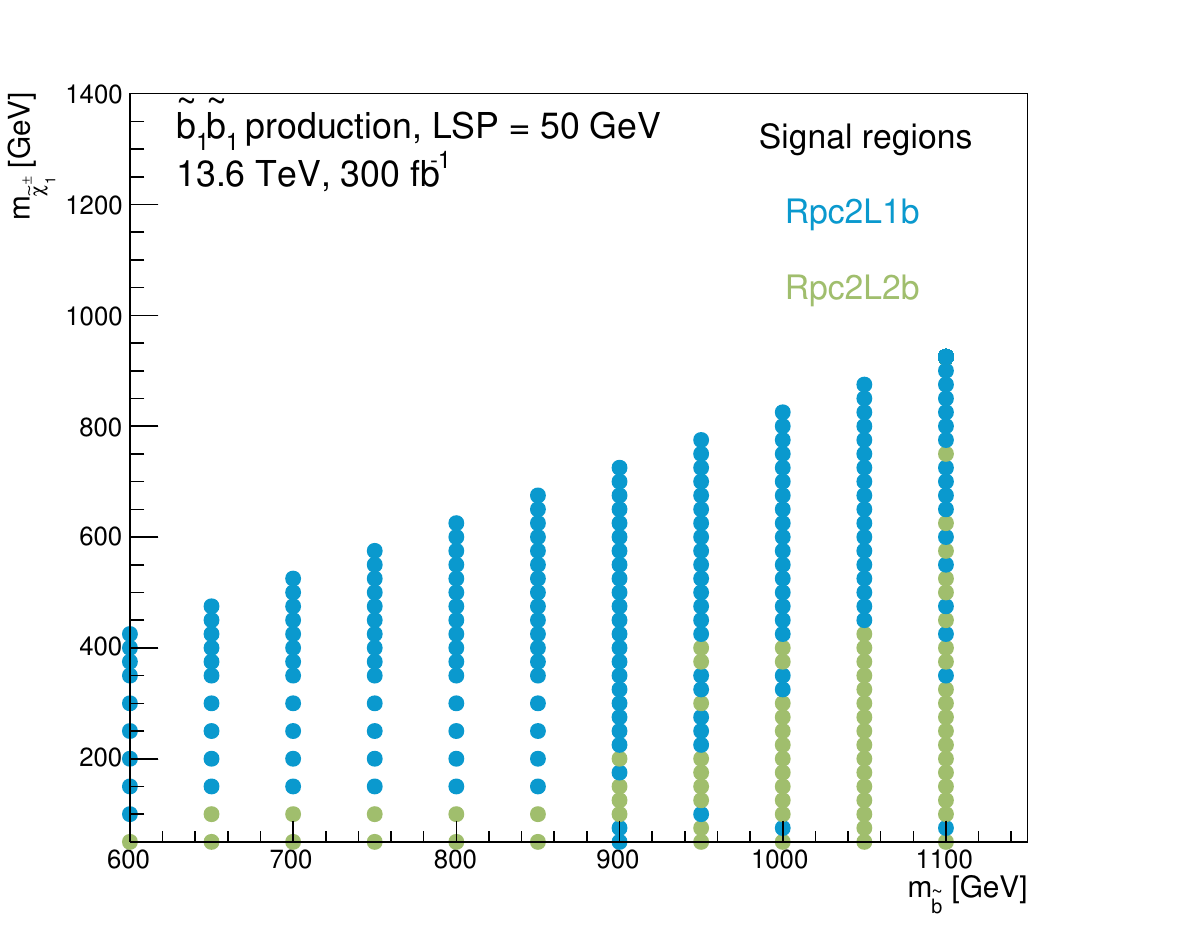}
        \caption{}
    \end{subfigure}
\caption{
     (a) Number of events in the Rpc2L1b signal region obtained using the \textit{Scenario~2} \sbsbModel signal samples (see $z$ axis), normalized to 300~fb$^{-1}$ at $\sqrt{s} = 13.6$~TeV. (b) The signal region yielding the best signal significance $Z$ for each mass point.
}
\label{fig:SignalEvnt_13TeV_SS3L_ATLAS_Scenario2M}
\end{figure}

\cref{tab:Sbottom13TeV_SS3LSRSel_CMSGrid} presents the event counts and acceptance values corresponding to the various pre-selection steps and the Rpc2L1b and Rpc2L2b signal region requirements, for the four benchmark signal mass points within the \textit{Scenario~2} \sbsbModel model. In contrast, \cref{fig:SignalEvnt_13TeV_SS3L_ATLAS_Scenario2M} shows results across a broader set of mass points, illustrating (a) the number of signal events passing the Rpc2L1b region and (b) identifying the signal region with the highest significance $Z$ for each point. 

For this model, Rpc2L2b provides better sensitivity in the region where the mass splitting between $\sbottom_1$ and $\chinoonepm$ is very high, while Rpc2L1b dominates elsewhere. When examining the results, one can see that for $\chinoonepm$ masses below 150~GeV, the signal yields drop significantly in both regions, leading to $Z$ values well below the exclusion threshold of 1.64. This is due to the small mass gap between $\chinoonepm$ and the LSP, resulting in soft $W$ bosons from $\chinoonepm \to W \ninoone$ decays and correspondingly low \met\ (see also \cref{fig:Distrib}). Since Rpc2L1b and Rpc2L2b both require relatively large \met, their acceptance is strongly suppressed in this part of the mass plane. This highlights a well-known challenge in detecting signals with soft kinematics, making this region difficult to probe using the signal regions defined in ATLAS Ref.~\citen{ATLAS:2019fag}. Future analyses could improve sensitivity by employing ISR-based tagging or by introducing dedicated signal regions with looser \met requirements.

\section{Expected background contributions}
\label{sec:ATLAS_BkgSection}

ATLAS Ref.~\citen{ATLAS:2021yyr} estimated the total background event yields to be 11.5$^{+2.9}_{-2.2}$, 3.5$^{+2.0}_{-2.2}$, and 1.19$^{+0.3}_{-0.28}$ in the SR0$^{\mathrm{loose}}_{\mathrm{bveto}}$, SR0$^{\mathrm{tight}}_{\mathrm{bveto}}$, and SR0$_{\mathrm{breq}}$ signal regions, respectively. Similarly, background estimates of 6.5$^{+1.5}_{-1.6}$ and 7.8$^{+2.1}_{-2.3}$ events were reported for the Rpc2L1b and Rpc2L2b signal regions in ATLAS Ref.~\citen{ATLAS:2019fag}. These estimations, along with their uncertainties, are adopted for the analysis at 139~fb$^{-1}$ integrated luminosity presented in this paper. For the projections to higher luminosities of 300~fb$^{-1}$ and 3000~fb$^{-1}$, the background uncertainties are conservatively reduced to 20\% and 10\%, respectively. This reflects the expectation that the precision of background measurements will improve as more data becomes available, primarily because the dominant systematic uncertainties are currently limited by the low statistics of data-driven background estimation methods~--~such as those for electron charge misidentification and fake/non-prompt lepton contributions\cite{ATLAS:2019fag}. These improvements in background precision are essential for enhancing the sensitivity of future searches at the LHC and HL-LHC.

In addition, the total number of background events is scaled by factors of 2.16 (for the 300~fb$^{-1}$ projections) and 21.60 (for the 3000~fb$^{-1}$ projections) to account for the increased data statistics relative to the 139~fb$^{-1}$ baseline. For projections at center-of-mass energies of 13.6~TeV and 14~TeV, the background estimates from ATLAS are further multiplied by factors of 1.1 and 1.2, respectively, to incorporate the expected rise in background production cross-sections at these higher energies.

\begin{figure}[t!]
    \begin{subfigure}[t]{0.45\columnwidth}
        \centering
        \includegraphics[width=\linewidth]{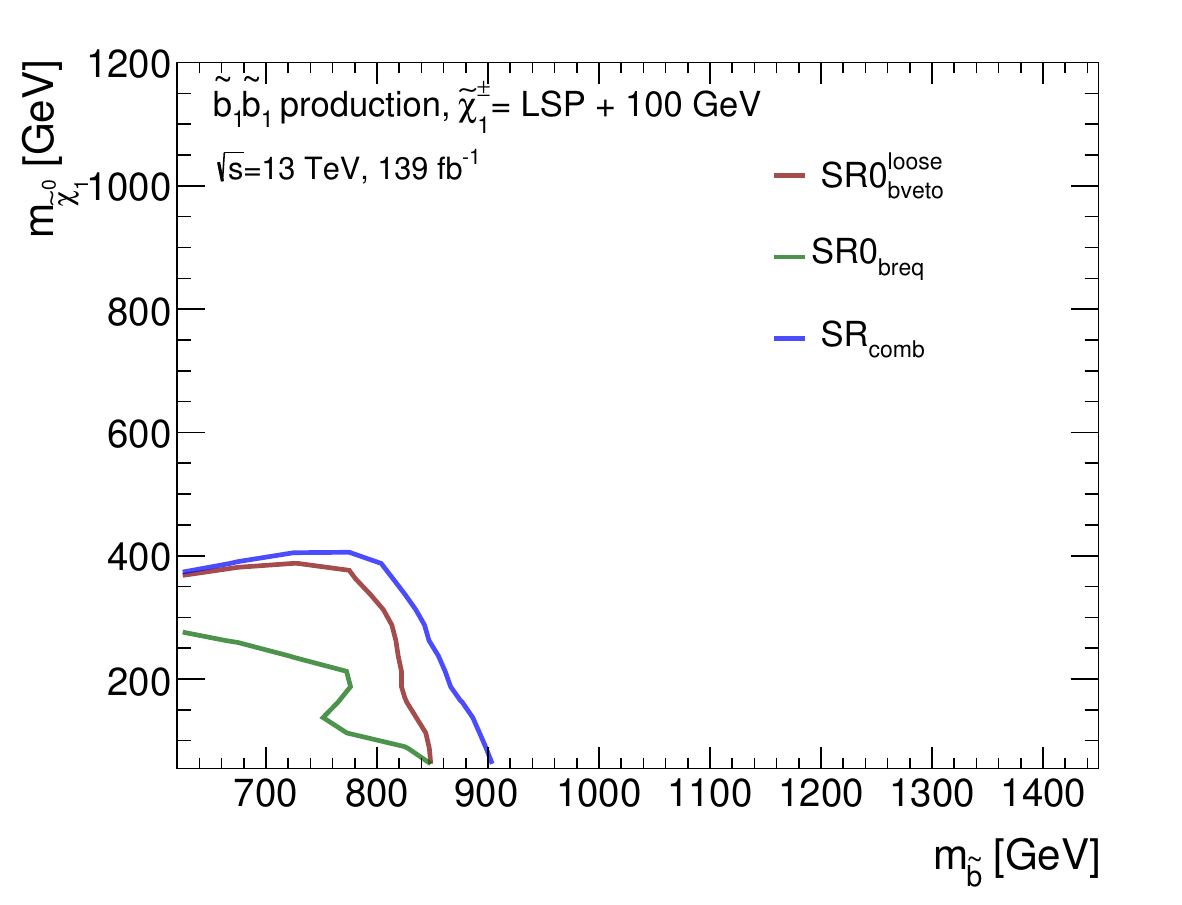}
        \includegraphics[width=\linewidth]{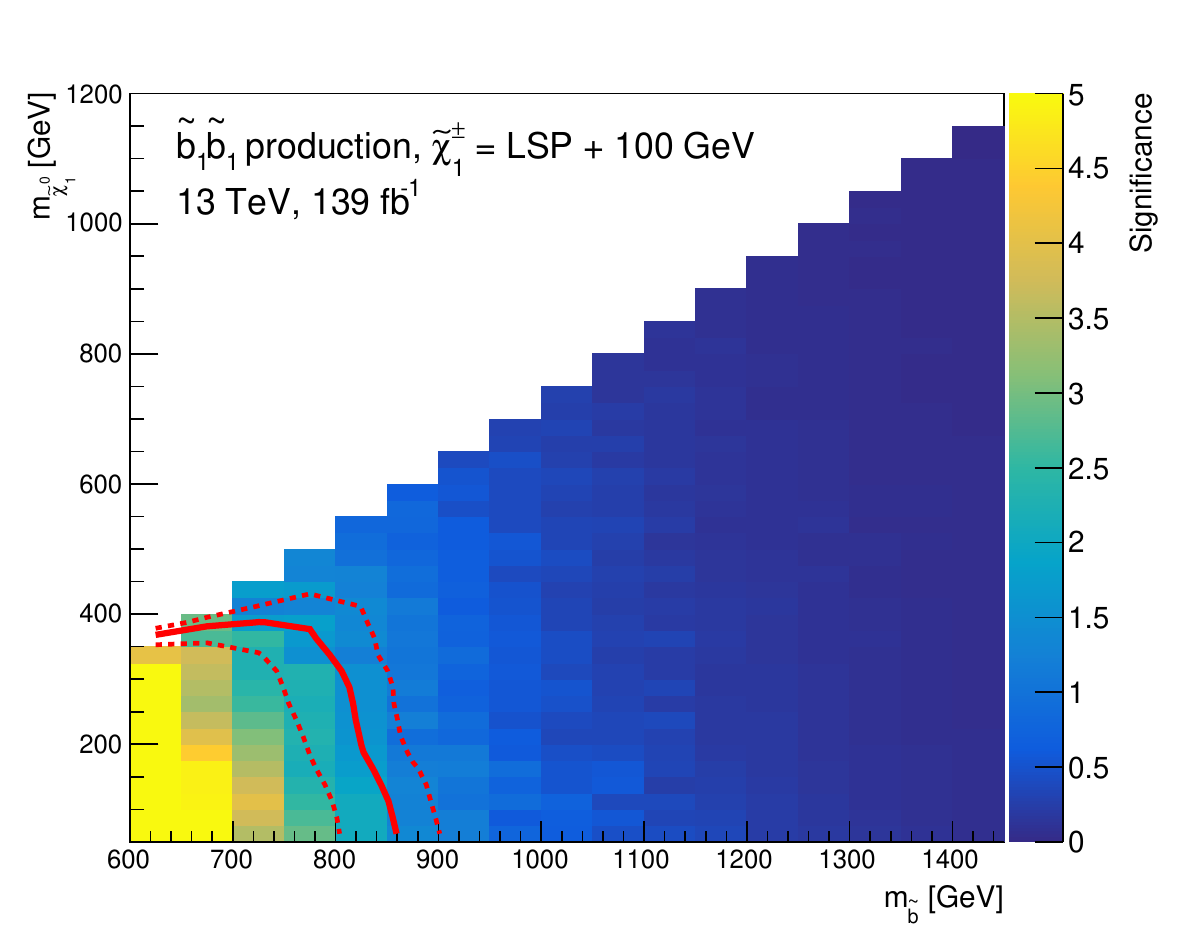}
        \caption{}
        \label{fig:Limits_13TeV_139ifb_Scenario1M_4L}
    \end{subfigure}%
    \hfill
    \begin{subfigure}[t]{0.45\columnwidth}
        \centering
        \includegraphics[width=\linewidth]{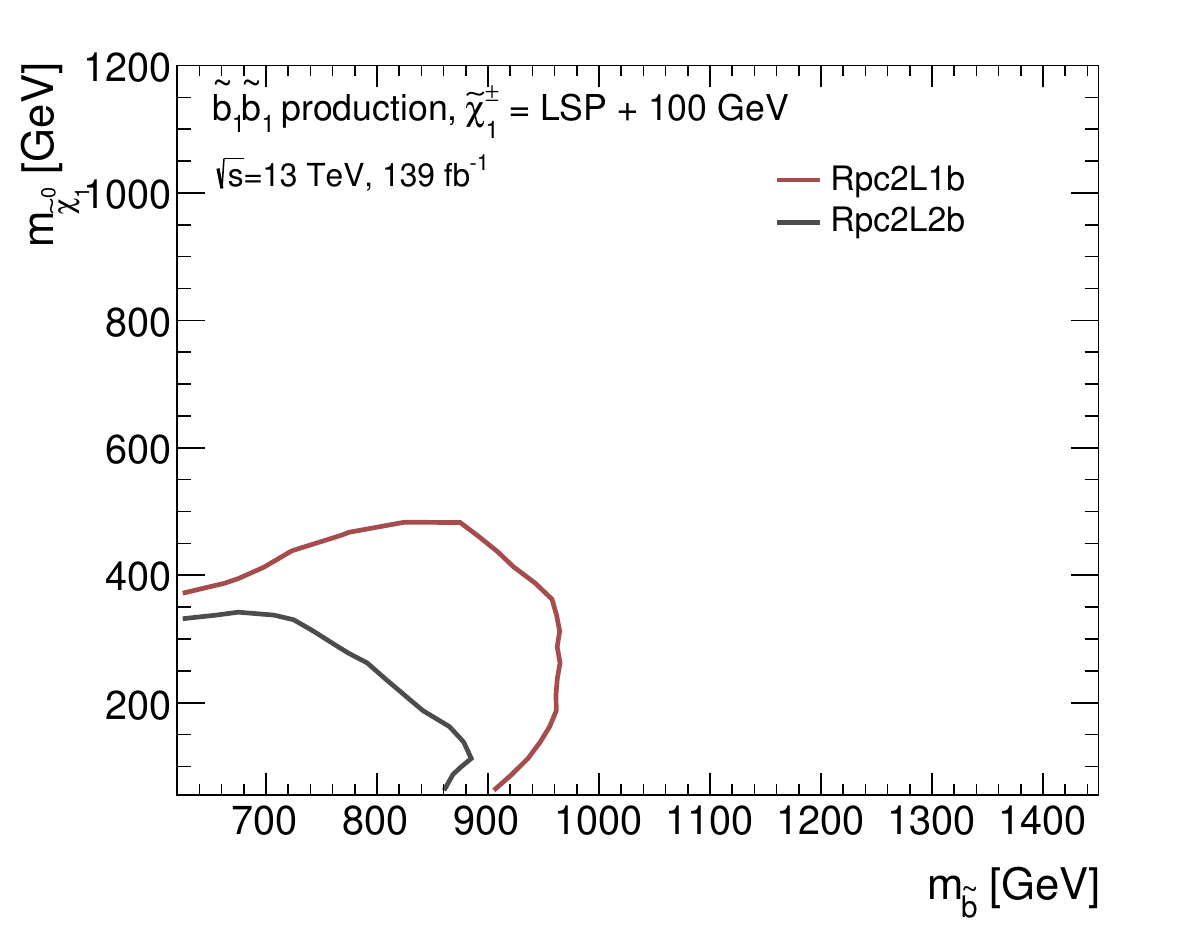}
        \includegraphics[width=\linewidth]{Limits_4L_13TeV/ATLASGrid_Limit_SR0_ECMLumiConfig0.pdf}
        \caption{}
        \label{fig:Limits_13TeV_139ifb_Scenario1M_SS3L}
    \end{subfigure}
\caption{
    \textit{Scenario~1} \sbsbModel model, assuming $\chinoonepm = \text{LSP} + 100$~GeV. 
    Exclusion mass limits obtained with (a) ATLAS Ref.~\citen{ATLAS:2021yyr} ($4\ell$) and (b) Refs.~\citen{ATLAS:2019fag,Ducu:2024arr} (\llSC/\lll) analyses.
    Top: exclusion mass limits obtained with the individual signal regions; Bottom: exclusion mass limits obtained with the signal region that gives the best signal significance $Z$ for each mass point. 
    Results are shown for $\sqrt s = 13$~TeV and an integrated luminosity of 139~fb$^{-1}$. 
    When present, the dashed lines correspond to the $\pm 1\sigma$ uncertainty on the signal event yield.
}
\label{fig:Limits_13TeV_139ifb_Scenario1M}
\end{figure}

\section{Projected Exclusion Limits at 13~TeV}

\begin{figure}[t!]
    \begin{subfigure}[t]{0.45\columnwidth}
        \centering
        \includegraphics[width=\linewidth]{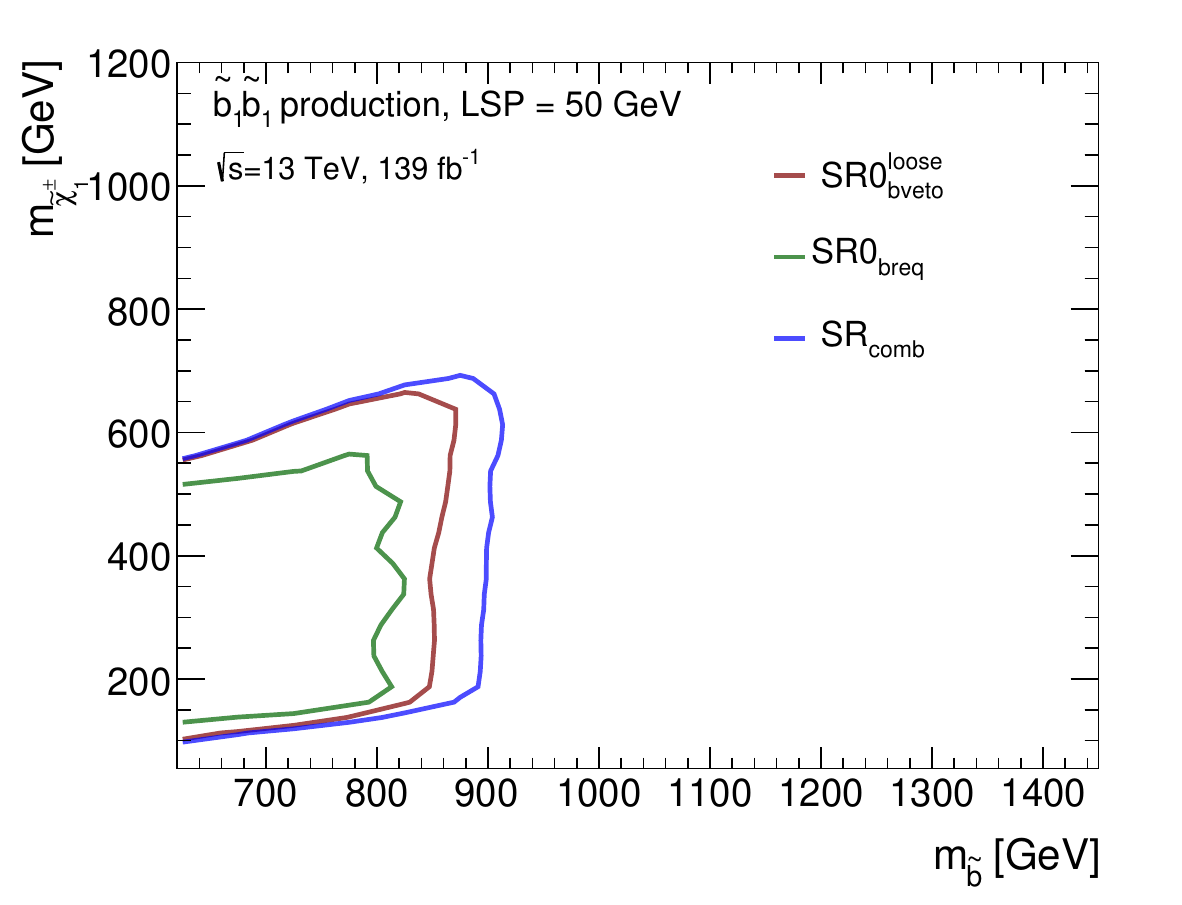}
        \includegraphics[width=\linewidth]{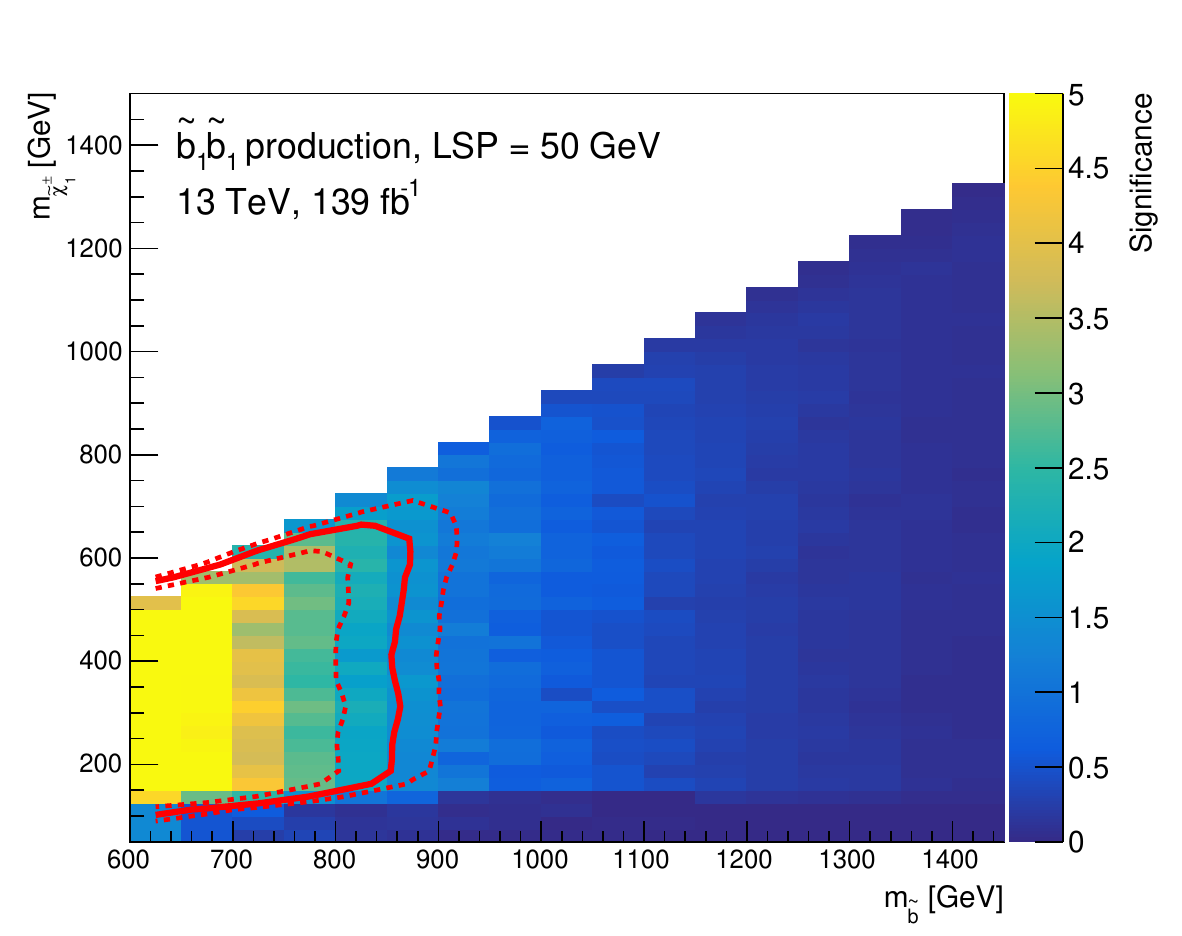}
        \caption{}
        \label{fig:Limits_13TeV_139ifb_Scenario2M_4L}
    \end{subfigure}%
    \hfill
    \begin{subfigure}[t]{0.45\columnwidth}
        \centering
        \includegraphics[width=\linewidth]{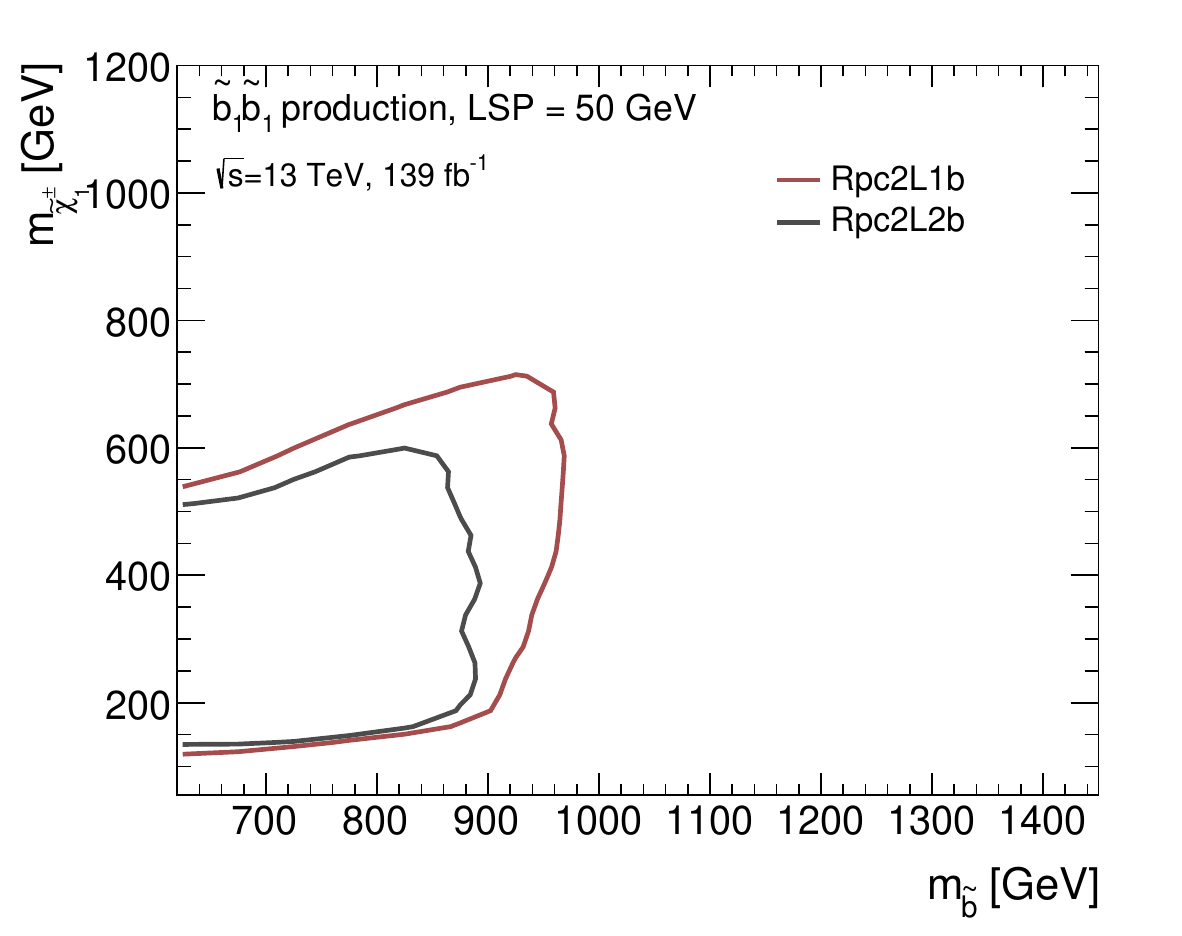}
        \includegraphics[width=\linewidth]{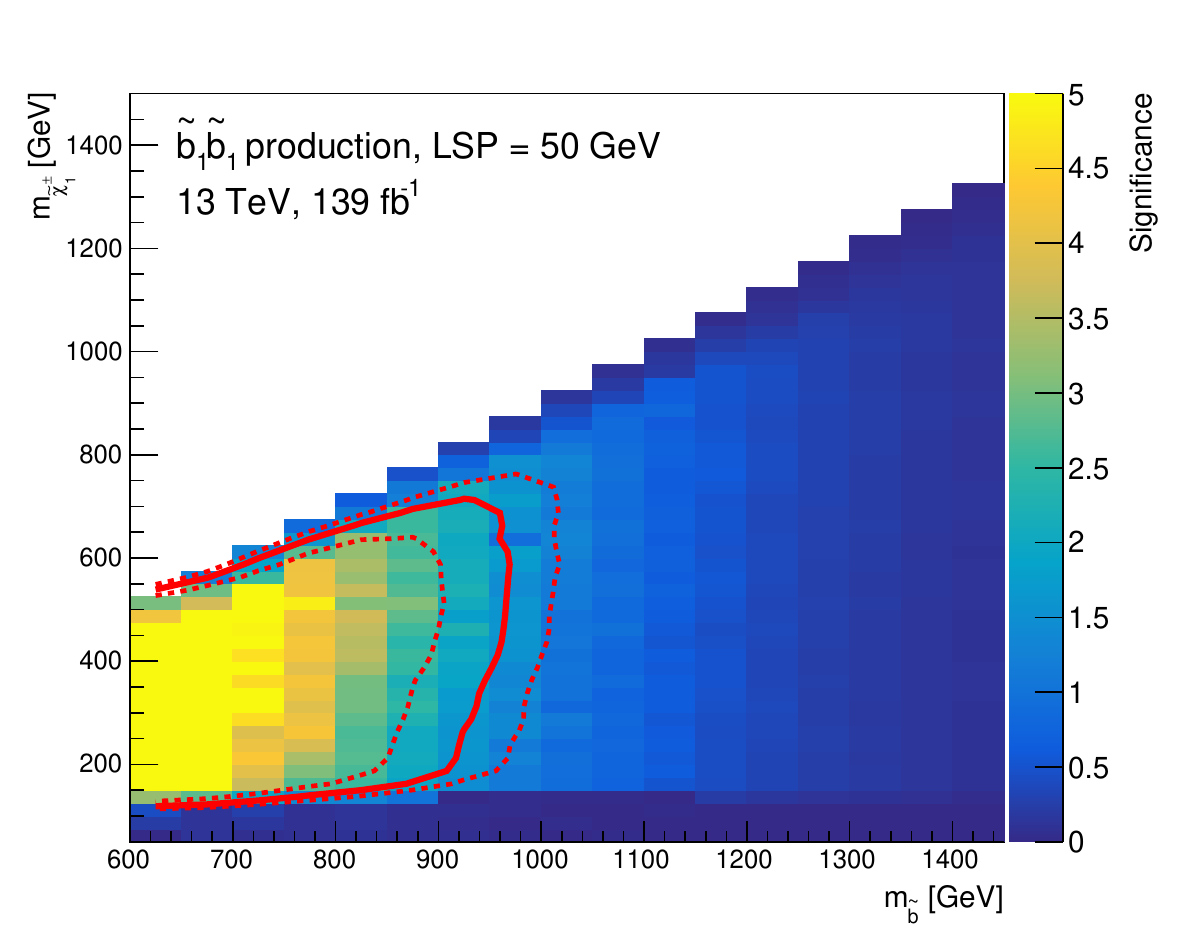}
        \caption{}
        \label{fig:Limits_13TeV_139ifb_Scenario2M_SS3L}
    \end{subfigure}
\caption{
    \textit{Scenario~2} \sbsbModel model, assuming an LSP mass of 50~GeV. 
    Exclusion mass limits obtained with (a) ATLAS Ref.~\citen{ATLAS:2021yyr} ($4\ell$) and (b) ATLAS Ref.~\citen{ATLAS:2019fag} (\llSC/\lll) analyses.
    Top: exclusion mass limits obtained with the individual signal regions; Bottom: exclusion mass limits obtained with the signal region that gives the best signal significance $Z$ for each mass point. 
    Results are shown for $\sqrt s = 13$~TeV and an integrated luminosity of 139~fb$^{-1}$. 
    When present, the dashed lines correspond to the $\pm 1\sigma$ uncertainty on the signal event yield.
}
\label{fig:Limits_13TeV_139ifb_Scenario2M}
\end{figure}

\Cref{fig:Limits_13TeV_139ifb_Scenario1M,fig:Limits_13TeV_139ifb_Scenario2M} show the exclusion limits for the two scenarios considered for the \sbsbModel model: variable LSP mass and fixed LSP mass, respectively. The limits are obtained for $\sqrt{s} = 13$~TeV and an integrated luminosity of 139~fb$^{-1}$. The red dashed lines represent the $\pm 1\sigma$ uncertainty on the number of signal events. 

These figures also display the limits obtained with each individual signal region, highlighting those with the best sensitivity: SR0$^{\mathrm{loose}}_{\mathrm{bveto}}$, SR0$_{\mathrm{breq}}$, and Rpc2L1b. If not present, the signal region has no sensitivity. A direct comparison with the best-performing individual signal region results (shown in the bottom row of \cref{fig:Limits_13TeV_139ifb_Scenario1M_4L,fig:Limits_13TeV_139ifb_Scenario2M_4L}) reveals that the combination of the SR0$^{\mathrm{loose}}_{\mathrm{bveto}}$ and SR0$_{\mathrm{breq}}$ signal regions (SR$_\mathrm{comb}$, shown in the top row of the same figures) improves the exclusion reach by up to 50~GeV. Depending on the LSP mass scenario considered, sbottom masses of up to 960~GeV can generally be excluded. As expected, the limits from the ATLAS Ref.~\citen{ATLAS:2021yyr} ($4\ell$) analysis are generally weaker than those from the signal regions defined in ATLAS Ref.~\citen{ATLAS:2019fag} (\llSC/\lll).

\begin{figure}[t!]
    \begin{subfigure}[t]{0.45\columnwidth}
        \centering
        \includegraphics[width=\linewidth]{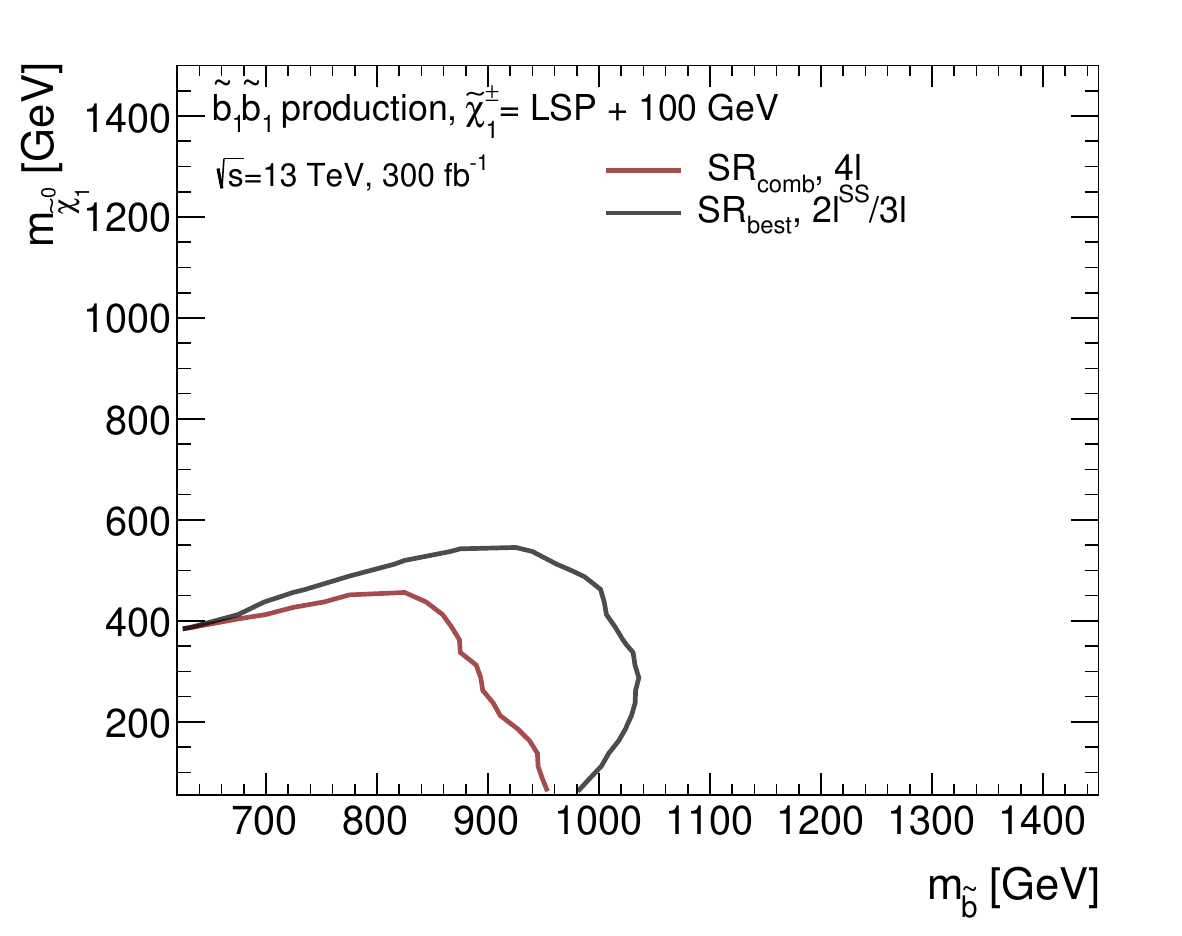}
        \includegraphics[width=\linewidth]{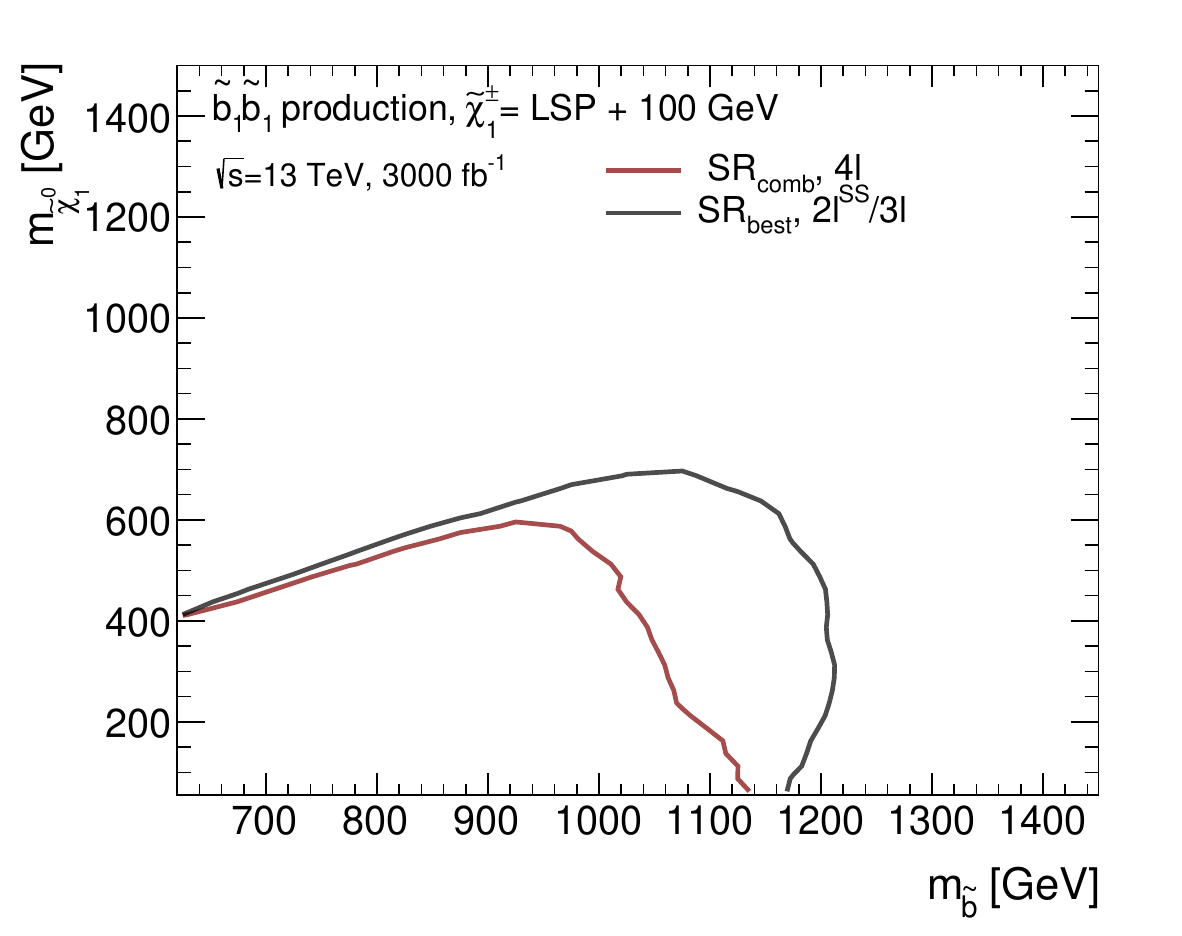}
        \caption{}
    \end{subfigure}%
    \hfill
    \begin{subfigure}[t]{0.45\columnwidth}
        \centering
        \includegraphics[width=\linewidth]{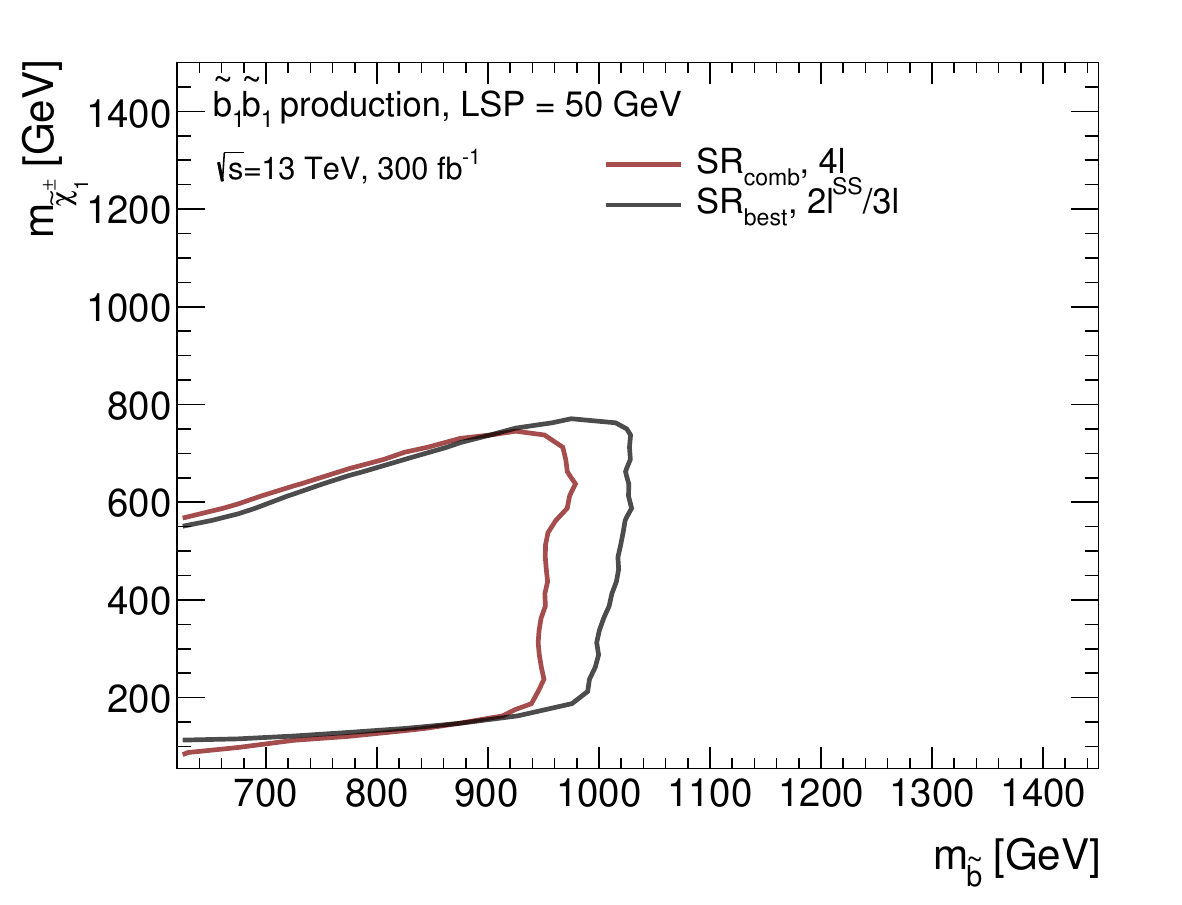}
        \includegraphics[width=\linewidth]{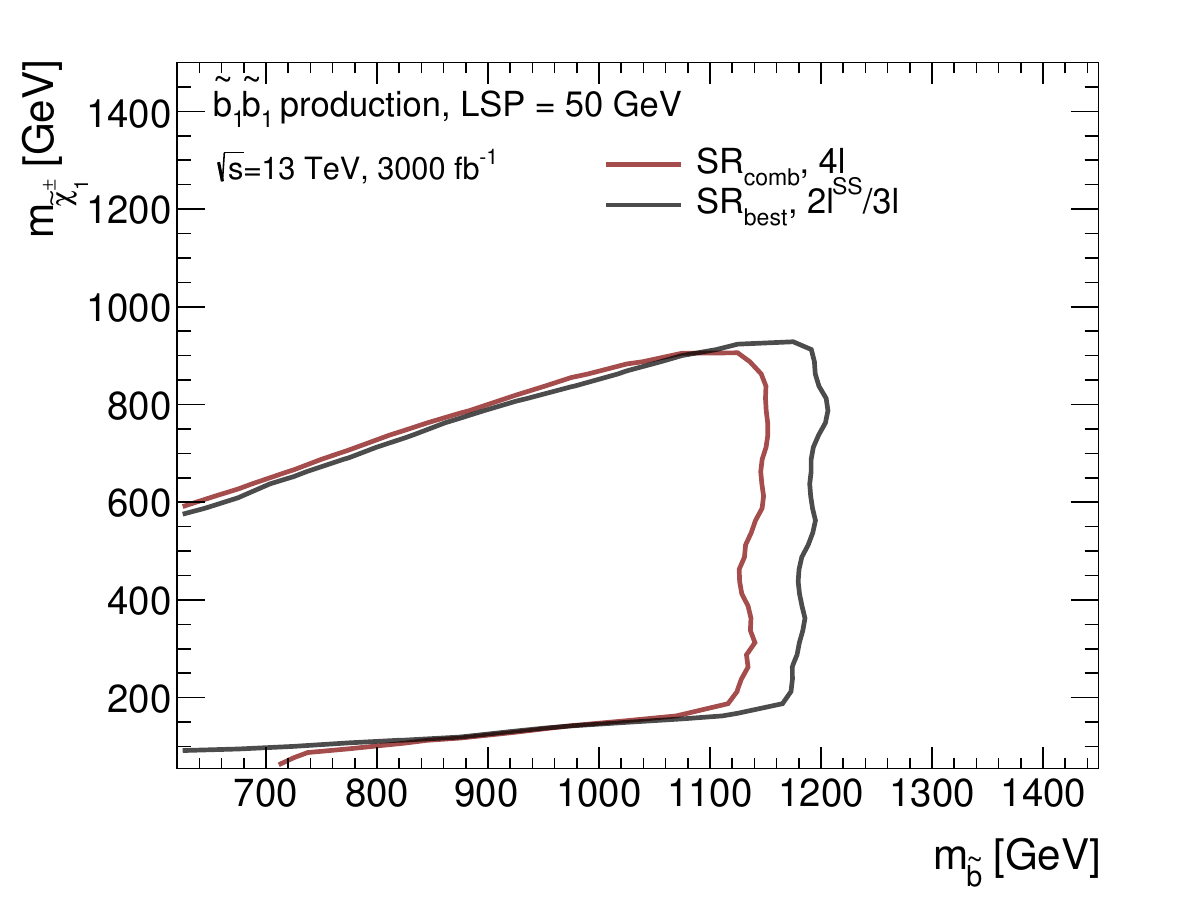}
        \caption{}
    \end{subfigure}
\caption{
     Exclusion mass limits at $\sqrt{s} = 13$~TeV obtained for (a) \textit{Scenario~1} \sbsbModel model and (b) \textit{Scenario~2} \sbsbModel model, showing the results obtained with the ATLAS Ref.~\citen{ATLAS:2021yyr} ($4\ell$) analysis and Refs.~\citen{ATLAS:2019fag,Ducu:2024arr} (\llSC/\lll) analyses.
    The considered integrated luminosities are 300~fb$^{-1}$ (top) and 3000~fb$^{-1}$ (bottom).
}
\label{fig:Limits_13TeV_Scenario1and2M}
\end{figure}

\cref{fig:Limits_13TeV_Scenario1and2M} presents the projected exclusion limits obtained with the two ATLAS analyses at $\sqrt{s} = 13$~TeV, for integrated luminosities of 300~fb$^{-1}$ and 3000~fb$^{-1}$. For the ATLAS Ref.~\citen{ATLAS:2021yyr} ($4\ell$) analysis, the combination of the SR0$^{\mathrm{loose}}_{\mathrm{bveto}}$ and SR0$_{\mathrm{breq}}$ signal regions (SR$_{\textrm{comb}}$) is used, whereas for the ATLAS Ref.~\citen{ATLAS:2019fag} (\llSC/\lll) analysis, the best-performing signal region (SR$_{\textrm{best}}$) is selected. This configuration is consistently adopted throughout the remainder of the paper for all exclusion limits presented. 

Overall, the \textit{Scenario~1} \sbsbModel model yields weaker constraints on the sbottom mass compared to \textit{Scenario~2}. A significant improvement in reach is observed with the increase in luminosity. Depending on the dataset size, sbottom masses of up to 1.2~TeV can be excluded, highlighting the enhanced sensitivity achievable with larger statistics. Notably, the ATLAS Ref.~\citen{ATLAS:2021yyr} ($4\ell$) analysis is also sensitive to the region with $\chinoonepm < 150$~GeV, where the ATLAS Ref.~\citen{ATLAS:2019fag} (\llSC/\lll) analysis exhibits no sensitivity.

The ATLAS Ref.~\citen{ATLAS:2021yyr} ($4\ell$) analysis performs surprisingly well for both \sbsbModel model configurations, despite its signal regions being originally optimized for different SUSY scenarios. Sensitivity is notably higher in the \textit{Scenario~2} configuration, primarily due to the larger mass splitting between the $\chinoonepm$ and the LSP, which produces more energetic leptons from the $W$ boson decays and enhances signal acceptance. Notably, this analysis retains sensitivity even in the \(\tilde{\chi}_1^\pm < 150\)~GeV region, thanks to the absence of a stringent \met requirement. In this phase space, sbottom masses up to 740~GeV can be excluded with an integrated luminosity of 3000~fb$^{-1}$. These results highlight the added value of four-lepton final states and support their continued use in future sbottom searches, especially for challenging kinematic regimes.

\begin{figure}[t!]
    \begin{subfigure}[t]{0.45\columnwidth}
        \centering
        \includegraphics[width=\linewidth]{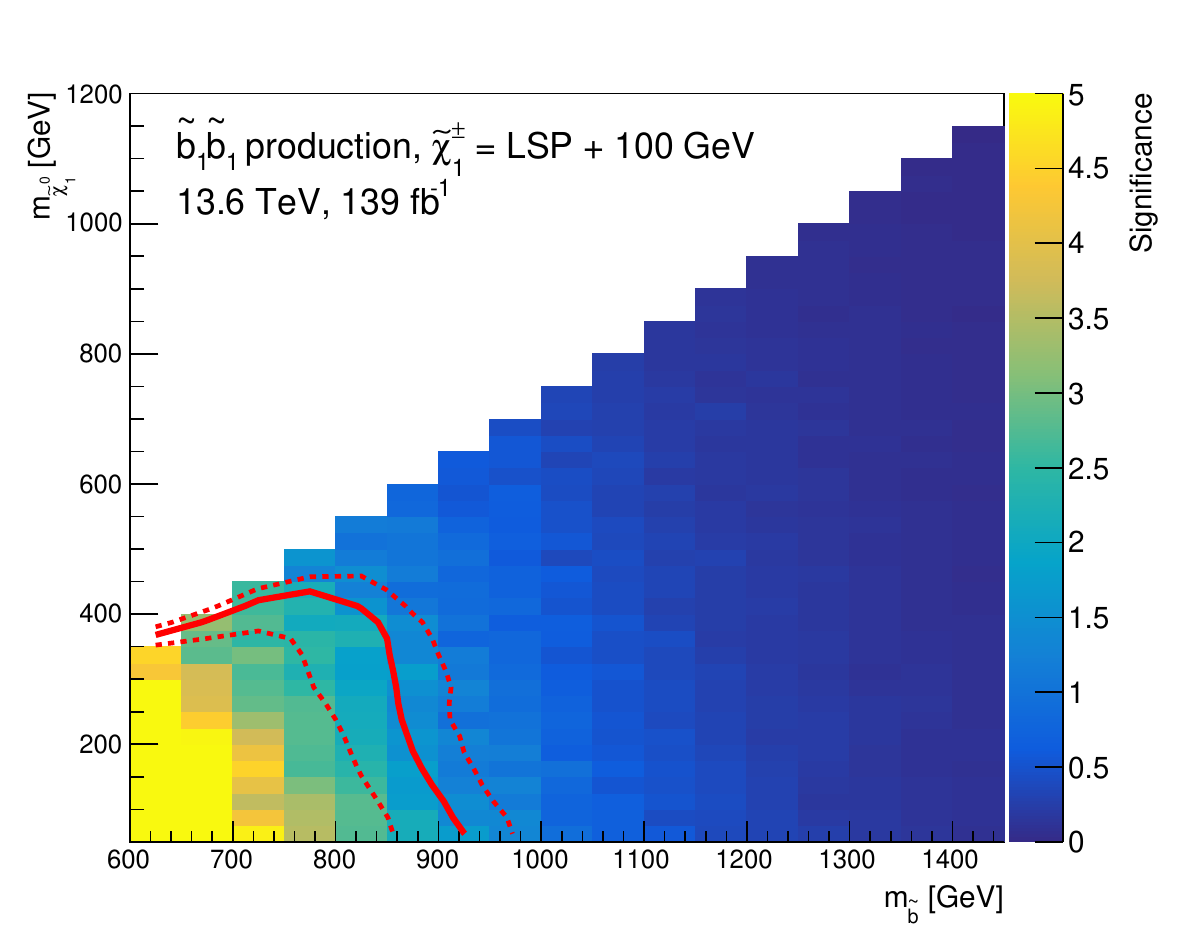}
        \includegraphics[width=\linewidth]{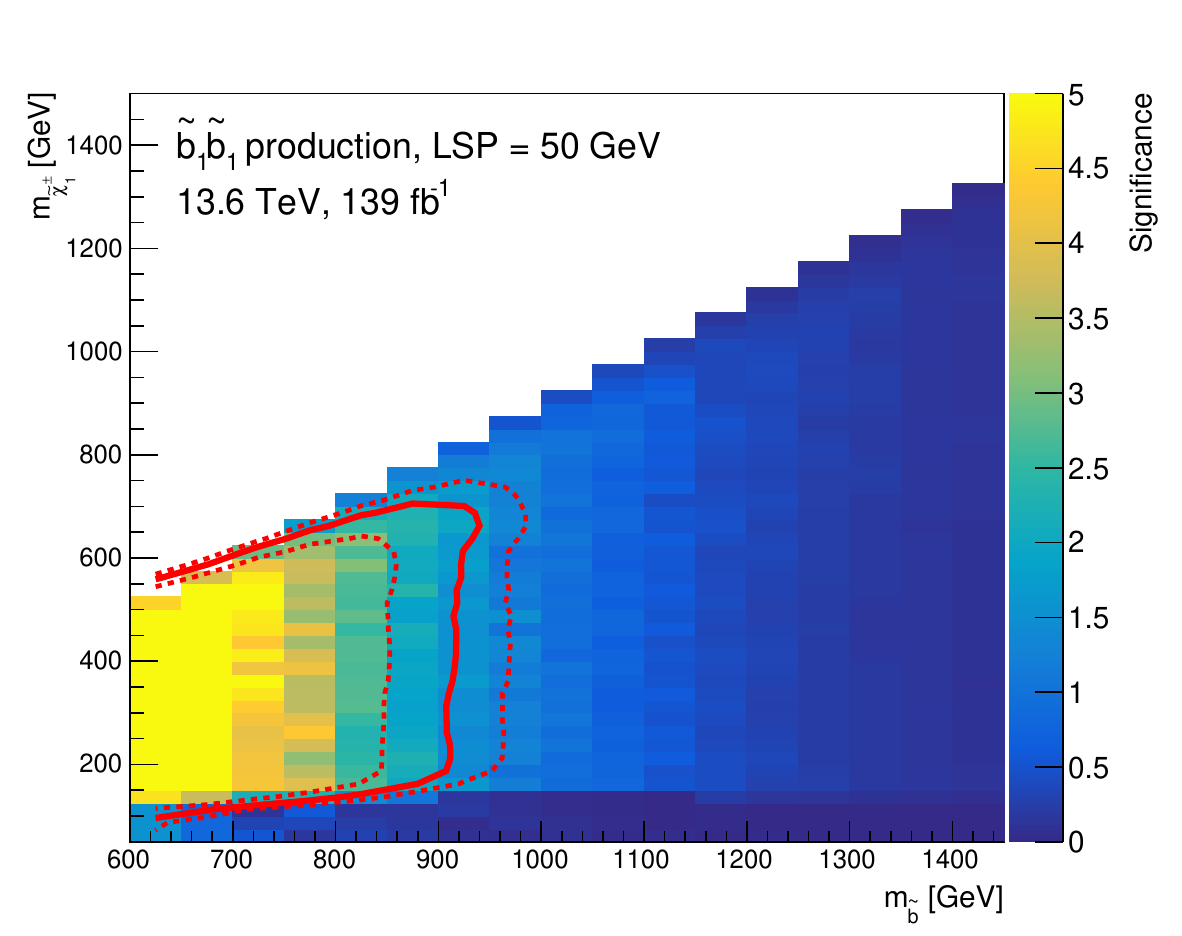}
        \caption{}
    \end{subfigure}%
    \hfill
    \begin{subfigure}[t]{0.45\columnwidth}
        \centering
        \includegraphics[width=\linewidth]{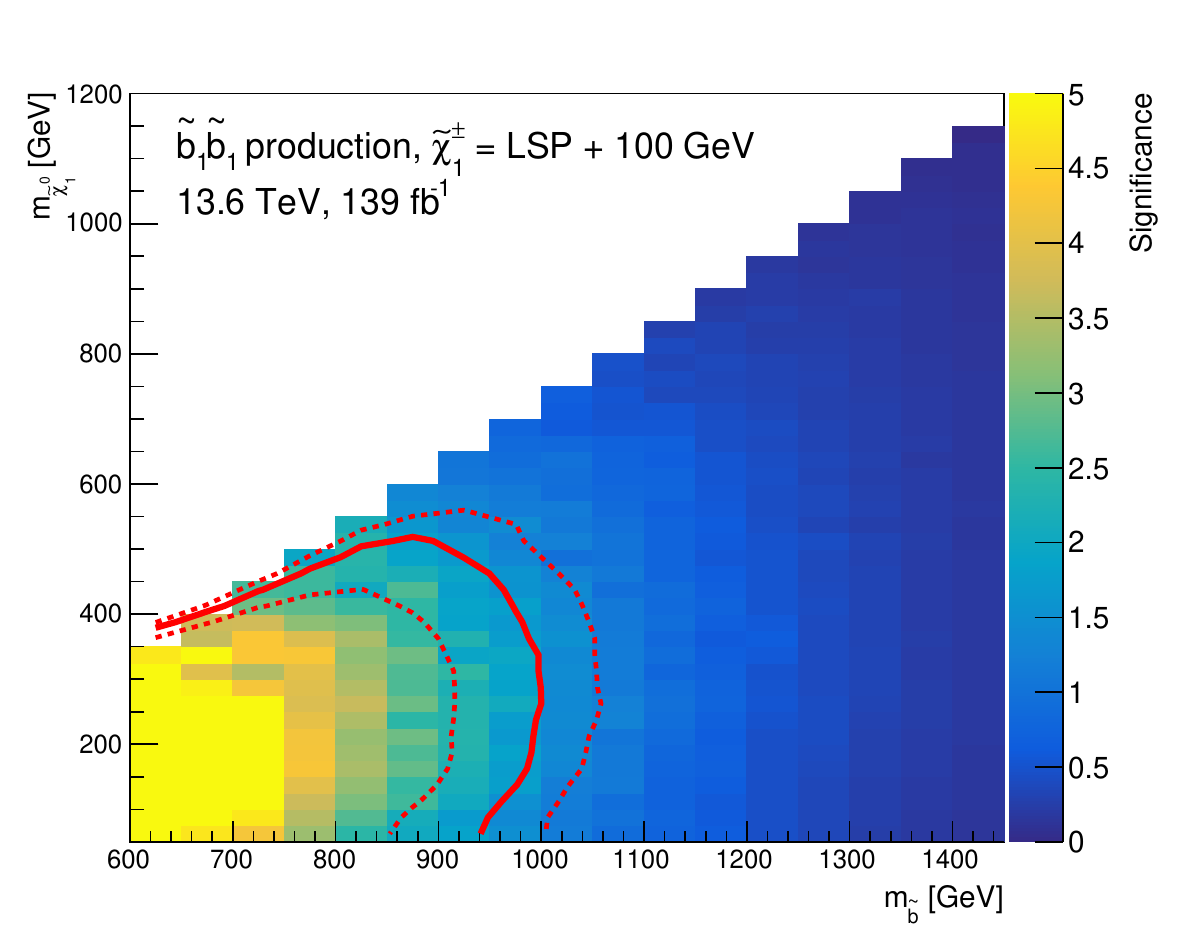}
        \includegraphics[width=\linewidth]{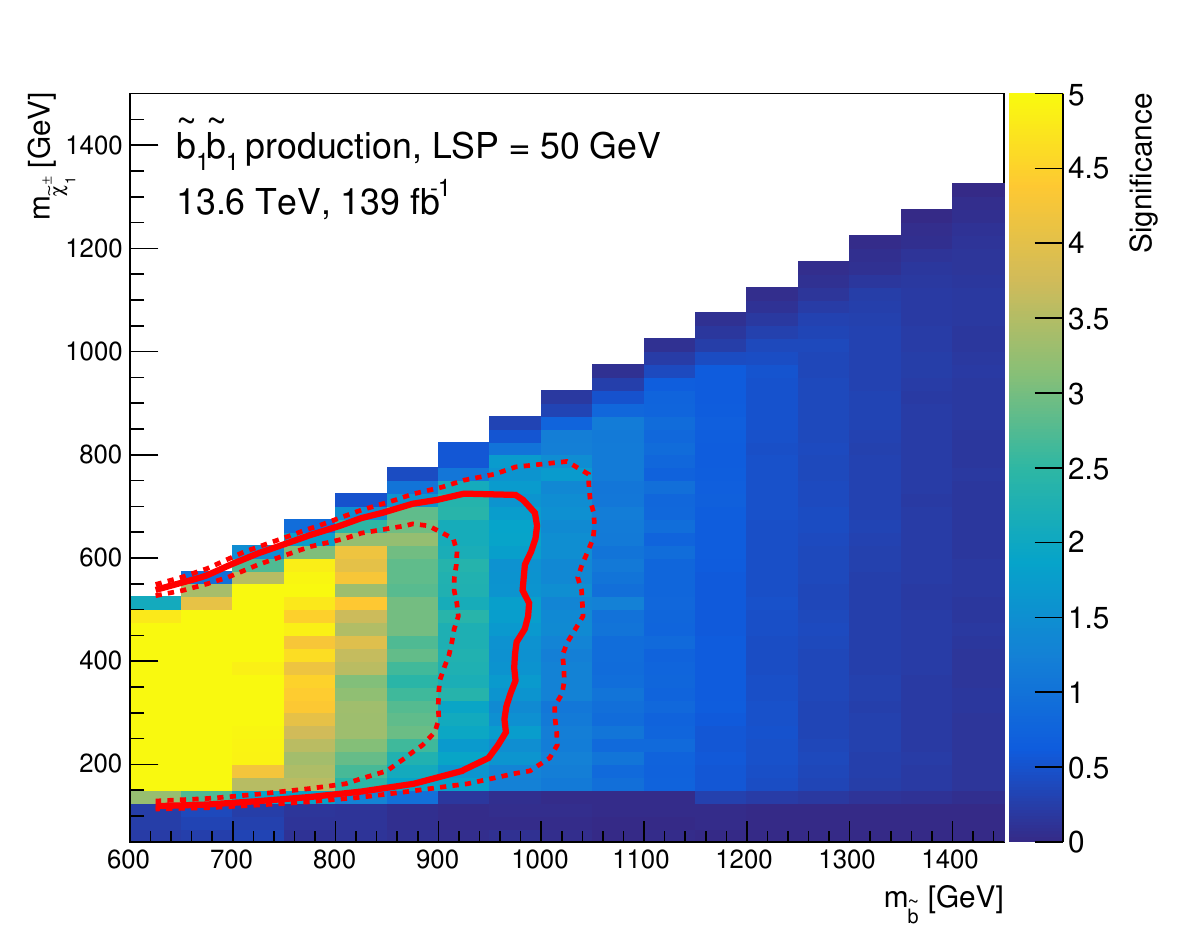}
        \caption{}
    \end{subfigure}
\caption{
    Exclusion mass limits obtained with (a) ATLAS Ref.~\citen{ATLAS:2021yyr} ($4\ell$) and (b) Refs.~\citen{ATLAS:2019fag,Ducu:2024arr} (\llSC/\lll) analyses.
    Top: \textit{Scenario~1} \sbsbModel model; Bottom: \textit{Scenario~2} \sbsbModel model.
    Results are shown for $\sqrt{s} = 13.6$~TeV and an integrated luminosity of 139~fb$^{-1}$.
    The dashed lines correspond to the $\pm 1\sigma$ uncertainty on the signal event yield.
}
\label{fig:Limits_13p6TeV_Scenario1and2M_139ifb}
\end{figure}

\section{Projected Exclusion Limits at 13.6~TeV}

\begin{figure}[t!]
\centering
    \begin{subfigure}[t]{0.45\columnwidth}
        \centering
        \includegraphics[width=\linewidth]{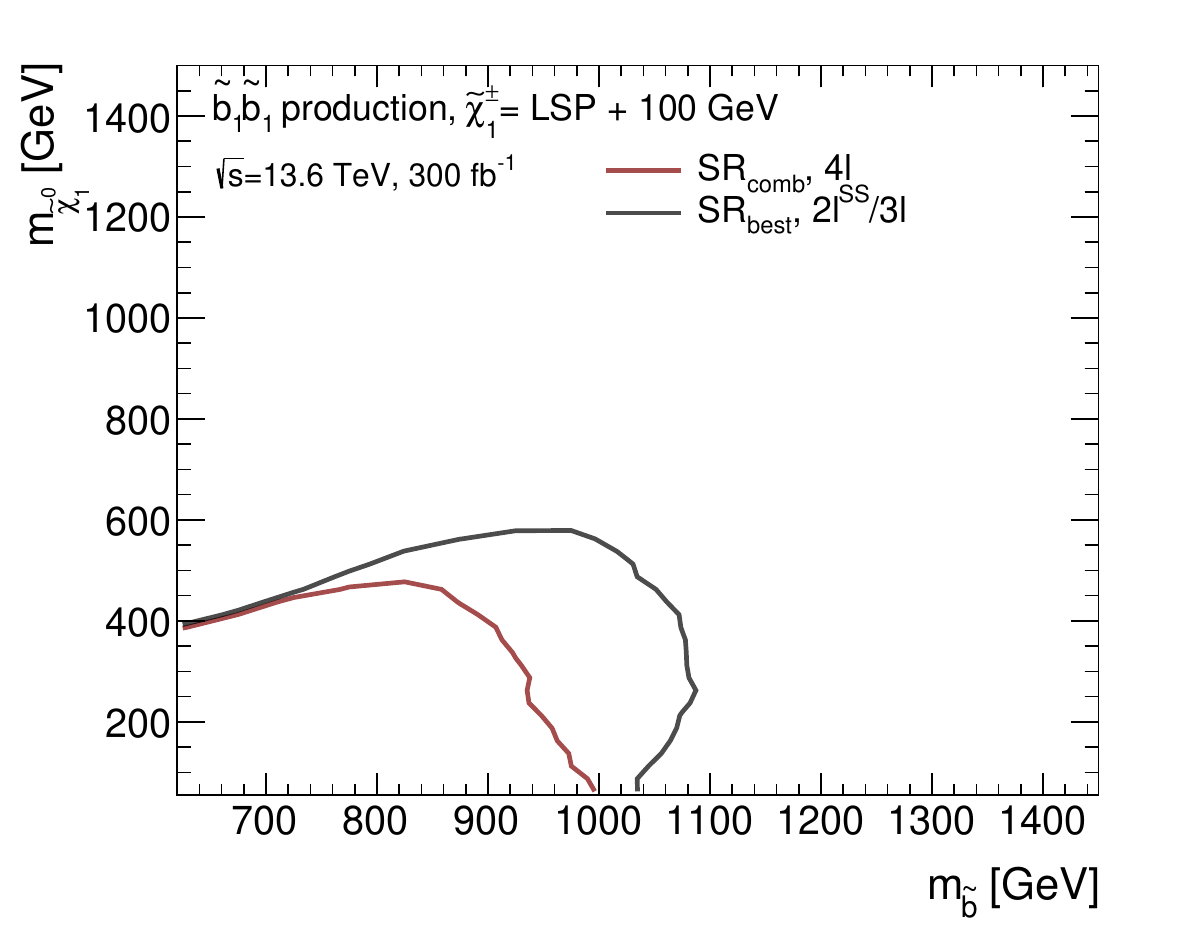}
        \includegraphics[width=\linewidth]{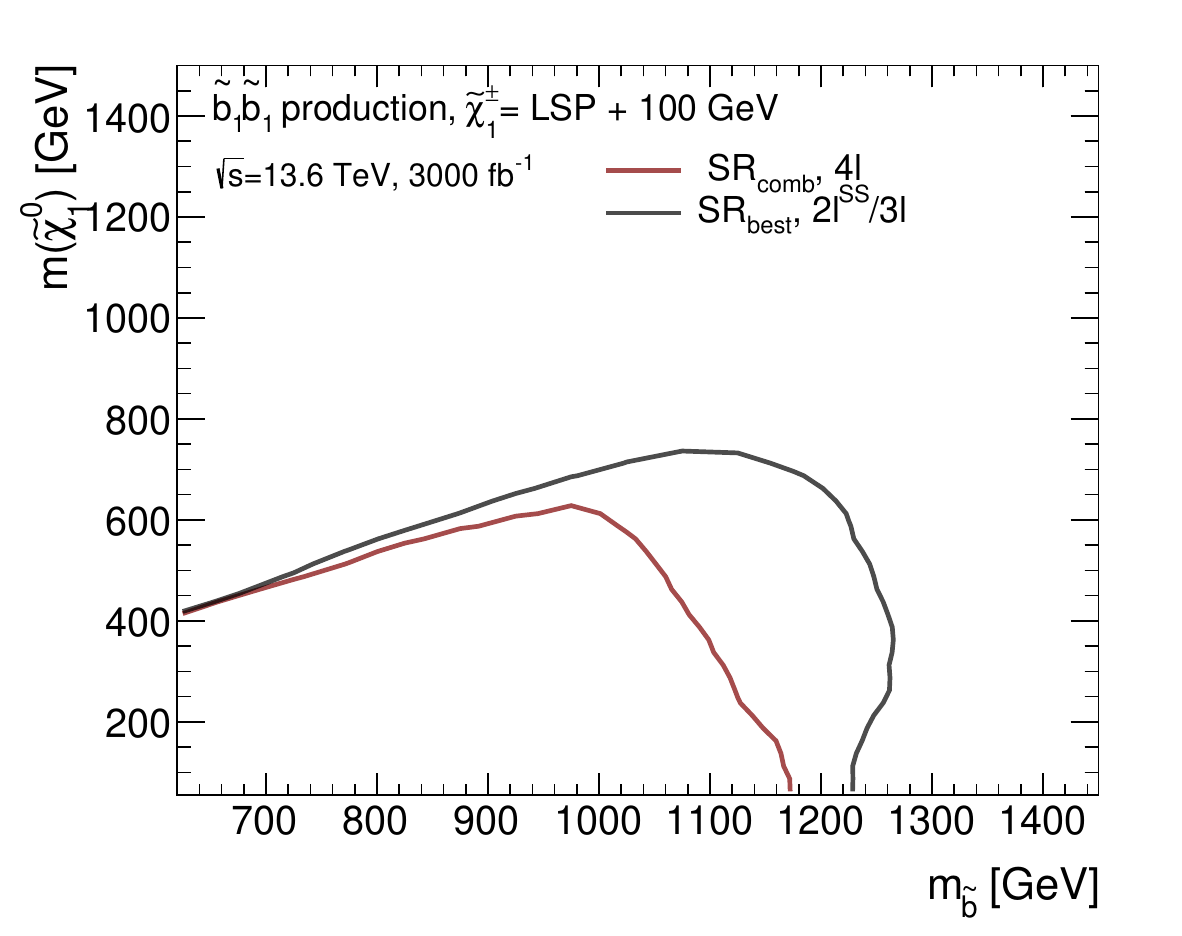}
        \caption{}
    \end{subfigure}%
    \hfill
    \begin{subfigure}[t]{0.45\columnwidth}
        \centering
        \includegraphics[width=\linewidth]{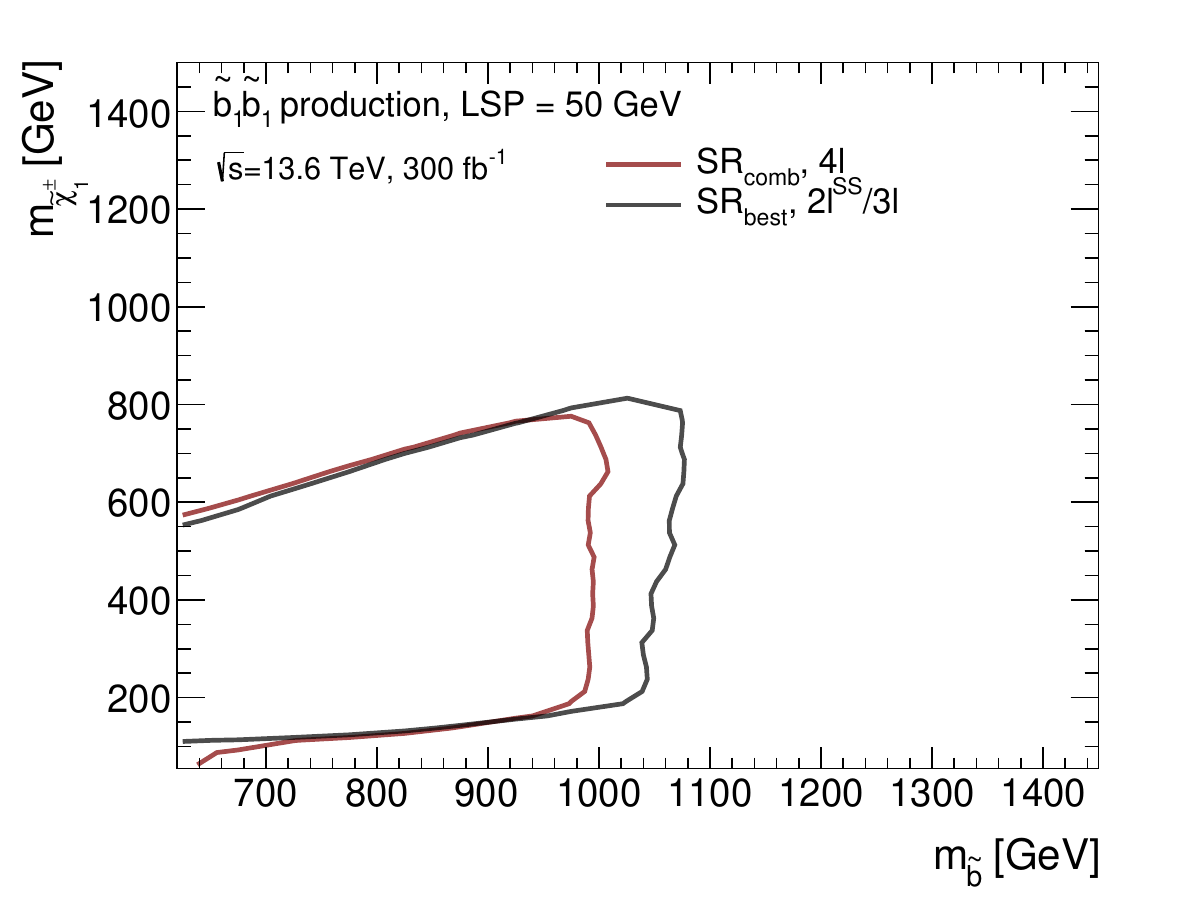}
        \includegraphics[width=\linewidth]{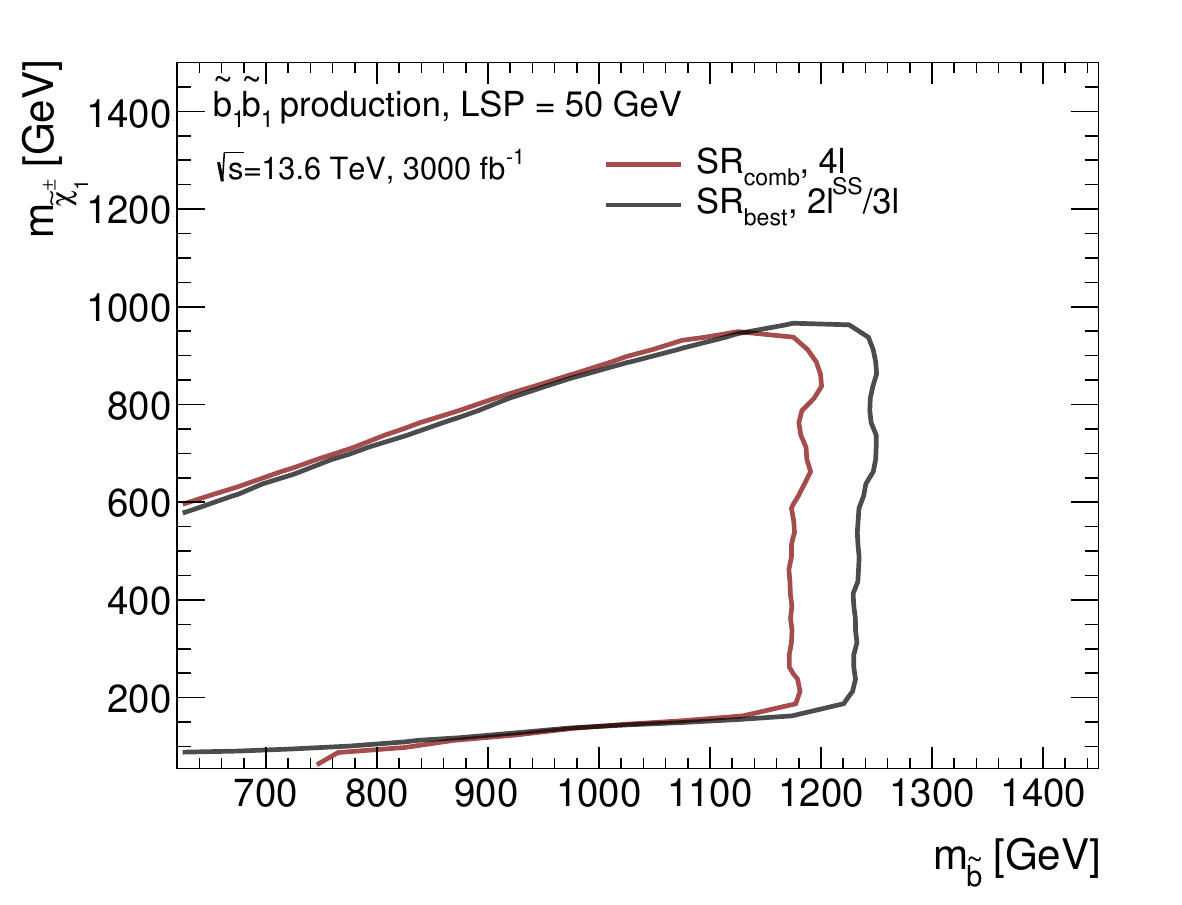}
        \caption{}
    \end{subfigure}
\caption{
    Exclusion mass limits at $\sqrt{s} = 13.6$~TeV obtained for (a) \textit{Scenario~1} \sbsbModel model and (b) \textit{Scenario~2} \sbsbModel model, showing the results obtained with the ATLAS Ref.~\citen{ATLAS:2021yyr} ($4\ell$) analysis and Refs.~\citen{ATLAS:2019fag,Ducu:2024arr} (\llSC/\lll) analyses.
    The considered integrated luminosities are 300~fb$^{-1}$ (top) and 3000~fb$^{-1}$ (bottom).
}
\label{fig:Limits_13p6TeV_Scenario1and2M}
\end{figure}

\cref{fig:Limits_13p6TeV_Scenario1and2M_139ifb} shows the projected exclusion limits for the two scenarios considered for the \sbsbModel model, obtained at $\sqrt{s} = 13.6$~TeV with an integrated luminosity of 139~fb$^{-1}$. The previous conclusion holds: the limits obtained with the ATLAS Ref.~\citen{ATLAS:2021yyr} ($4\ell$) analysis are generally weaker than those obtained using the signal regions defined in ATLAS Ref.~\citen{ATLAS:2019fag} (\llSC/\lll). \cref{fig:Limits_13p6TeV_Scenario1and2M} presents the corresponding projections for integrated luminosities of 300~fb$^{-1}$ and 3000~fb$^{-1}$. As expected, the increase in sensitivity due to the rise in center-of-mass energy from 13~TeV to 13.6~TeV is modest, as the increase in the production cross section is insignificant. Depending on the luminosity, sbottom masses of up to 1.25~TeV can be excluded, generally with tighter limits obtained for the \textit{Scenario~2} \sbsbModel model. This improved sensitivity for the second scenario  holds across most of the phase space, driven by the more favorable kinematic configurations and larger mass splittings, which enhance the signal acceptance in the relevant search regions.

\begin{figure}[t]
\centering
    \begin{subfigure}[t]{0.45\columnwidth}
        \centering
        \includegraphics[width=\linewidth]{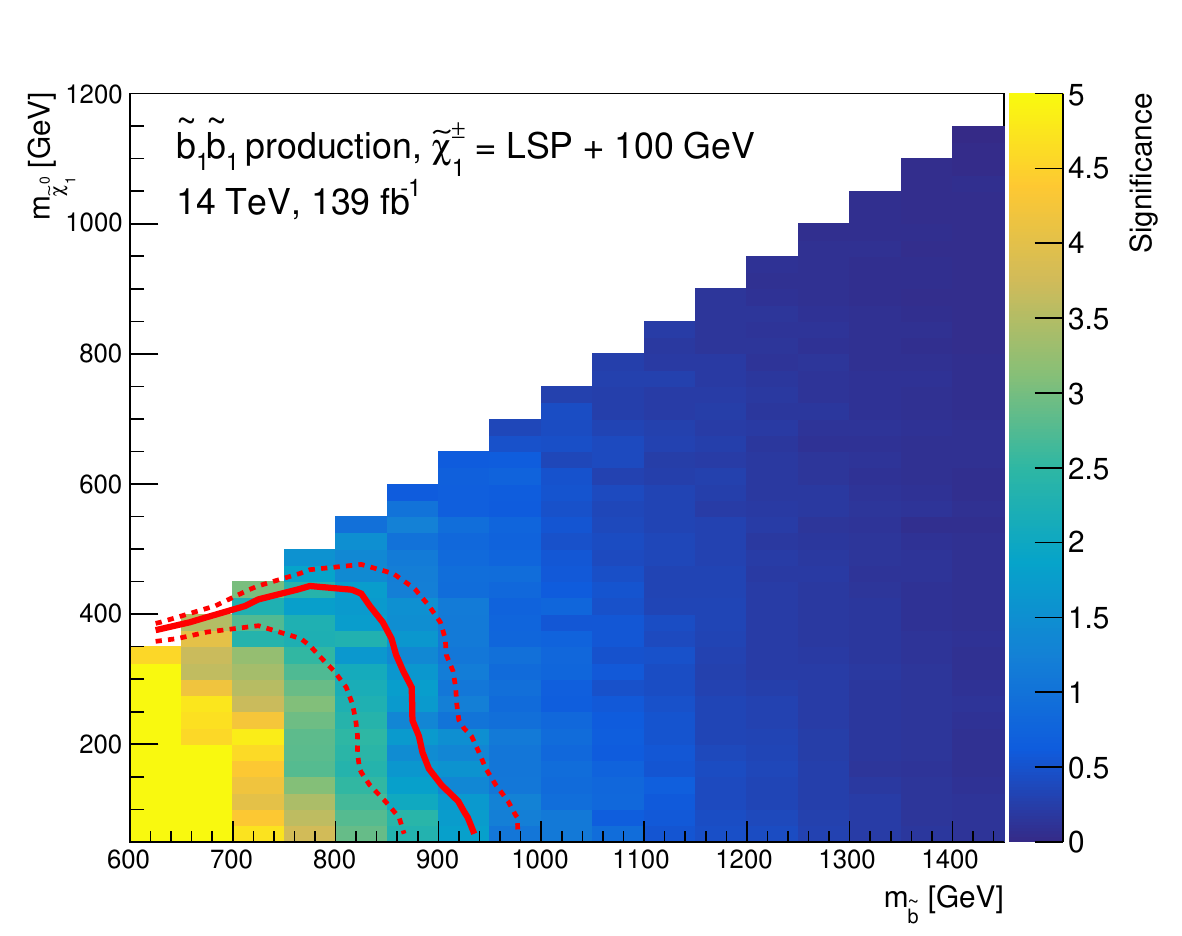}
        \includegraphics[width=\linewidth]{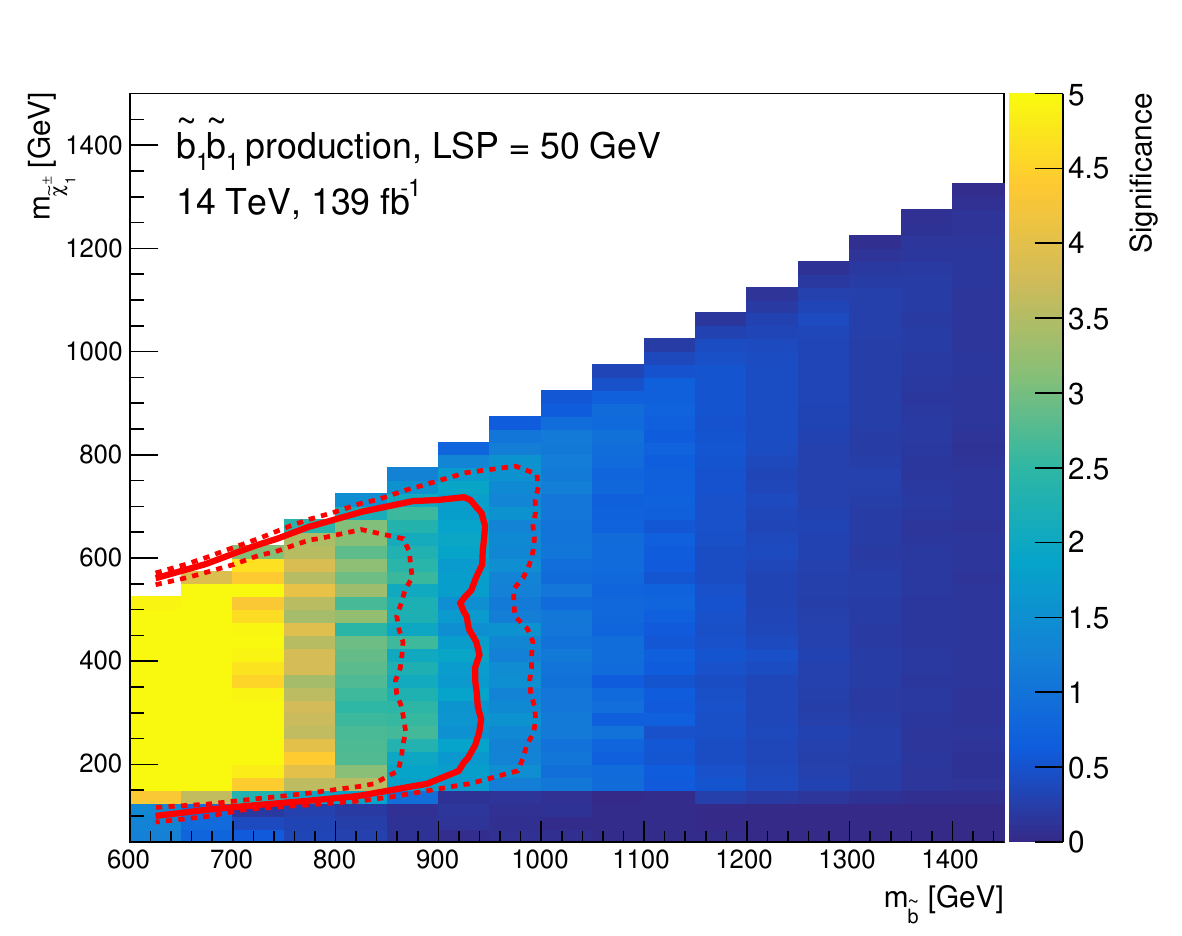}
        \caption{}
    \end{subfigure}%
    \hfill
    \begin{subfigure}[t]{0.45\columnwidth}
        \centering
        \includegraphics[width=\linewidth]{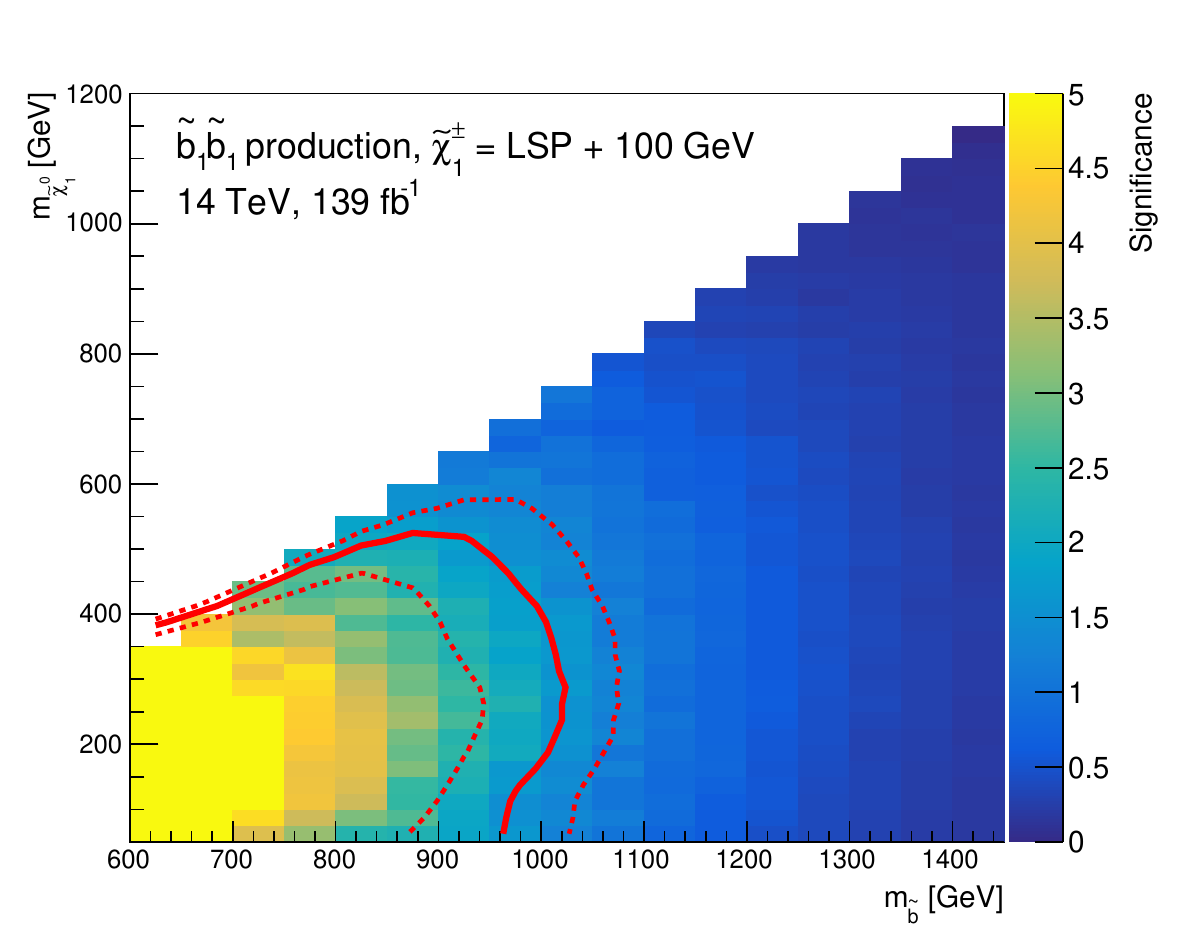}
        \includegraphics[width=\linewidth]{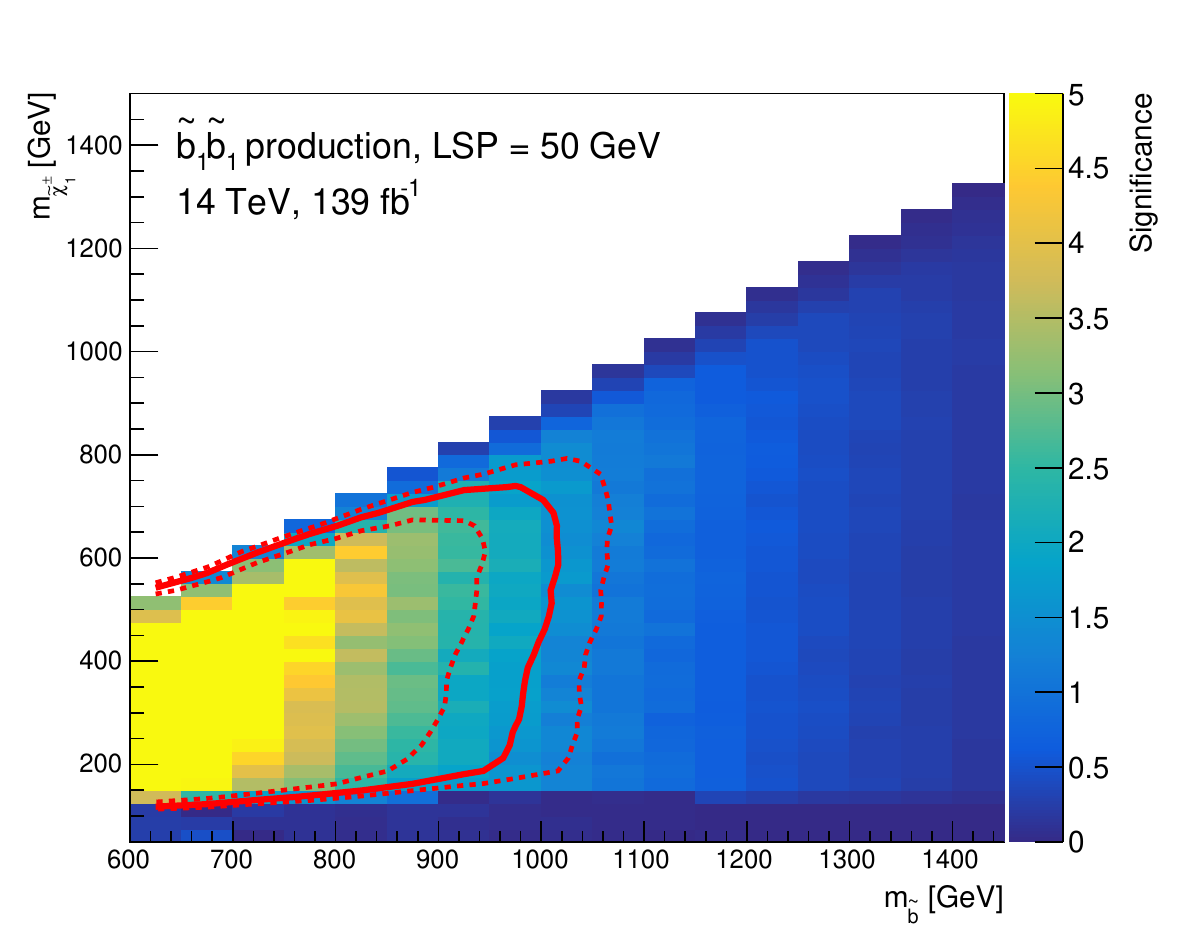}
        \caption{}
    \end{subfigure}
\caption{
    Exclusion mass limits obtained with (a) ATLAS Ref.~\citen{ATLAS:2021yyr} ($4\ell$) and (b) Refs.~\citen{ATLAS:2019fag,Ducu:2024arr} (\llSC/\lll) analyses.
    Top: \textit{Scenario~1} \sbsbModel model; Bottom: \textit{Scenario~2} \sbsbModel model.
    Results are shown for $\sqrt{s} = 14$~TeV and an integrated luminosity of 139~fb$^{-1}$.
    The dashed lines correspond to the $\pm 1\sigma$ uncertainty on the signal event yield.
}
\label{fig:Limits_14TeV_Scenario1and2M_139ifb}
\end{figure}

\section{Projected Exclusion Limits at 14~TeV}

\begin{figure}[t!]
\centering
    \begin{subfigure}[t]{0.45\columnwidth}
        \centering
        \includegraphics[width=\linewidth]{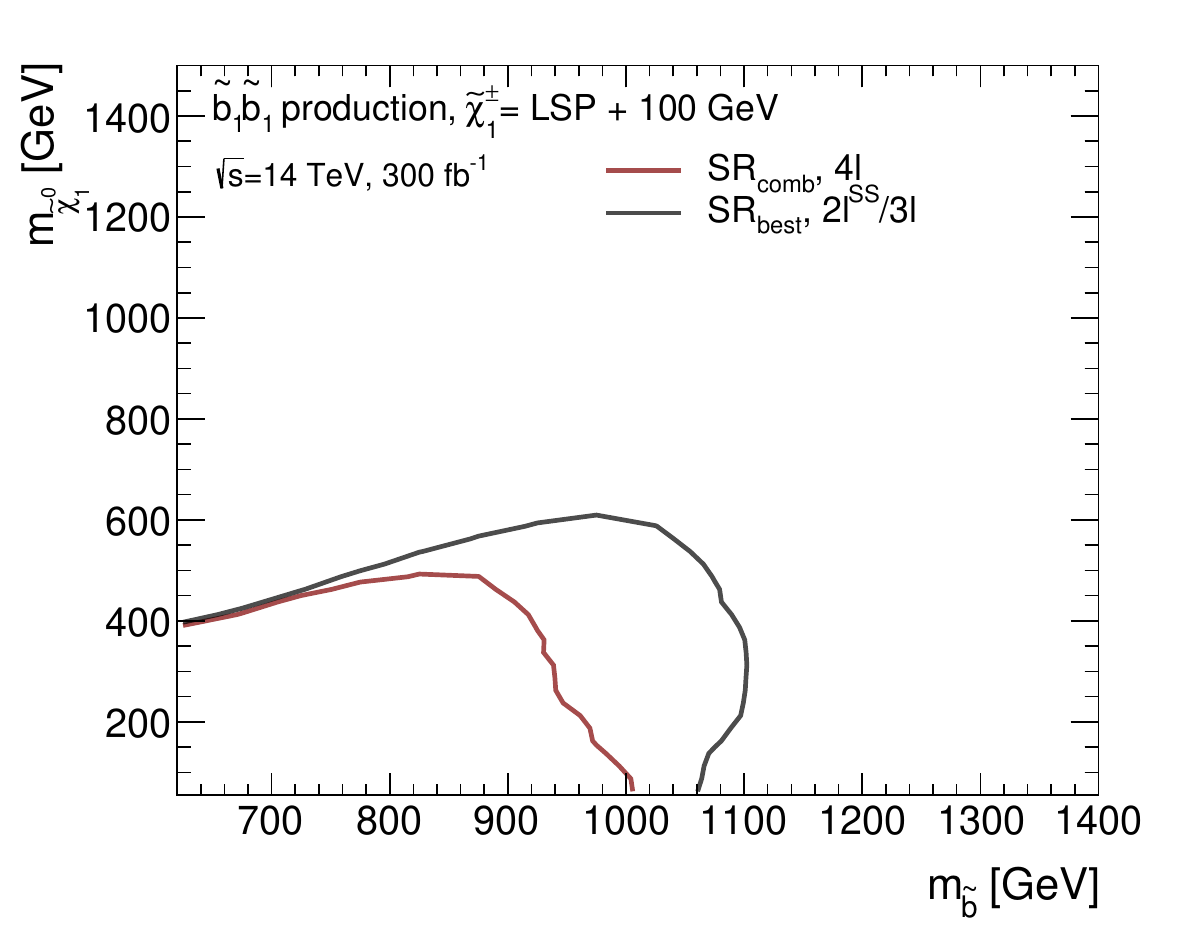}
        \includegraphics[width=\linewidth]{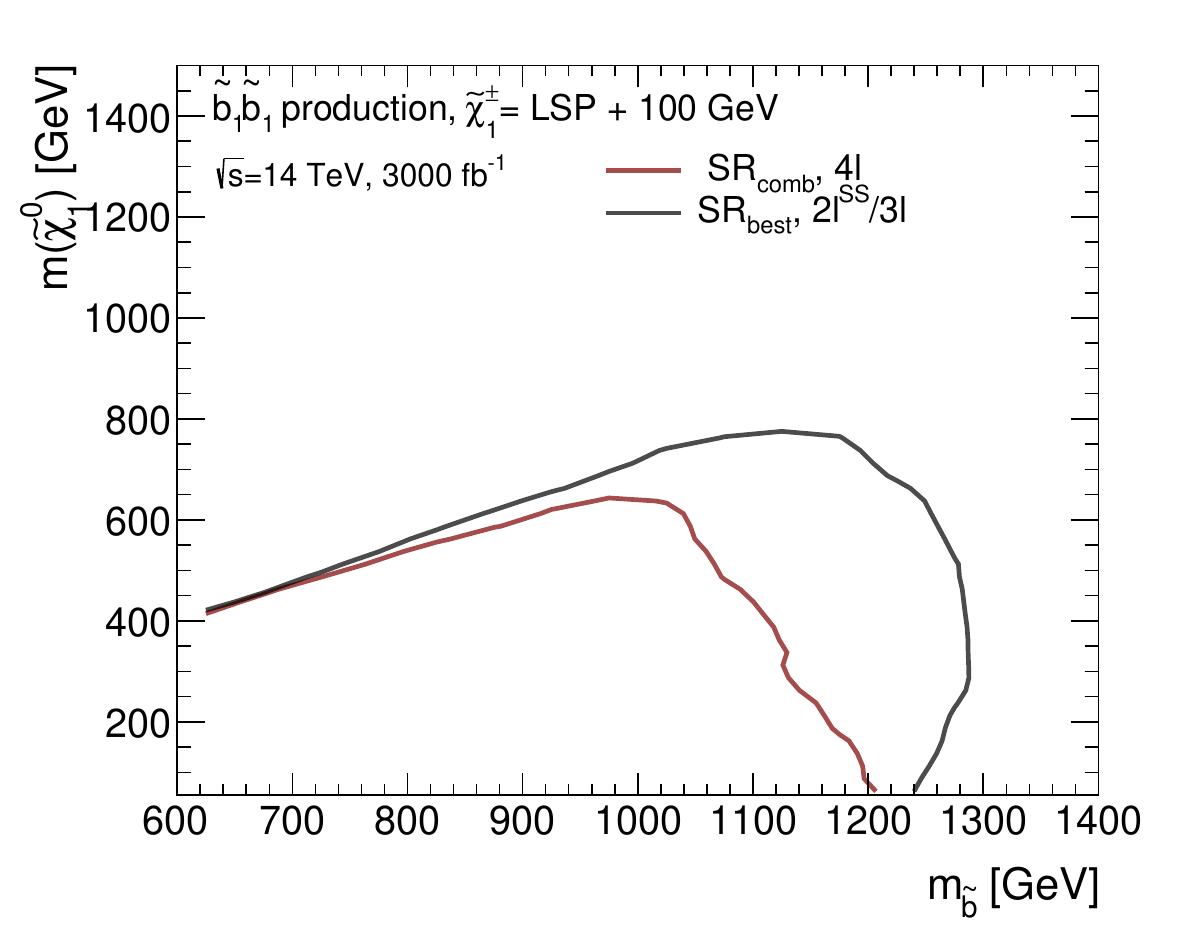}
        \caption{}
    \end{subfigure}%
    \hfill
    \begin{subfigure}[t]{0.45\columnwidth}
        \centering
        \includegraphics[width=\linewidth]{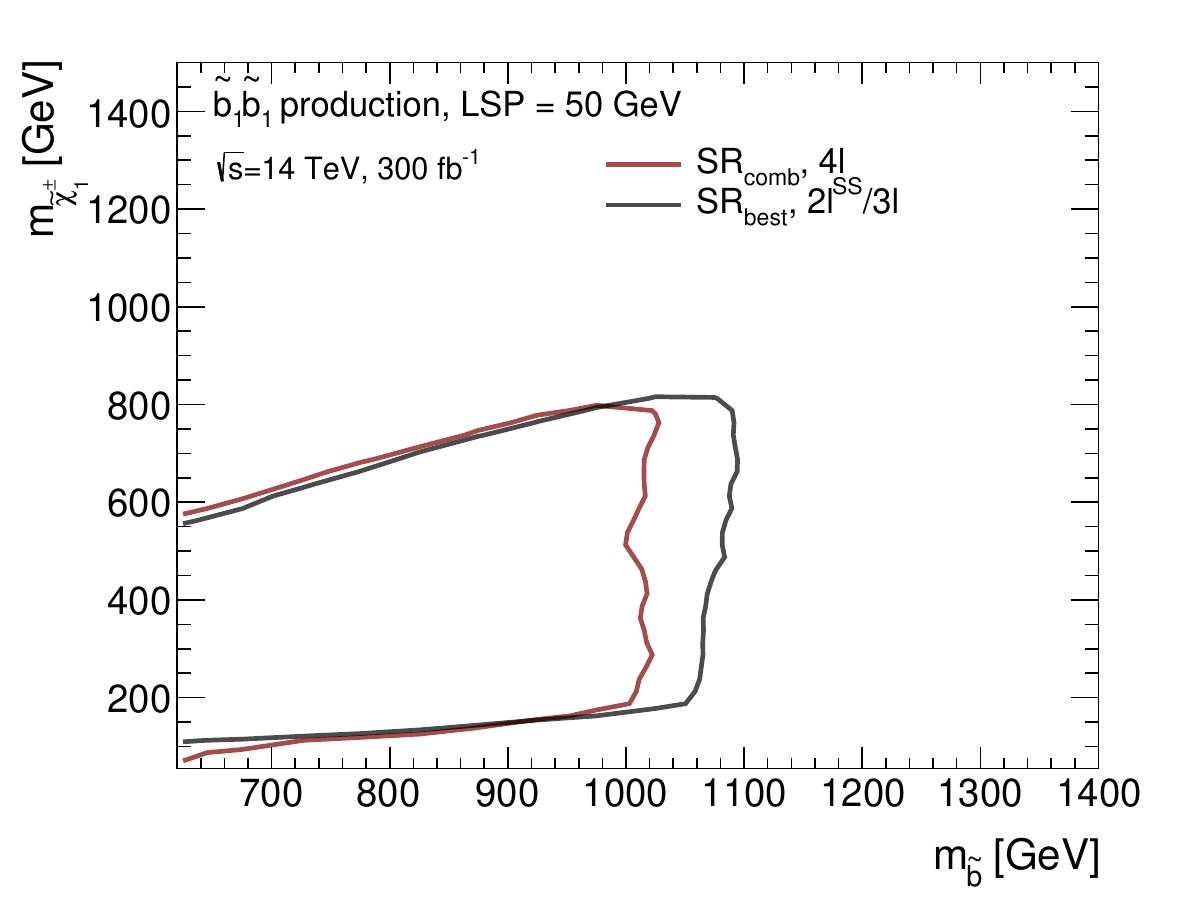}
        \includegraphics[width=\linewidth]{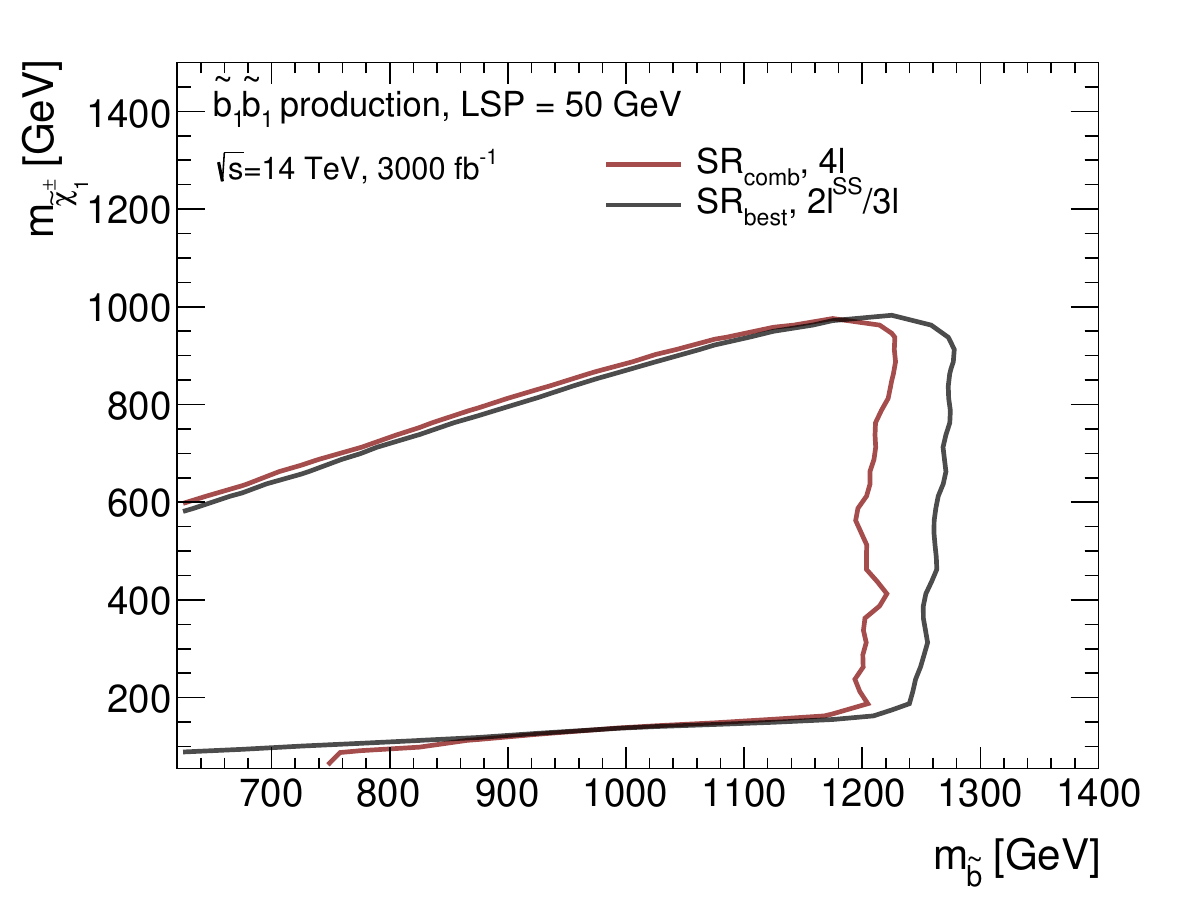}
        \caption{}
    \end{subfigure}
\caption{
    Exclusion mass limits at $\sqrt{s} = 14$~TeV obtained for (a) \textit{Scenario~1} \sbsbModel model and (b) \textit{Scenario~2} \sbsbModel model, showing the results obtained with the ATLAS Ref.~\citen{ATLAS:2021yyr} ($4\ell$) analysis and Refs.~\citen{ATLAS:2019fag,Ducu:2024arr} (\llSC/\lll) analyses.
    The considered integrated luminosities are 300~fb$^{-1}$ (top) and 3000~fb$^{-1}$ (bottom).
}
\label{fig:Limits_14TeV_Scenario1and2M}
\end{figure}

\cref{fig:Limits_14TeV_Scenario1and2M_139ifb} shows the projected exclusion limits for the two scenarios considered for the \sbsbModel model, obtained at $\sqrt{s} = 14$~TeV with an integrated luminosity of 139~fb$^{-1}$, while \cref{fig:Limits_14TeV_Scenario1and2M} presents the corresponding projections for integrated luminosities of 300~fb$^{-1}$ and 3000~fb$^{-1}$. Depending on the luminosity, sbottom masses of up to 1.28~TeV can be excluded, with generally tighter limits obtained for the \textit{Scenario~2} \sbsbModel model.

\section{Prospects for improved sensitivity}	

The results presented in this paper are based on the existing signal regions defined in the two ATLAS analyses Refs.~\citen{ATLAS:2021yyr,ATLAS:2019fag}, which were originally optimized for LHC Run-2 data. However, substantial improvements in sensitivity are expected to be achievable at LHC Run-3 and the HL-LHC through dedicated re-optimization of these signal regions.

In particular, in ATLAS Ref.~\citen{ATLAS:2021yyr}, for scenarios with compressed spectra, it would be beneficial to lower the selection thresholds on the transverse momentum of the leading leptons (e.g., considering leptons with $\pt < 20$~GeV), and to retain events with reduced values of \meff and \met. In both ATLAS analyses, it would also be advantageous to explore the use of soft $b$-taggers, which allow for the identification of $b$-tagged jets with $\pt$ below the currently applied threshold of 20~GeV. While such strategies may lead to an increased contribution from fake/non-prompt leptons, machine learning techniques could be employed to suppress background contributions and maintain sensitivity. The use of machine learning classifiers trained on dedicated kinematic variables could help define signal regions that are both inclusive and robust to such backgrounds.

For the very boosted regime, it would be advantageous to relax the selection on the $\met/\meff$ ratio, particularly in ATLAS Ref.~\citen{ATLAS:2021yyr}. Given the typically low SM background in four-lepton final states, further optimization of the signal regions in this channel appears particularly promising. This could involve combining looser kinematic cuts with machine learning techniques to improve signal discrimination without compromising background control. Additionally, the use of binned signal regions~--~either as standalone selections or in combination with machine learning~based classifiers~--~can further enhance sensitivity by exploiting the full shape information of relevant observables.

In summary, re-optimizing analysis strategies with machine learning tools and tailoring them to specific kinematic regimes could significantly enhance the discovery or exclusion potential for sbottom searches in the next phases of the LHC program.

\section{Conclusions}	

The results presented in this paper demonstrate the exclusion potential of various ATLAS analyses for the \sbsbModel model under different mass assumptions and luminosity scenarios, extending the study presented in Ref.~\citen{Ducu:2024arr}. The sbottom is assumed to decay with a 100\% branching ratio via a one-step decay through a chargino, $\tilde{b}_1 \to t \tilde{\chi}_1^\pm$. The chargino subsequently decays into a $W$ boson and the lightest neutralino, $\tilde{\chi}_1^\pm \to W \tilde{\chi}_1^0$, also with a 100\% branching ratio. Two distinct configurations were considered: \textit{Scenario~1}, featuring a variable LSP mass, and \textit{Scenario~2}, where the LSP mass is fixed at 50~GeV. Dedicated Monte Carlo simulations were performed at three center-of-mass energies (13~TeV, 13.6~TeV, and 14~TeV) and for three integrated luminosity scenarios (139~fb$^{-1}$, 300~fb$^{-1}$, and 3000~fb$^{-1}$), using \texttt{MadGraph}+\texttt{Pythia} for event generation and \texttt{DELPHES} for ATLAS detector simulation.

Overall, \textit{Scenario~2} consistently leads to stronger exclusion limits due to the enhanced kinematics of the final-state particles. These improvements can offer a very good sensitivity and a robutstness of the analysis. The reinterpretation of the ATLAS $4\ell$ analysis\cite{ATLAS:2021yyr} has shown a surprisingly good performance, especially in regions where traditional searches lose sensitivity, such as for low $\chinoonepm$ masses ($<150$~GeV). This is primarily due to the relaxed \met requirement and the clean signature of multiple isolated leptons.

Furthermore, projections at $\sqrt{s} = 14$~TeV indicate that, with increased integrated luminosities, sbottom masses up to 1.28~TeV can be excluded. The complementary nature of the signal regions defined by 0~$b$-tagged jets versus those with $\geq 1$~$b$-tagged jets, as well as multilepton versus dilepton final states, was found to be especially important. This complementarity enhances sensitivity when combining multiple channels or optimizing selections based on final-state multiplicities and kinematic thresholds.

These results emphasize the continued relevance of multilepton final states in future sbottom searches and suggest that further gains can be achieved through dedicated signal region optimization, potentially incorporating machine learning–based approaches and more inclusive object selections.

Looking ahead to LHC Run-3 and the HL-LHC, the dominant uncertainties are expected to arise from theoretical predictions and experimental systematics, while statistical uncertainties will be substantially reduced due to the larger data samples. Improved modeling of SM backgrounds is also anticipated, as the increased statistics will enable more precise control regions to better constrain these backgrounds. On the experimental front, notable advancements  in the reconstruction, identification, and isolation of physics objects are foreseen, largely driven by the upgraded ITk inner tracking detector, as highlighted in Ref.~\citen{ATLAS:2019mfr}.

The upcoming upgrades to the ATLAS detector for LHC Run-3 and the HL-LHC, alongside anticipated breakthroughs in $b$- and $c$-jet tagging and the application of machine learning to lepton reconstruction, identification and isolation, open an exhilarating new chapter in the search for physics beyond the Standard Model. Although supersymmetric particles have not yet been observed, the prospects for SUSY searches remain strong. With vastly increased luminosity, state-of-the-art detector capabilities, and innovative analysis techniques, we are well-positioned to probe deeper into unexplored regions of parameter space and potentially reveal subtle signals of new physics.

\section*{Acknowledgments}
This work received support from IFIN-HH under Contract ATLAS++ / CERN-RO with the Romanian Ministry of Education and Research.


\bibliographystyle{utphys}
\bibliography{MyBibliography.bib}

\end{document}